\documentclass{ws-m3as}
\usepackage{epsfig,graphics,empheq,graphicx,amsmath,amssymb,amsfonts}
\usepackage{epstopdf}

\usepackage{color}
\usepackage{cases}
\usepackage[position=bottom]{subfig}  
\usepackage{paralist}
\usepackage{enumitem}
\usepackage[percent]{overpic}
\usepackage{tikz}

\newsavebox{\largestimage}

\newcommand{\rev}[2][black]{\textcolor{#1}{#2}}

\def\u{{\bm u}}
\def\v{{\bm{v}}}
\def\x{{\bm{x}}}

\def\0{{\bm 0}}

\def\cl {\nonumber \\}
\def\el {\nonumber}

\DeclareMathOperator*{\argmin}{arg\,min}
 
\begin{document}

\title{A kinetic theory approach for 2D crowd dynamics with emotional contagion}

\author{Daewa Kim}
\address{Department of Mathematics, West Virginia University, \\
94 Beechurst Ave, Morgantown, WV 26506 \\ daewa.kim@mail.wvu.edu}

\author{Kaylie O'Connell}
\address{Department of Mathematics, University of Houston, \\
3551 Cullen Blvd, Houston TX 77204 \\ koconnell@uh.edu}

\author{William Ott}
\address{Department of Mathematics, University of Houston, \\
3551 Cullen Blvd, Houston TX 77204 \\ ott@math.uh.edu}

\author{Annalisa Quaini}
\address{Department of Mathematics, University of Houston, \\
3551 Cullen Blvd, Houston TX 77204 \\ aquaini@uh.edu}

\maketitle

\begin{abstract}
In this paper, we present a computational modeling approach 
for the dynamics of human crowds, where the spreading of an emotion (specifically fear) has an
influence on the pedestrians' behavior. Our approach is based on the methods of the kinetic theory of
active particles. 
The model allows us to weight between two competing behaviors depending on fear level: the search for less congested 
areas and the tendency to follow the stream unconsciously (herding).
The fear level of each pedestrian influences her walking speed and is influenced by the fear levels of her neighbors. 
Numerically, we solve our pedestrian model with emotional contagion using an operator splitting scheme.
We simulate evacuation scenarios involving two groups of interacting pedestrians to assess how domain geometry 
and the details of fear propagation impact evacuation dynamics.
Further, we reproduce the evacuation dynamics of an experimental study involving distressed ants.
\end{abstract}



\section{Introduction} \label{sec:Intro}


Due to the complexity of human behavior, the mathematical modeling of crowd motion has attracted a lot of attention. 
While there exists a large body of literature on this topic, the interest in designing more and more realistic models 
is not declining. This is due to the fact that realistic
modeling of human crowd dynamics could bring great benefit to society. 
In fact, the quantitative and computational analysis associated  with a mathematical model could support e.g.
crisis managers in emergency situations. In this paper, we focus on one important aspect of crowd behavior: 
the effect on crowd dynamics of an emotional contagion (specifically fear) that spreads within the crowd itself.
  
In broad terms, the models of crowd dynamics that have been proposed can be grouped into three categories by scale.
One category is the microscopic, individual-based model inspired by Newtonian mechanics focusing on
inter-particle interaction. For a survey on the \emph{microscopic} approach, we refer to Ref.~\refcite{RevModPhys.73.1067}.
The second category treats crowd dynamics as a continuum flow, an approach that works well for large-scale, dense crowds.
For a survey on this \emph{macroscopic} approach, see e.g. Refs.~\refcite{Hughes2003} and \refcite{Cristiani2014_book}. 

The third category, called \emph{mesoscopic} or \emph{kinetic}, uses a scale of observation in between the previous two.
For introductory concepts, we refer to Refs.~\refcite{Bellomo2013_new} and \refcite{BellomoBellouquid2015}.
Further developed in Refs.~\refcite{Bellomo2017_book,Bellomo2015_new,Bellomo2019_new} and \refcite{kim_quaini},
the mesoscale approach derives a Boltzmann-type evolution equation for the statistical
distribution function of the position and velocity of the pedestrians,
in a framework close to that of the kinetic theory of gases.
There is one key difference though: the interactions in Boltzmann equations 
are conservative and reversible, while the interactions in the kinetic theory of active particles are irreversible, 
non-conservative and, in some cases, nonlocal and nonlinearly additive. 
An important consequence is that often for active particles the
Maxwellian equilibrium does not exist.~\cite{Aristov_2019}

Multiscale approaches have also been investigated. See e.g. Ref.~\refcite{Bellomo2020}
for a multiscale view of human crowds that is consistent at the three possible modeling scales.
Each modeling scale has its advantages and pitfalls. We choose to work with a kinetic-type model because of its flexible ability to account for multiple interactions and heterogeneous behavior among pedestrians.

By and large, existing mathematical models assume that pedestrians behave rationally. However, in emergency situations
people may behave irrationally when responding to fear.
A majority of the current models therefore do not yet effectively support
crowd managers. Several works have shown that social phenomena such as fear transmission are pervasive in crowds
and modify the interaction rules.~\cite{Helbing2000,Helbing2009,PhysRevE.75.046109,Moussad2755,RONCHI201611,WIJERMANS2016142,WIJERMANS2013} 
One of the first works on a multiscale (from microscopic to macroscopic) approach to crowd dynamics
with emotional contagion is Ref.~\refcite{Bertozzi2015}. Therein, fear is propagated by a BGK-like model
(see e.g. Ref.~\refcite{Cercignani1994_book}) and results are limited to one space dimension. 
Studies on the impact of social dynamics on individual interactions and their influence at a higher scale
are carried out in Refs.~\refcite{Degond2013,Degond2017}.
More recently, a kinetic approach to modeling pedestrian dynamics in the presence of social phenomena
(e.g. propagation of stress) is presented in Ref.~\refcite{Bellomo2019_new}. The numerical results in Ref.~\refcite{Bellomo2019_new}
show that stress propagation significantly affects crowd density patterns and overall crowd dynamics. 

In this paper, we extend the kinetic-type model in Ref.~\refcite{kim_quaini} to account for the propagation 
of stress conditions in time and space. To this end, we draw inspiration from the 1D model in 
Ref.~\refcite{Bertozzi2015}. The results produced by the model in Ref.~\refcite{kim_quaini} 
have been compared against data collected in a recent empirical study reported in Ref.~\refcite{Kemloh}.
We have shown that for medium-sized groups of people, the average population
density and flow rate computed with our kinetic-type model agree with the respective measured quantities in the absence of fear.
The model in Ref.~\refcite{Bertozzi2015} adapts the model
in Refs.~\refcite{Bertozzi2014ContagionSI,688c8c70f9c748e486b48806e7ce2fdb} and has been designed to track the level of fear within individuals under the assumption that fear influences motion.
The numerical results produced by the agent-based model in Ref.~\refcite{688c8c70f9c748e486b48806e7ce2fdb}
have compared more favorably to actual crowd footage than other pedestrian models. See Ref.~\refcite{inproceedings}. 

We assess our kinetic model with emotional contagion through three test cases. 
Our first two tests involve two groups of pedestrians with different initial fear distributions.
We have set up an evacuation scenario wherein the two groups interact as they move toward the exit.
We study how the evacuation dynamics depend on domain geometry (specifically the exit size) 
and on the spatial length scale over which fear can be directly transmitted.
Our results for these two tests confirm what observed in  Ref.~\refcite{Bellomo2019_new}:
the spreading of a stress condition significantly affects the overall evacuation dynamics and leads to overcrowding
in certain areas of the domain. 
Our third test demonstrates that our model can recapitulate the results of experimental ant panic studies.

The paper is organized as follows. In Sec.~\ref{sec:simplified_m}, we present a 2D extension of the model 
in Ref.~\refcite{Bertozzi2015} under one simplifying assumption: pedestrians change walking speed 
because of the spreading of fear, but they do not alter walking direction.
We discretize the simplified model and present numerical results.
In Sec.~\ref{sec:model}, we remove the simplifying assumption and present our full 2D model, where fear affects both walking speed and direction.
Sec.~\ref{sec:num_meth} describes a splitting algorithm for the model in Sec.~\ref{sec:model} and 
Sec.~\ref{sec:num_res} presents our numerical results for the three test cases mentioned above. 
We draw conclusions in Sec.~\ref{sec:concl}.

\section{A simplified two-dimensional kinetic model}\label{sec:simplified_m}

We extend the one-dimensional emotional contagion model in Ref.~\refcite{Bertozzi2015} to 2D
with one important simplifying assumption: the pedestrians have a common walking direction $\theta$ that does not change over time. This assumption will be subsequently removed (see Section~\ref{sec:model}).

We start from an agent-based model at the microscopic level. Let particle $n$ represent a pedestrian, whose
position and fear level are given by $\x_n(t)=(x_n(t), y_n(t))^T$ and $q_n(t)$, respectively. 
In addition, let $q_n^*$ be an average fear level \rev{perceived by pedestrian $n$}. 
The microscopic model reads:
\begin{align}
&\frac{d\x_n}{dt}=q_n(\cos\theta, \sin\theta)^T, \quad \frac{dq_n}{dt}=\gamma(q_n^*-q_n), \cl 
&q_n^* =\frac{\sum_{m=1}^N \kappa_{n,m}q_m}{\sum_{m=1}^N\kappa_{n\rev{,}m}}, \quad n= 1,2,3, \dots, N, \label{eq:diseasef}
\end{align}
where $N$ is the total number of particles. This model assumes that: 
(i) \rev{at a given time an agent} walk\rev{s} faster if more scared \rev{(see
the first equation in \eqref{eq:diseasef} where the walking speed is equal to the fear level)} and 
(ii) \rev{agents} tend to equilibrate their fear level with the average fear level
\rev{they perceive}.
In model \eqref{eq:diseasef}, the interaction kernel $\kappa_{n,m}$ serves as the weights in the average $q_n^{\ast}$. This 
kernel is a decreasing function of mutual distance between two particles and is parametrized by an interaction distance $R$:
\begin{align} \label{kappa}
\kappa_{n,m} = \kappa(|\x_n-\x_m|) = \dfrac{R}{(|\x_n-\x_m|^2+R^2) \pi}.
\end{align}
Parameter $\gamma$ in \eqref{eq:diseasef} describes the contagion interaction strength 
and it may vary from particle to particle for more general cases. 

From the agent-based model \eqref{eq:diseasef}, we derive a model at the kinetic level. 
Denote the empirical distribution by
\begin{equation}
h^N=\frac{1}{N}\sum_{n=1}^N\delta(\x-\x_n(t))\delta(q-q_n(t)), \cl
\end{equation}
where $\delta$ is the Dirac delta measure.
We assume that the particles remain in a fixed compact domain $(\x_n(t), q_n(t)) \in \Omega \subset \mathbb{R}^3$ for all $n$ and for the entire time interval under consideration.
Prohorov's theorem implies that the sequence $\{h^N\}$ is relatively compact in the weak$^{\ast}$ sense. 
Therefore, there exists a subsequence $\{h^{N_k}\}_k$ such that $h^{N_k}$ converges to $h$ with weak$^{\ast}$-convergence in $\mathcal{P}(\mathbb{R}^3$) and pointwise convergence in time as $k \rightarrow \infty$. 
Here, $\mathcal{P}(\mathbb{R}^3$) denotes the space of probability measures on $\mathbb{R}^3$.

Consider a test function $\psi \in C_0^1(\mathbb{R}^3)$. We have
\begin{align}\label{eq:testfunction}
\frac{d}{dt} \langle h^N, \psi \rangle_{\x,q} 
&=\frac{d}{dt} \biggl< \frac{1}{N}\sum_{n=1}^{N}\delta(\x-\x_n(t))\delta(q-q_n(t)), \psi \biggr>_{\x,q} \cl
&=\frac{d}{dt} \frac{1}{N}\sum_{n=1}^{N}\psi(\x_n(t), q_n(t)) \cl
&=\frac{1}{N}\sum_{n=1}^{N} \psi_x q_n \cos \theta +\psi_y q_n \sin \theta + \psi_q\gamma(q_n^{\ast}-q_n) \cl
&= \langle \psi_x q \cos \theta, h^N \rangle_{\x,q} + \langle \psi_y q \sin \theta, h^N \rangle_{\x,q} \cl
&\quad + \frac{\gamma}{N}\sum_{n=1}^{N}\psi_q \biggl( \frac{\sum_{m=1}^{N} \kappa_{n,m}q_n}{\sum_{m=1}^{N}\kappa_{n,m}}- q_n \biggr),
\end{align} 
where $\langle \cdot \rangle_{\x,q}$ means integration against both $\x$ and $q$
\rev{and the subindex of $\psi$ denotes the variable with respect to which we take the partial derivative
of $\psi$}. 

Let us define
\begin{equation}
\rho_{h^N}(\x)=\frac{1}{N}\sum_{n=1}^{N}\delta(\x-\x_n) \cl
\end{equation}
and
\begin{equation}
m_{h^N}(\x)=\biggl< q, \frac{1}{N}\sum_{m=1}^{N}\delta(\x-\x_m)\delta(q-q_m) \biggr>_{\x,q}
=\frac{1}{N}\sum_{m=1}^{N}\delta(\x-\x_m)q_m, \cl
\end{equation}
We have
\begin{equation}
\frac{1}{N}\sum_{m=1}^{N}\kappa(|\x_n-\x_m|)= \biggl< \kappa(|\x_n-\tilde{\x}|), \frac{1}{N}\sum_{m=1}^{N}\delta(\tilde{\x}-\x_m) \biggr>_{\x}
=\kappa \ast\rho_{h^{N}}(\x_n), \cl
\end{equation}
\begin{equation}
\frac{1}{N}\sum_{m=1}^{N}\kappa(|\x_n-\x_m|)q_m= \biggl< \kappa(|\x_n-\tilde{\x}|), \frac{1}{N}\sum_{m=1}^{N}\delta(\tilde{\x}-\x_m)q_m \biggr>_{\x}
=\kappa \ast m_{h^{N}}(\x_n), \cl
\end{equation}
where $\langle \cdot \rangle_\x$ means integration only in $\x$.
Then, we can rewrite eq.~($\ref{eq:testfunction}$) as
\begin{equation}
\frac{d}{dt} \langle h^{N}, \psi \rangle_{\x,q}= \langle \psi_x q \cos \theta, h^N \rangle_{\x,q} + \langle \psi_y q \sin \theta, h^N \rangle_{\x,q}
+ \gamma \biggl< h^{N}, \frac{\kappa \ast m_{h^N}}{\kappa \ast \rho_{h^N}}\psi_{q}-q\psi_{q} \biggr>_{\x,q}, \cl
\end{equation}
which leads to
\begin{equation}\label{eq:Nkineticsystem}
h_t^N+\nabla \cdot (q(\cos\theta, \sin\theta)^T h^N)=\gamma((q-q^\ast)h^N)_q, 
\end{equation}
via integration by parts. 

Now letting $k \rightarrow \infty$, the subsequence $h^{N_k}$ formally leads to the limiting kinetic equation
\begin{equation} \label{eq:2kineticeq}
h_t+\nabla \cdot (q(\cos\theta, \sin\theta)^T  h)=\gamma((q-q^\ast)h)_q, 
\end{equation}
where $h(t,\x,q)$ is the probability of finding a person \rev{in an infinitesimal neighborhood of $(\x,q)$}.
The quantity $q^*(t,\x)$ is the local \emph{average} emotional contagion level weighted by the distance to $\x$:
\begin{equation} \label{q_act}
q^{\ast}(t,\x)= \frac{\iint\kappa(|\x-\overline{\x}|) h(t,\overline{\x},q)qdqd\overline{\x}}{\iint \kappa(|\x-\overline{\x}|)h(t,\overline{\x},q)dqd\overline{\x}}.
\end{equation}

\subsection{Full discretization}\label{sec:discr_Bertozzi}

In this section, we present the numerical discretization of the kinetic equation (\ref{eq:2kineticeq}).

Let us start from the space discretization. 
Divide the spatial and velocity domain into a number of cells 
$[x_{i-\frac{1}{2}}, x_{i+\frac{1}{2}}] \times [y_{j-\frac{1}{2}}, y_{j+\frac{1}{2}}]$ 
of the length $\Delta x$ and $\Delta y$. 
The discrete mesh points $x_{i}$ and $y_j$ are given by
\begin{align}\label{eq:x_i_y_j} 
x_i =i \Delta x, \quad x_{i+1/2}=x_{i}+ \frac{\Delta x}{2}=\Big(i+\frac{1}{2}\Big)\Delta x, \el \\
y_j =j \Delta y, \quad y_{j+1/2}=y_{j}+ \frac{\Delta y}{2}=\Big(j+\frac{1}{2}\Big)\Delta y,
\end{align}
for $i= 0, 1, \dots, N_x$ and $j = 0, 1, \dots, N_y$,
The contagion level domain is partitioned into subdomains $[q_{k-\frac{1}{2}}, q_{k+\frac{1}{2}}]$ and the length of $\Delta q$ with $k \in 1,2,\dots, N_q$.
\begin{align}
q_k =k \Delta q, \quad q_{k+1/2}=q_{k}+ \frac{\Delta q}{2}=\Big(k+\frac{1}{2}\Big)\Delta q. \label{eq:q_k} 
\end{align}
Here $N_x$, $N_y$ and $N_q$ are the total number of points in $x-$, $y-$ and $q-$ directions, respectively. 
For simplicity, we assume each cell is centered at $x_{i}$, $y_{j}$ or $q_k$ with a uniform length $\Delta x$, $\Delta y$ and $\Delta q$.

Let us denote $h_{i, j, k}=h(t, x_{i}, y_{j}, q_k)$. The average fear level 
is computed using a midpoint rule for the integrals in \eqref{q_act}:
\begin{align}
q^{\ast}(t,x_i, y_j) \approx q^{\ast}_{i,j} &= \frac{\sum_k \sum_{\overline{j}} \sum_{\overline{i}} \kappa_{\overline{i},\overline{j}} h_{\overline{i}, \overline{j},k} q_k \Delta q \Delta x \Delta y}{\sum_k \sum_{\overline{j}} \sum_{\overline{i}} \kappa_{\overline{i},\overline{j}} h_{\overline{i}, \overline{j},k} \Delta q \Delta x \Delta y}, \label{q_sum} \\
\kappa_{\overline{i},\overline{j}} &= \frac{R}{((x_i - x_{\overline{i}})^2+(y_j - y_{\overline{j}})^2+R^2)\pi}. \el
\end{align}

We consider the following scheme for eq.~\eqref{eq:2kineticeq}:
\begin{align} \label{eq:disease_modified}
\partial_t h_{i, j, k}&+\frac{\overline{\eta}_{i, j, k}-\overline{\eta}_{i-1, j, k}}{\Delta x}
+\frac{\overline{\phi}_{i, j, k}-\overline{\phi}_{i, j-1, k}}{\Delta y} \cl
&+\gamma \frac{\xi_{i, j, k+\frac{1}{2}}-\xi_{i, j, k-\frac{1}{2}}}{\Delta q}
+\gamma \frac{C_{i, j, k+\frac{1}{2}}-C_{i, j, k-\frac{1}{2}}}{\Delta q}= 0,
\end{align}
where
\begin{align}
\overline{\eta}_{i, j, k} & =\eta_{i, j, k} + \frac{\eta_{i+1, j, k} - \eta_{i, j, k}}{2} \varphi\left( \frac{\eta_{i, j, k} - \eta_{i-1, j, k}}{\eta_{i+1, j, k} - \eta_{i, j, k}} \right), ~\eta_{i, j, k}  =q_k \cos \theta \,h_{i,j,k}, \label{eta_bar}\\
\overline{\phi}_{i, j, k} &= \phi_{i, j, k} + \frac{\phi_{i, j+1, k} - \phi_{i, j, k}}{2} \varphi\left( \frac{\phi_{i, j, k} - \phi_{i, j-1, k}}{\phi_{i, j+1, k} - \phi_{i, j, k}} \right),~\phi_{i, j, k} =q_k \sin \theta\,h_{i,j,k}, \label{phi_bar} \\
\xi_{i, j, k+\frac{1}{2}} &
 = \dfrac{| q_{i, j}^\ast - q_{k+\frac{1}{2}} |+(q_{i, j}^\ast - q_{k+\frac{1}{2}})}{2}h_{i, j, k} 
+\dfrac{( q_{i, j}^\ast - q_{k+\frac{1}{2}} ) - | q_{i, j}^\ast - q_{k+\frac{1}{2}} |}{2}h_{i, j, k+1} \cl
C_{i, j, k+\frac{1}{2}} & =\dfrac{1}{2} \left| s_{i, j,k+\frac{1}{2}} \right| \left(1-\dfrac{\Delta t}{\Delta q} \left| s_{i, j, k+\frac{1}{2}} \right| \right)
W_{i, j,k-\frac{1}{2}}\varphi \left( \dfrac{W_{i,j,\textbf{k}-\frac{1}{2}}}{W_{i, j, k-\frac{1}{2}}} \right) 
\label{eq:eta_xi_C}
\end{align}
with  
\begin{align}\label{eq:s_W}
q_{k+\frac{1}{2}} = \frac{q_k + q_{k+1}}{2}, \quad
s_{i, j, k+\frac{1}{2}}= q^{\ast}_{i, j}-q_{k+\frac{1}{2}}, \quad W_{i, j, k-\frac{1}{2}}=h_{i, j,k}-h_{i, j,k-\frac{1}{2}}. 
\end{align}
In eq.~\eqref{eta_bar} and \eqref{phi_bar}, $\varphi$ is a slope limiter function. We choose the Van Leer function:
\begin{equation}\label{van_leer}
\varphi(\theta)=\frac{|\theta|+\theta}{1+|\theta|}.
\end{equation}
In eq.~\eqref{eq:eta_xi_C}, the subscript \textbf{k} is k-1 if $s_{i, j, k-\frac{1}{2}} > 0$ and k+1 if $s_{i, j, k-\frac{1}{2}} < 0$. 
This scheme is second-order in velocity thanks to the use of the flux limiter.

Next, let us discretize in time.
Let $\Delta t \in \mathbb{R}$, $t^n = t_0 + n \Delta t$, with $n = 0, ..., N_T$ and $T = t_0 + N_T \Delta t$,
where $T$ is the end of the time interval under consideration. 
Moreover, we denote by $y^n$ the approximation of a generic quantity $y$ at the time $t^n$. We adopt the forward Euler scheme for the time discretization of problem \eqref{eq:disease_modified}
\begin{align} \label{eq:full_discretization}
h_{i, j, k}^{l+1} = h_{i, j, k}^l - \Delta t
\Big( & \dfrac{\overline{\eta}^l_{i, j, k}-\overline{\eta}^l_{i-1, j, k}}{\Delta x}
+\dfrac{\overline{\phi}^l_{i, j, k}-\overline{\phi}^l_{i, j-1, k}}{\Delta y} \cl
+& \gamma \dfrac{\xi^l_{i, j, k+\frac{1}{2}}-\xi^l_{i, j, k-\frac{1}{2}}}{\Delta q}
+\gamma \dfrac{C^l_{i, j, k+\frac{1}{2}}-C^l_{i, j, k-\frac{1}{2}}}{\Delta q} \Big)
\end{align}

The time step $\Delta t$ is chosen as 
\begin{align}\label{eq:CFL}
\Delta t = \frac{1}{2}\min\Bigg\{ \frac{\Delta x}{\max_{k}q_{k}},  \frac{\Delta y}{\max_{k}q_{k}},
\frac{\Delta q}{2\gamma \max_{k}q_{k}} \Bigg\}
\end{align} 
to satisfy the Courant-Friedrichs-Lewy (CFL) condition.

\subsection{Numerical results}

The computational domain in the $xy$-plane is $[-10, 10] \times [-10, 10]$ and the fear level $q$ is in $[0, 1]$. 
We choose $\theta = \pi/4$ as walking direction. 
For the space discretization, we choose $\Delta x = \Delta y=0.05$ 
(we note that these are dimensional values) and $\Delta q=0.005$. 
We set $R = 0.0002$ and $\gamma = 100$. These relatively small interaction radii and 
rather large interaction strength (i.e.~quick interactions) are meant to model a dense crowd setting. 
The time interval under consideration is $[0, 1]$ and the time step $\Delta t$ is set to $1.25 \cdot 10^{-5}$.

We approximate the delta function as follows:
\begin{equation}
\delta(q) \sim E(q) = \cfrac{1}{\sqrt{\pi}R_0} e^{-\cfrac{q^2}{R_0^2}}, \quad R_0 =0.04.
\end{equation}

Following Ref.~\refcite{Bertozzi2015}, the initial condition for the distribution density is set as: 
\begin{align}
h(0, \x, q) & = \dfrac{1}{2}\left( \sin\left(\cfrac{\pi x}{10} \right)+2 \right)\left( \cfrac{1}{4}\,\,E(q-q_I(x)-0.5)+\cfrac{3}{4}\,\,E(q-q_I(x)+0.3) \right)  \nonumber \\
&+   \dfrac{1}{2}\ \left( \sin\left(\cfrac{\pi y}{10} \right)+2 \right)\left( \cfrac{1}{4}\,\,E(q-q_I(y)-0.5)+\cfrac{3}{4}\,\,E(q-q_I(y)+0.3) \right),
\nonumber 
\end{align}
where
\begin{equation}
q_I(0,x)=\cfrac{1}{2}\Big(3-\tanh x \Big).
\nonumber 
\end{equation}

\begin{figure}[h!t]
\centering
\begin{overpic}[width=0.32\textwidth, grid=false]{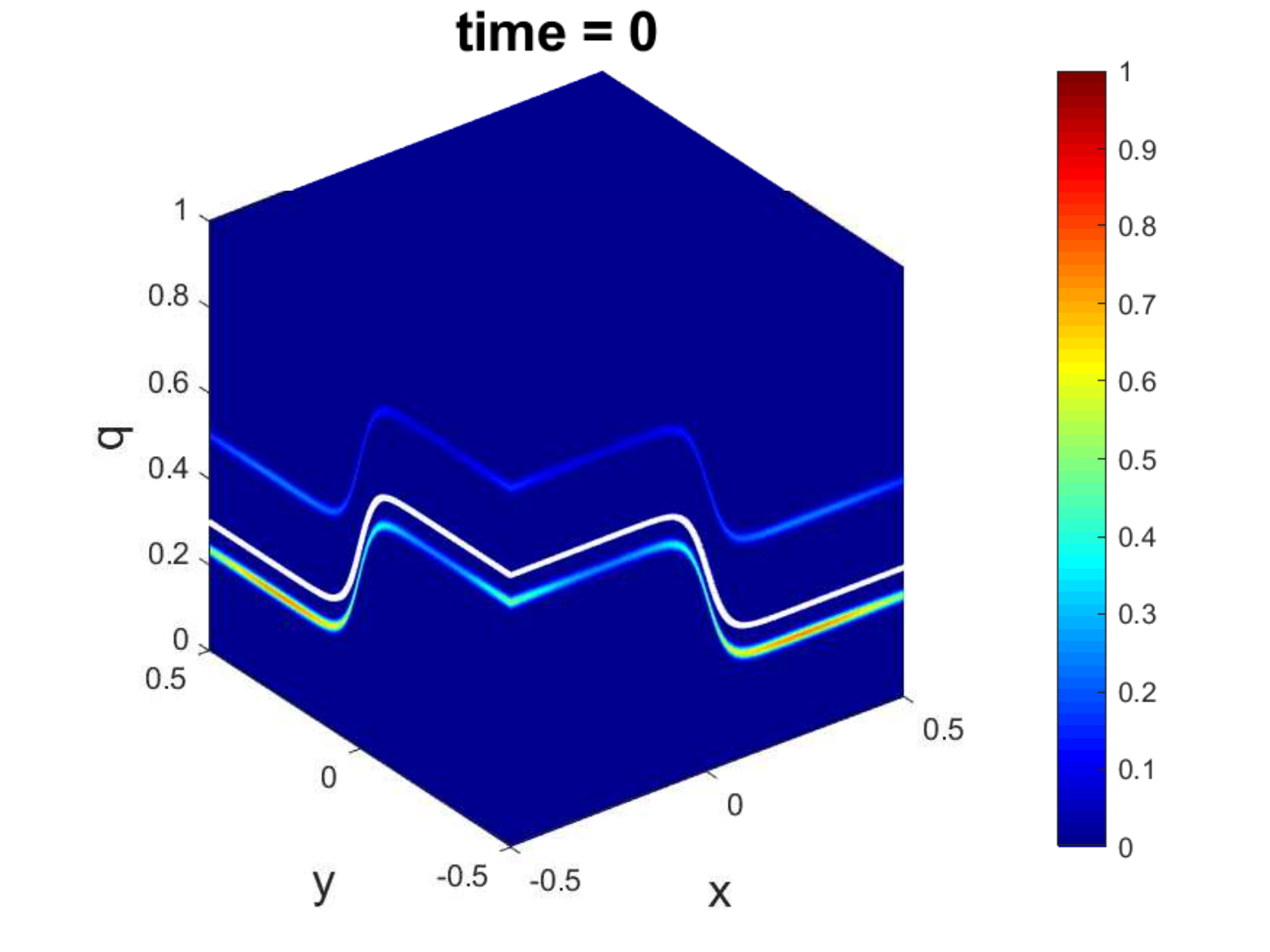}\end{overpic}
\begin{overpic}[width=0.32\textwidth, grid=false]{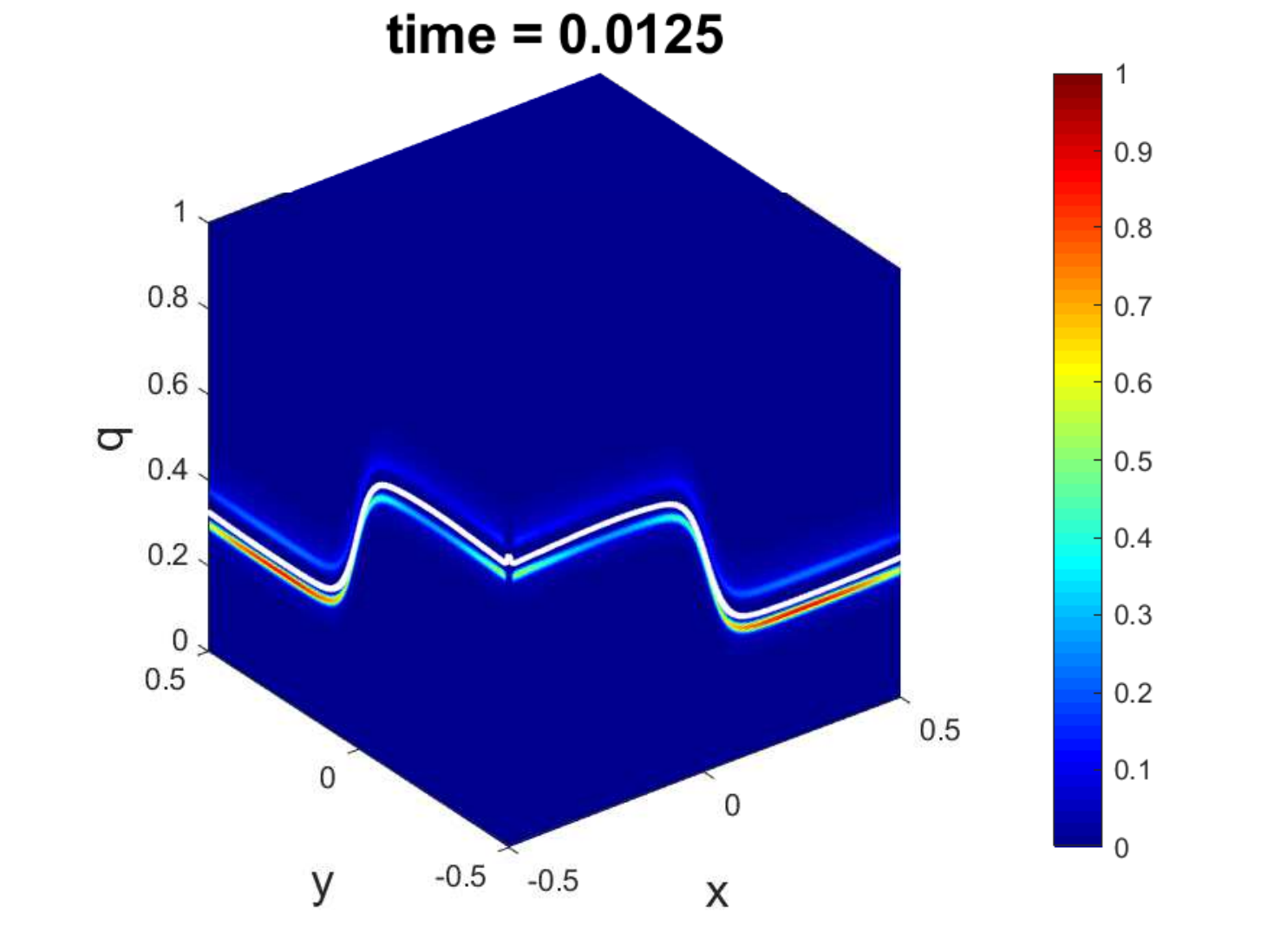}\end{overpic}
\begin{overpic}[width=0.32\textwidth, grid=false]{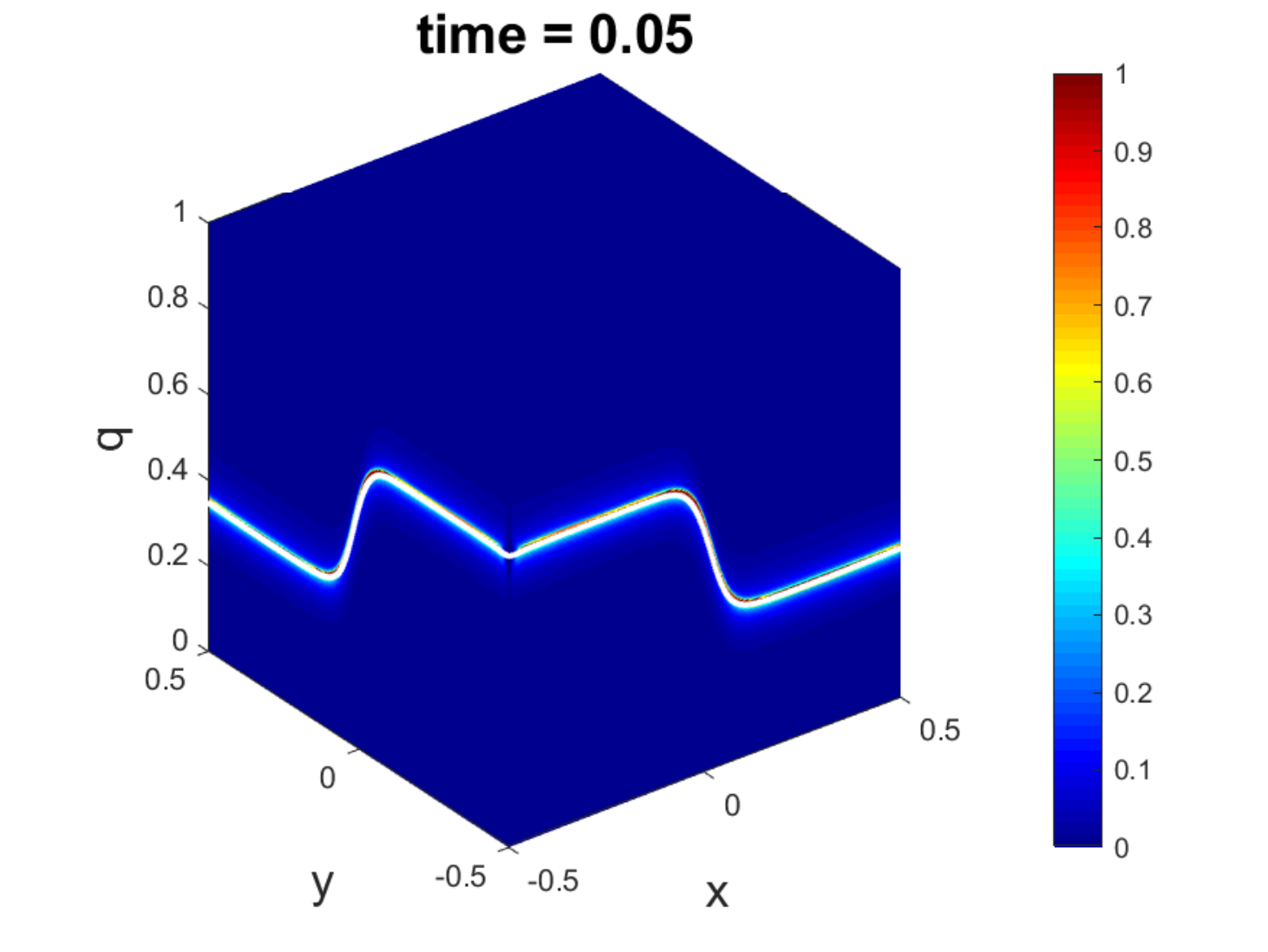}\end{overpic}
\begin{overpic}[width=0.32\textwidth, grid=false]{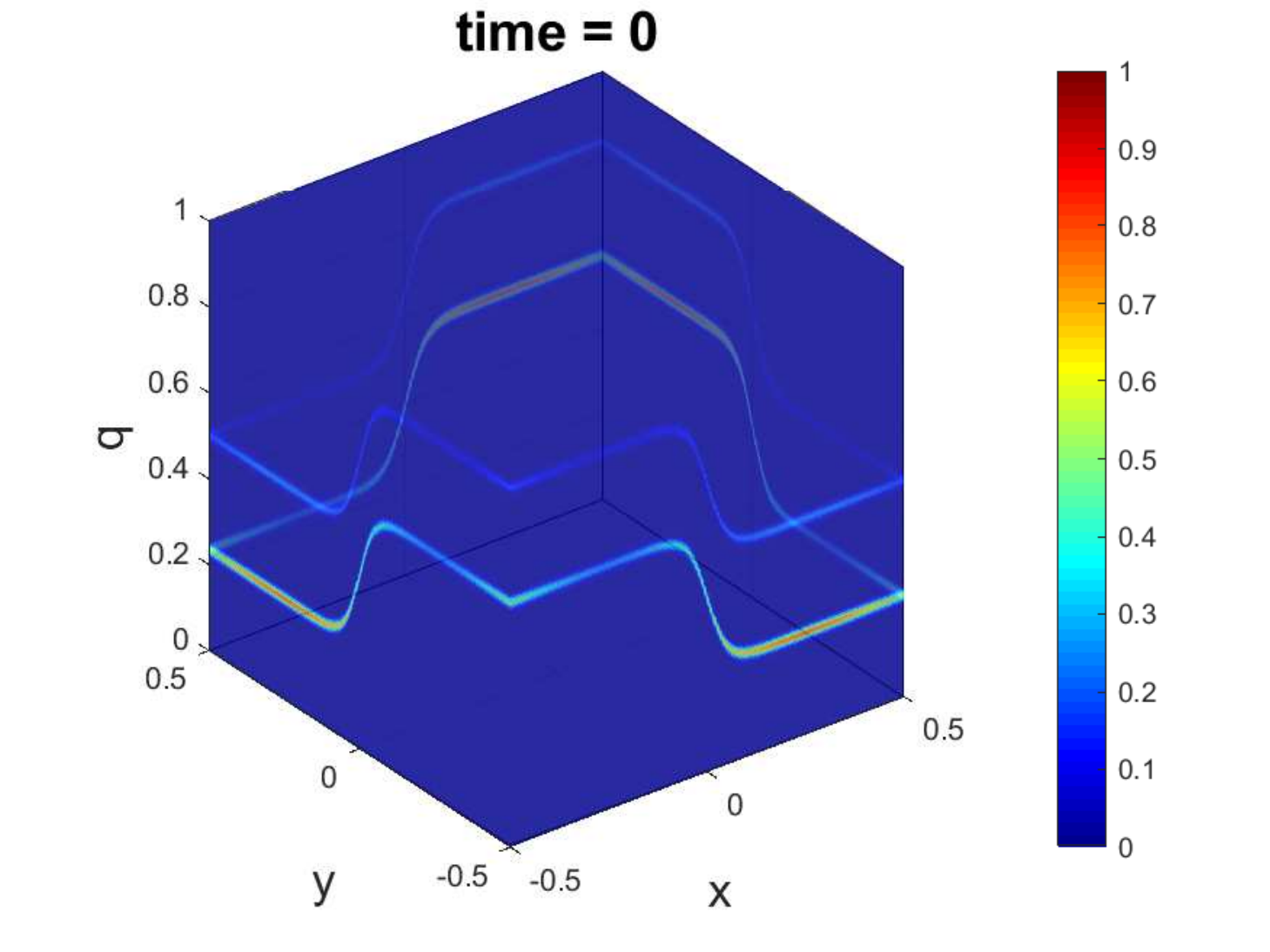}
\put(47.4,35.6){\color{green}\line(0,1){33.7}}
\put(47.2,69.2){\color{green}\line(-1,-4){6.9}}
\put(40.3,8.3){\color{green}\line(0,1){33.5}}
\put(47.4,35.6){\color{green}\line(-1,-4){6.8}}
\end{overpic}
\begin{overpic}[width=0.32\textwidth, grid=false]{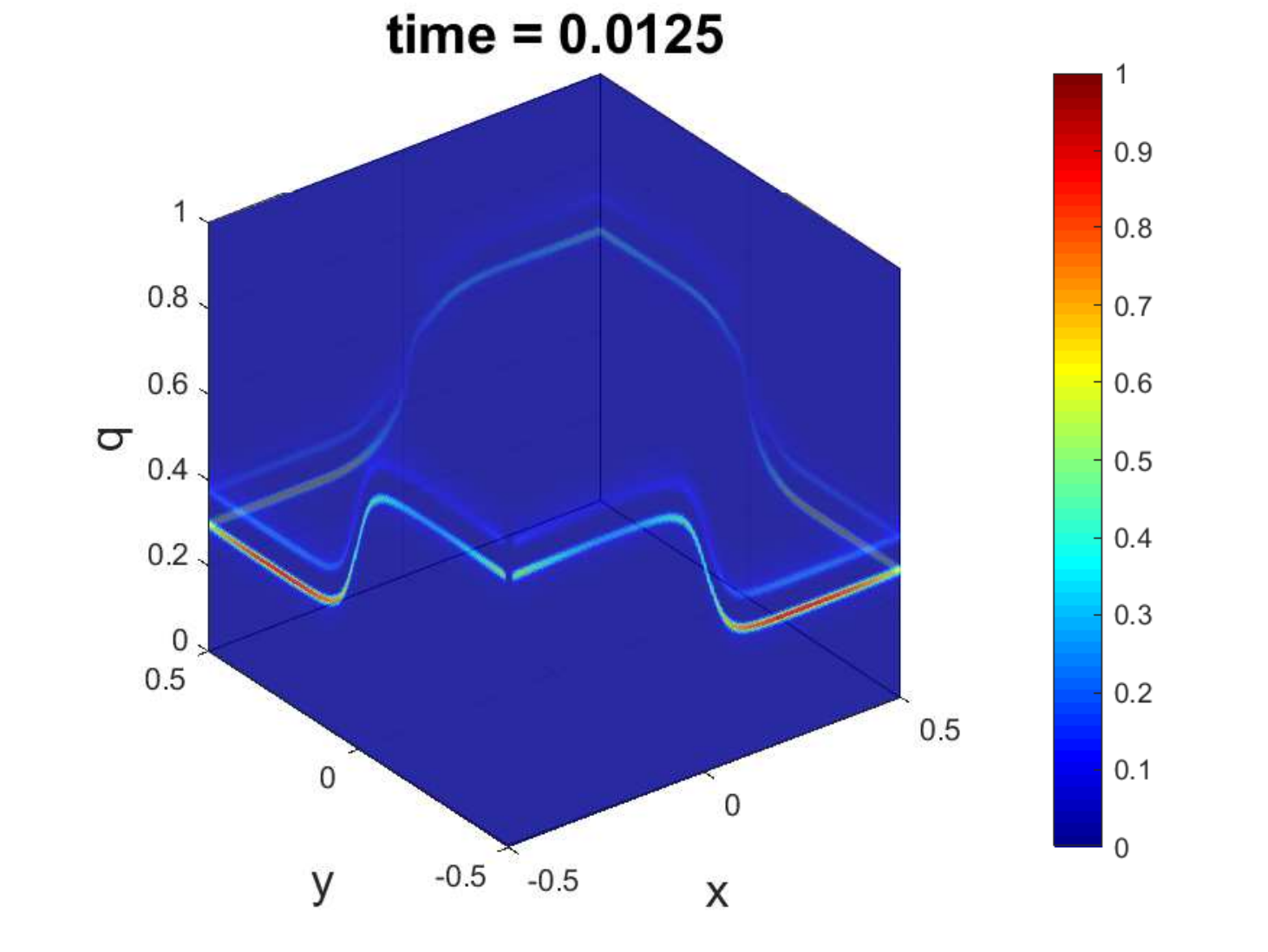}
\put(47.4,35.6){\color{green}\line(0,1){33.7}}
\put(47.2,69.2){\color{green}\line(-1,-4){6.9}}
\put(40.3,8.3){\color{green}\line(0,1){33.5}}
\put(47.4,35.6){\color{green}\line(-1,-4){6.8}}
\end{overpic}
\begin{overpic}[width=0.32\textwidth, grid=false]{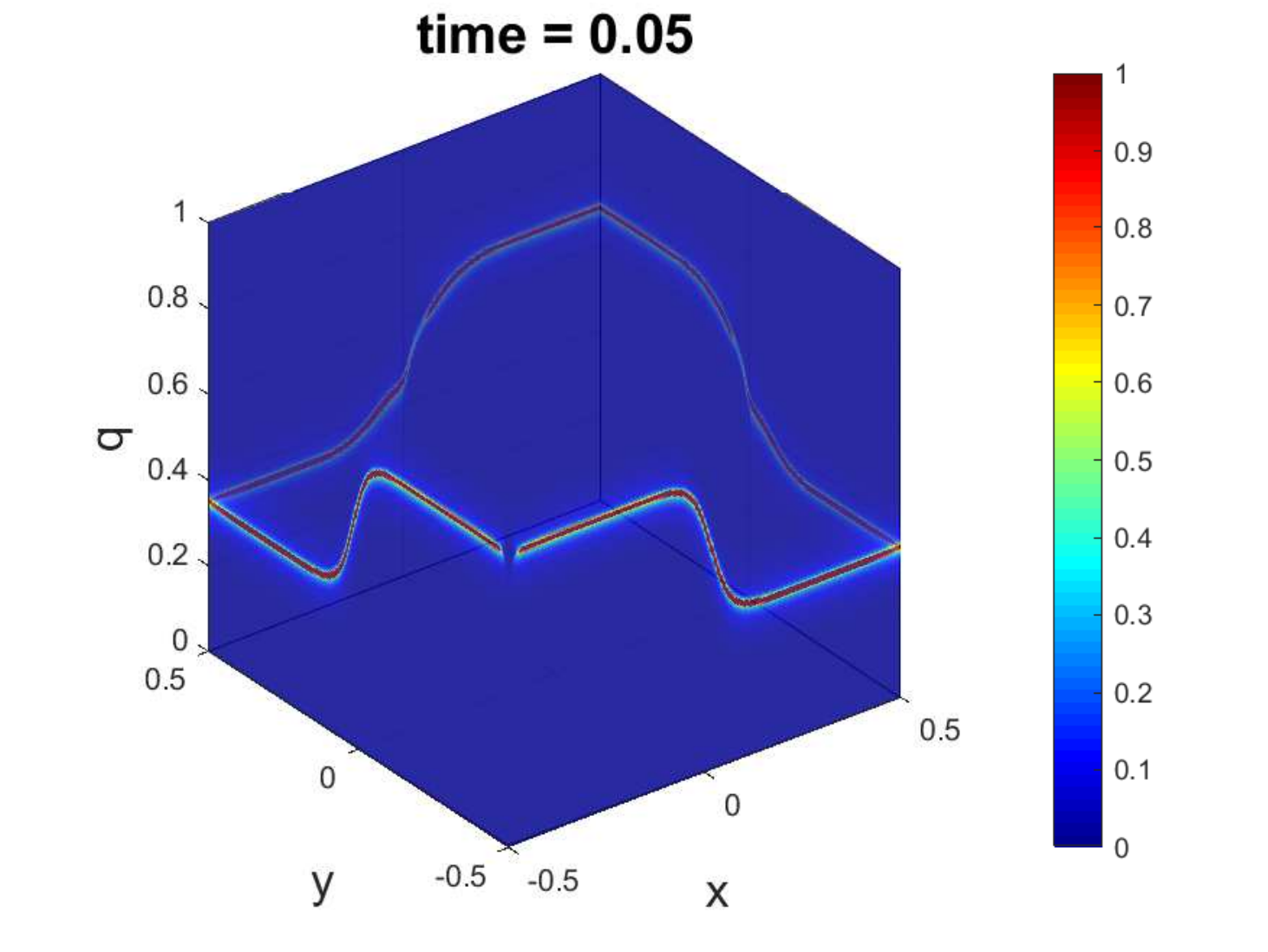}
\put(47.4,35.6){\color{green}\line(0,1){33.7}}
\put(47.2,69.2){\color{green}\line(-1,-4){6.9}}
\put(40.3,8.3){\color{green}\line(0,1){33.5}}
\put(47.4,35.6){\color{green}\line(-1,-4){6.8}}
\end{overpic}
\begin{overpic}[width=0.32\textwidth, grid=false]{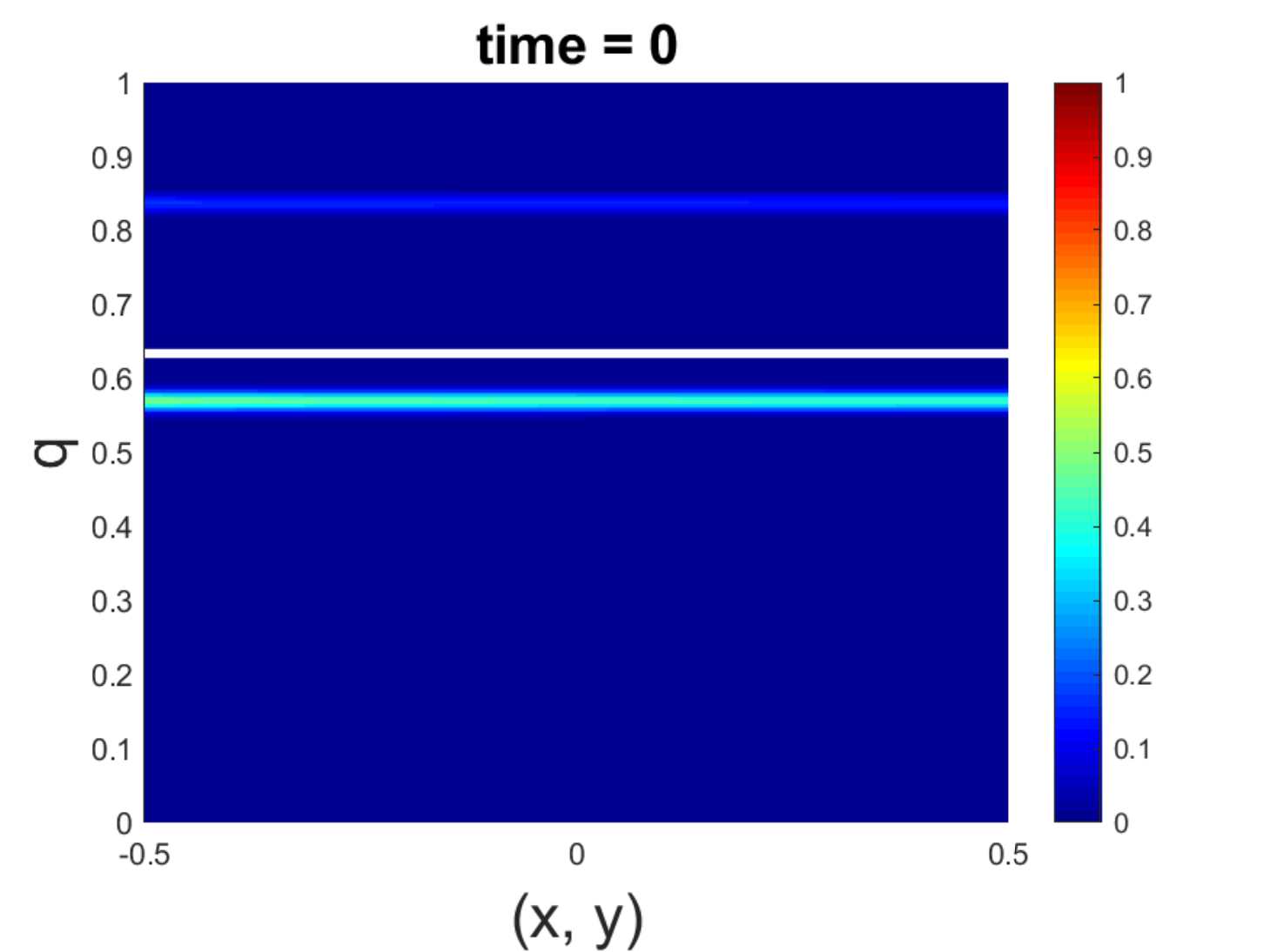}
\end{overpic}
\begin{overpic}[width=0.32\textwidth, grid=false]{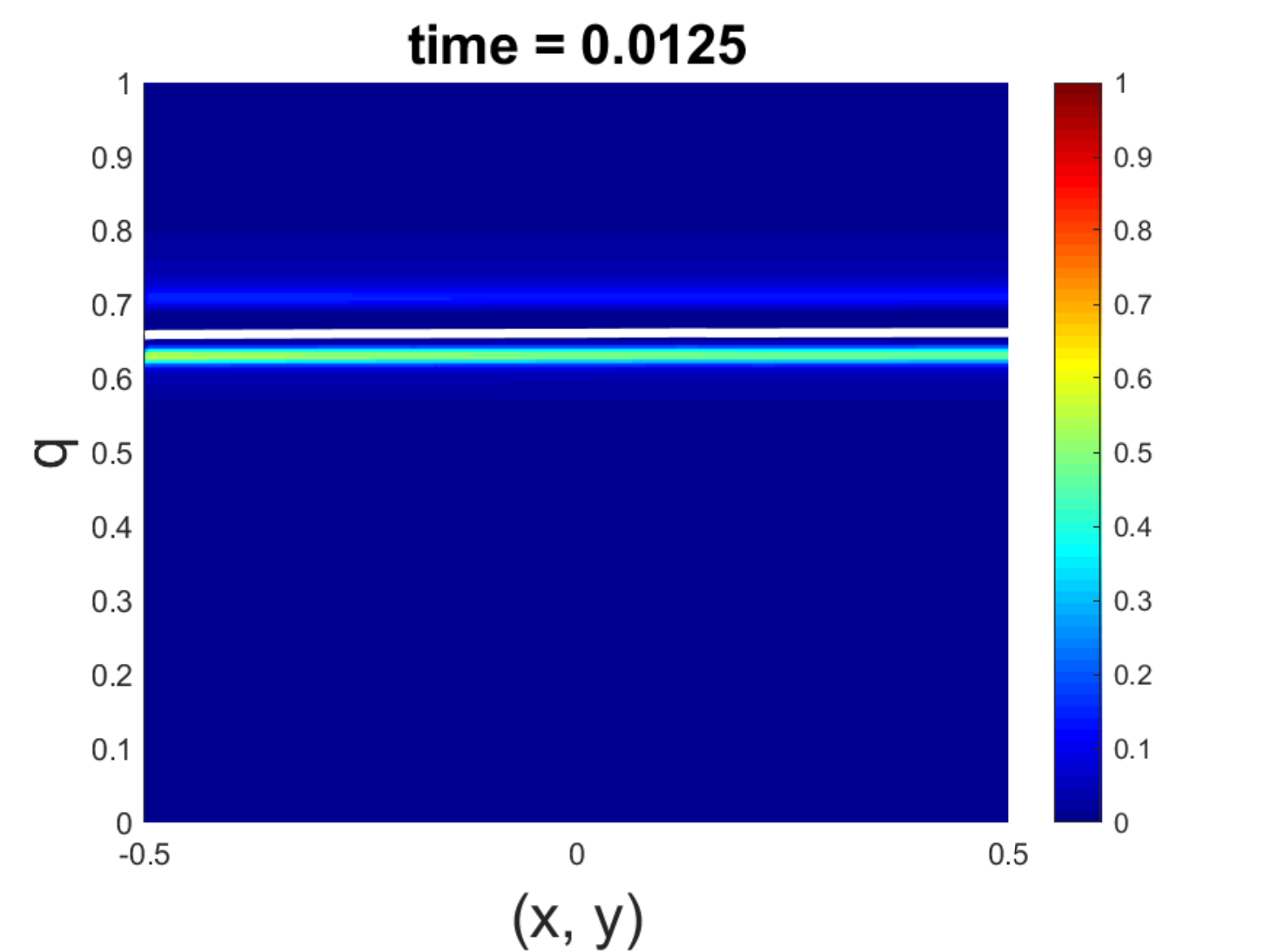}\end{overpic}
\begin{overpic}[width=0.32\textwidth, grid=false]{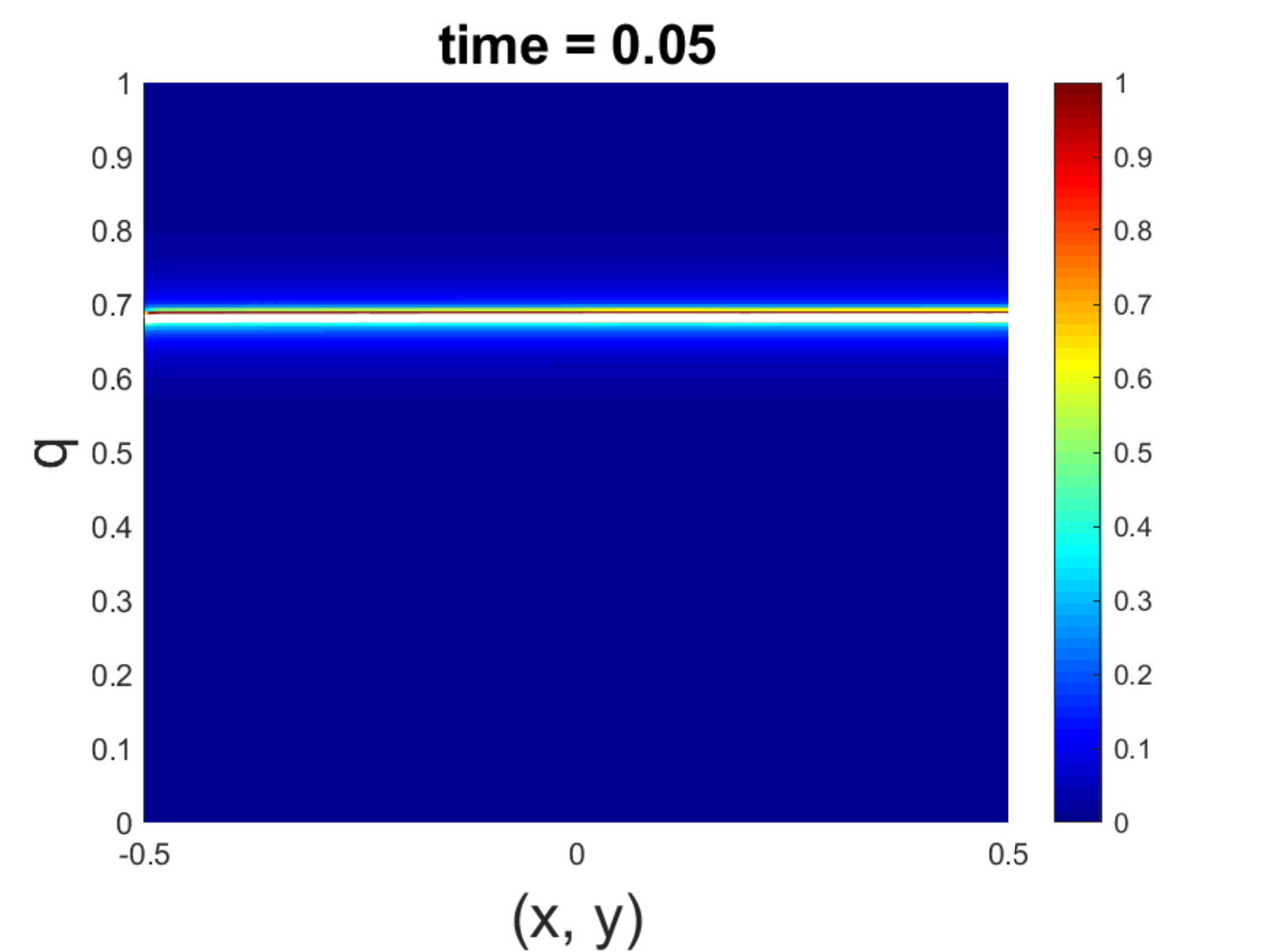}\end{overpic}\caption{Computed distribution density $h$ at time $t = 0$ (left), $t = 0.0125$ (center), $t = 0.5$ (right). 
The top panels show one view of the computed $h$, the middle panels show the same view but with 
a reduced the level of opacity to see through and highlight\rev{ed} in green the section whose results are 
reported in the bottom panels. 
The white line in all the views represents the average $ q^\ast$.}
\label{2d_two_initial_contagion}
\end{figure}
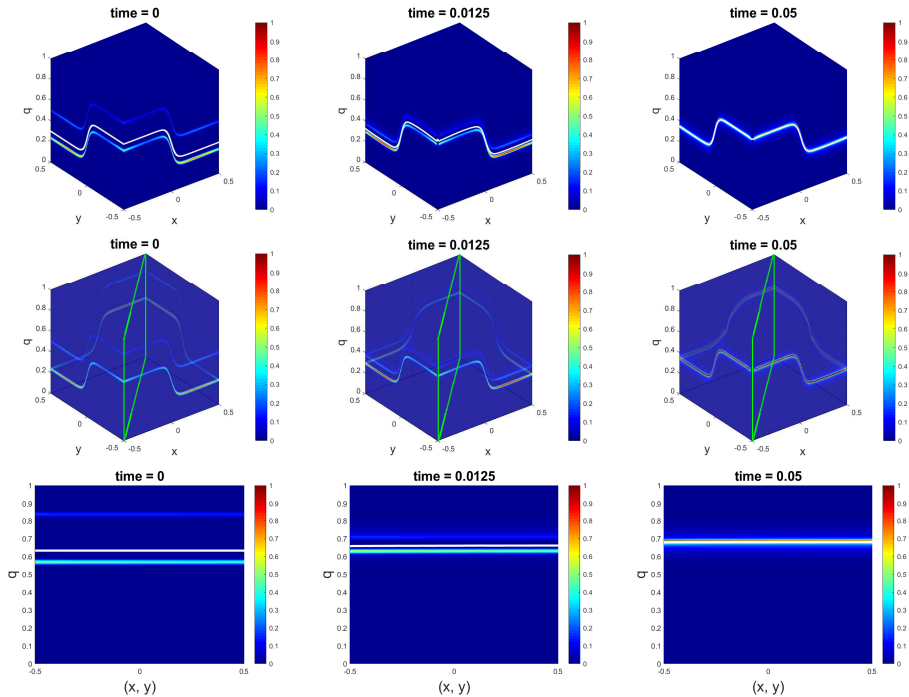

Figure \ref{2d_two_initial_contagion} shows the simulated evolution of $h(t,\x,q)$. Initially, $h$ has two bumps in $q$ for every $\x$, as
shown in the panels on the left. As time evolves, $h$ starts to concentrate on $q^*(t,\x)$, as confirmed
by the panels on the right.

\section{Our full two-dimensional emotional contagion model}\label{sec:model}

The model we develop is based on the model proposed in Ref.~\refcite{Agnelli2015}.

Let  $\Omega \subset \mathbb{R}^2$ denote a bounded domain. 
We assume that pedestrians head to the exit $E$ within the domain.
The case of multiple exits (i.e., $E$ is the finite union of disjoint sets)
can be handled too. See, e.g., Ref.~\refcite{kim_quaini}. At the boundary 
of the domain and within the domain, we have walls $W$.
Let $\x=(x,y)$ denote position and ${\v}=q (\cos \theta, \sin \theta)$ denote  
velocity, where the velocity modulus $q$ corresponds to the fear level
(like in Section~\ref{sec:simplified_m}), and $\theta$ is the velocity direction.
Unlike Section~\ref{sec:simplified_m}, here pedestrians can change their walking direction
as explained in what follows.
For a system composed \rev{of} a large number of pedestrians distributed inside  
$\Omega$, the distribution function is given by 
\[ 
f= f(t, \x, q, \theta)\quad \text{for all} \,\,\, t \ge 0,  \,\, \x \in \Omega,  \,\,  q \in [0,1],  \,\,  \theta \in [0, 2\pi).
\]
Under suitable integrability conditions, $f(t, \x, q, \theta)d \x d q d \theta$ represents the number of individuals who, at time $t$, 
are located in the infinitesimal rectangle $[x, x+dx] \times [y, y+dy]$ and have a velocity belonging to $[q, q + dq] 
\times [\theta, \theta+d\theta]$. 

Following Ref.~\refcite{Agnelli2015}, we assume that the crowd size is not enough to
justify the continuity of the distribution function over the variable $\theta$. Thus, 
variable $\theta$ is discrete and it can take values in the set:
\[ I_{\theta}=\left \{ \theta_{i}= \frac{i-1}{N_d} 2\pi : i = 1, \dots, N_d \right \}, \]
where $N_d$ is the maxim number of possible directions. Then, we can write
the kinetic type representation
\begin{equation}\label{eq:f}
f(t, \x, q, \theta)= \sum_{i=1}^{N_d} f^i(t, \x,q)\delta(\theta - \theta_i),
\end{equation}
where $f^i(t, \x,q)=f(t, \x, q, \theta_i)$ represents the active particles that, at time $t$  and position $\x$, move with
speed $q$ and direction $\theta_i$. In equation~\eqref{eq:f}, $\delta$ denotes the Dirac delta measure.
Notice that a discrete set of directions is also convenient from the computational point of view.
Our model could be extended to handle a continuum of directions. See, e.g., Ref.~\refcite{Bellomo2019_new}.


Let us introduce some reference quantities that will be use to make the variables non-dimensional. 
We define: 
\begin{itemize}
\item[-] $D$: the largest distance a pedestrian can cover in domain $\Omega$;
\item[-] $V_M$: the highest velocity modulus a pedestrian can reach;
\item[-] $T$: a reference time given by $D/V_M$;
\item[-] $\rho_M$: the maximal admissible number of pedestrians per unit area.
\end{itemize}
The dimensionless variables are then: position $\hat{\x}=\x/D$, time $\hat{t}=t/T$,  and distribution function $\hat{f}=f/ \rho_M$. 
Notice that the fear level $q \in [0, 1]$ is already a non-dimensional variable. Therefore, the velocity modulus
is a non-dimensional variable and the dimensional speed is $q V_M$.
From now on, all the variables will be 
non-dimensional and hats will be omitted to simplify notation. 

Due to the normalization of $f$, and of each $f^i$, the non-dimensional local density is given by: 
\begin{align}\label{eq:rho}
\rho(t, \x)=\sum_{i=1}^{N_d} \int_{0}^1 f^i(t, \x,q) dq.
\end{align} 


\subsection{Modeling interactions} \label{modelinginteractions}

\begin{figure}[h!]
\begin{center}
\begin{overpic}[width=0.30\textwidth, height=0.20\textheight, grid=false,angle=-90, tics=10]{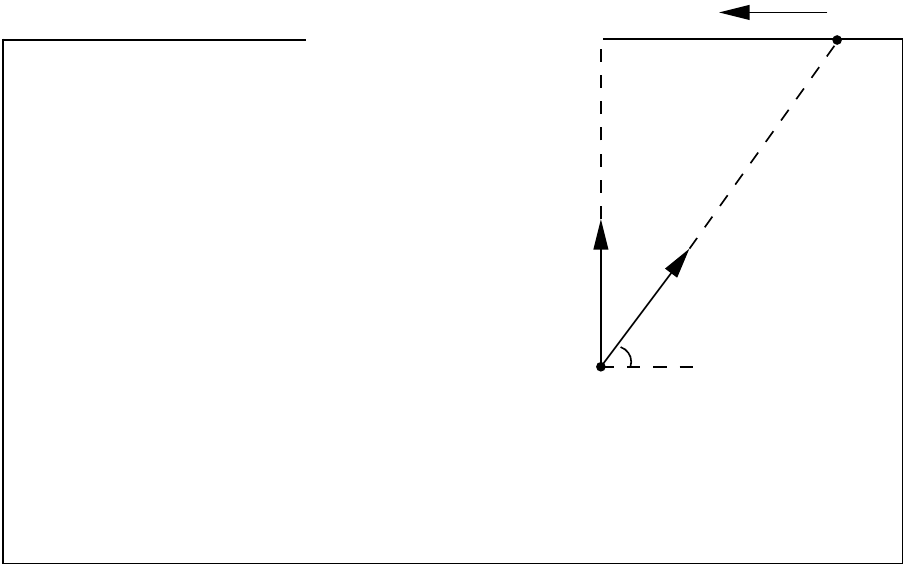}
      \put(15,79){\text{$\Omega$}}
      \put(96,47){\text{$E$}}
      \put(30,35){\scriptsize{\text{$\x$}}}
      \put(95, 3){\scriptsize{\text{$\x_{W}$}}}
      \put(60,40){\scriptsize{\text{$d_E(\x)$}}}
      \put(55,10){\scriptsize{\text{$d_W(\x)$}}}
      \put(43,35){\scriptsize{\text{$\u_E$}}}
      \put(100,12){\scriptsize{\text{$\u_W$}}}
      \put(38,23){\scriptsize{\text{$\theta_h$}}}
\end{overpic}

\caption{Sketch of computational domain $\Omega$ with exit $E$ and \rev{candidate particle $h$} located at $\x$, moving
with direction $\theta_h$. The pedestrian should choose direction $\u_E$ to reach the exit, 
while direction $\u_W$ is to avoid collision with the wall. The distances form the exit and from the wall
are $d_E$ and $d_W$, respectively.
}
\label{velocity}
\end{center}
\end{figure}
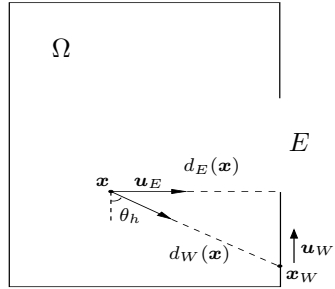

Each pedestrian is modeled as a particle. Interactions involve three types of particles: 

\begin{itemize}
\item[-] \textit{test particles} with state $(\x, \theta_i)$: they are representative of the whole system;
\item[-] \textit{candidate particles} with state $(\x, \theta_h)$: they can reach in probability the state of the test particles after individual-based interactions with the environment or with field particles; 
\item[-] \textit{field particles} with state $(\x, \theta_k)$: their presence triggers the interactions of the candidate particles.
\end{itemize}
The process through which a pedestrian decides the direction to take is the results of several factors. 
We take into account four factors:

\begin{enumerate}[label={(F\arabic*})]
\item \textit{The goal to reach the exit.}\\
Given a candidate particle at the point $\x$, we define its distance to the exit as
\[d_E(\x)= \min_{\x_E \in E} || \x-\x_E ||,\]
and we consider the unit vector $\u_E(\x)$, pointing from $\x$ to the exit. See Figure~\ref{velocity}.

\item \textit{The need to avoid the collision with the walls.}\\
Given a candidate particle at point $\x$ moving with direction $\theta_h$, we define the distance $d_W(\x, \theta_h)$ from the particle to a wall at a point $\x_W(\x, \theta_h)$ where the particle is expected to collide with the wall.
The unit tangent vector $\u_W(\x, \theta_h)$ to $\partial \Omega$ at $\x_W$
points to the direction of the the exit. Vector $\u_W$ is used to avoid a collision with the walls. See Figure~\ref{velocity}.

\item \textit{The tendency to look for less congested area.}\\
A candidate particle $(\x, \theta_h)$ may decide to change direction in order to avoid congested areas. 
This is achieved with the direction that gives the minimal directional derivative of the density at the point $\x$. 
We denote such direction by unit vector $\u_C(\theta_h, \rho)$. In the literature, this behavior is sometimes referred to as \emph{individualistic}. \cite{Llorca_2019}

\item \textit{The tendency to follow the stream.}\\
A candidate particle $h$ interacting with a field particle $k$  may decide to follow it and thus adopt its direction, denoted with unit vector $\u_F=(\cos\theta_k, \sin\theta_k)$. In the literature, this behavior is also referred to as \emph{herding}. \cite{Llorca_2019}
\end{enumerate}

Factors (F1) and (F2) are related to geometric aspects of the domain, while factors (F3) and (F4)
consider that people's behavior is influenced by surrounding crowd. 
Factors (F3) and (F4) compete with each other: (F4) is dominant
in a stress situation, while (F3) characterizes rational behavior. 
To weight between (F3) and (F4), we use the fear level $q \in [0,1]$. 

The interaction with the bounding walls is modeled with two terms:
\begin{compactitem}
\item[-] $\mu[\rho]$: the \textit{interaction rate} models the frequency of interactions between candidate particles and 
the walls and/or an exit. If the local density is low, it is easier for pedestrians to see the walls and doors. Thus, we set $\mu[\rho] =1-\rho$.  

\item[-] $\mathcal{A}_h(i)$: the \textit{transition probability} gives the probability that a candidate particle $h$, i.e.~with direction  $\theta_h$, adjusts its direction into that of the test particle $\theta_i$ due to the presence of the walls and/or an exit. 
The following constraint for $\mathcal{A}_h(i)$ has to be satisfied:
\[
\sum_{i=1}^{N_d} \mathcal{A}_h(i)=1 \quad \text{for all} \,\, h \in \{1, \dots, N_d\}.
\]
\end{compactitem}
We assume that particles change direction, in probability, only to an adjacent direction in the discrete set $I_\theta$
\rev{for each possible change step}. 
This means a candidate particle $h$ may end up into the states $h-1, h+1$ or remain in the state $h$. In the case $h=1$, we set $\theta_{h-1}=\theta_{N_d}$ and, in the case $h=N_d$, we set $\theta_{h+1}=\theta_1$.
The set of all transition probabilities $\mathcal{A}=\{\mathcal{A}_h(i) \}_{h,i= 1, \dots, N_d}$ forms the so-called \textit{table of games} that models the game played by active particles interacting with the walls.

To take into account factors (F1) and (F2), we define the vector
\begin{align}\label{eq:uG}
\u_G(\x, \theta_h) &= \frac{(1-d_E(\x))\u_E(\x) + (1-d_W(\x, \theta_h))\u_W(\x, \theta_h)}{|| (1-d_E(\x))\u_E(\x) + (1-d_W(\x, \theta_h))\u_W(\x, \theta_h) ||} \cl
&= (\cos \theta_G, \sin \theta_G).
\end{align}

Here $\theta_G$ is the  \textit{preferred geometrical direction}, which is the ideal direction that a pedestrian should take in order to reach the exit and avoid the walls in an optimal way. Notice that the closer a pedestrian is
to an exit (resp., a wall), the larger the weight for direction $\u_E$ (resp., $\u_W$). 

A candidate particle $h$ will update its direction by choosing
the angle closest to $\theta_G$ among the three allowed 
angles $\theta_{h-1}, \theta_{h}$ and $\theta_{h+1}$. The transition probability is given by:
\begin{equation}\label{eq:A}
\mathcal{A}_h(i)=\beta_h\delta_{s,i} + (1-\beta_h)\delta_{h, i}, \quad i=h-1, h, h+1,
\end{equation}
where
\[s=\argmin_{j \in \{h-1,h+1\}}\{d(\theta_G, \theta_j)\},\]
with
\begin{equation}\label{eq:distance}
 d(\theta_p, \theta_q)=
\begin{cases}
|\theta_p - \theta_q|  & \text{if} \,\,\, |\theta_p - \theta_q | \leq \pi, \\
2\pi - |\theta_p- \theta_q|  & \text{if} \,\,\, |\theta_p- \theta_q| > \pi .             
\end{cases}
\end{equation}
In \eqref{eq:A}, $\delta$ denotes the Kronecker delta function. Coefficient $\beta_h$ is defined by:
\[
\beta_h=
\begin{cases}
1 & \text{if} \,\,\, d(\theta_h, \theta_G) \geq \Delta\theta, \\
\dfrac{d(\theta_h, \theta_G)}{\Delta\theta}& \text{if} \,\,\, d(\theta_h, \theta_G)< \Delta \theta ,             
\end{cases}
 \]
where $\Delta\theta=2\pi/{N_d}$. The role of $\beta_h$ is to allow for a transition to $\theta_{h-1}$ or
$\theta_{h+1}$ even in the case that the 
geometrical preferred direction $\theta_G$ is closer to $\theta_h$. Such a transition is more likely 
to occur the more distant $\theta_h$ and $\theta_G$ are.
Notice that if $\theta_G=\theta_h$, then $\beta_h=0$ and $\mathcal{A}_h(h)=1$, meaning that a 
pedestrian keeps the same direction (in the absence of interactions other than with the 
environment) with probability 1.

The interaction with other pedestrians is modeled with two terms:
\begin{itemize}
\item[-] $\eta[\rho]$: the \textit{interaction rate} defines the number of binary encounters per unit time. If the local density increases, then the interaction rate also increases. For simplicity, we take $\eta[\rho]= \rho$. 

\item[-] $\mathcal{B}_{hk}(i)[\rho\rev{, q}]$: the \textit{transition probability} gives the probability that a candidate particle 
$h$ modifies its direction $\theta_h$ into that of the test particle $i$, i.e. $\theta_i$, due to the research 
of less congested areas (factor (F3)) and the interaction with a field particle $k$ that moves with direction $\theta_k$
(factor (F4)). 
The following constrain for $\mathcal{B}_{hk}(i)[\rev{\rho,q]}$ has to be satisfied:
\[
\sum_{i=1}^{N_d} \mathcal{B}_{hk}(i)[\rho,q]=1 \quad \text{for all} \,\, h, k  \in \{1, \dots, N_d\},
\]
where the square brackets denote the dependence on the density $\rho$
and fear level $q$.
\end{itemize}

The game consists in choosing the less congested direction among the three admissible ones.
This direction can be computed for a candidate pedestrian $h$ situated at $\x$, by taking 
\[C=\argmin_{j \in \{h-1, h, h+1\}}\{\partial_j\rho(t, \x)\},\]
where $\partial_j\rho$ denotes the directional derivative of $\rho$ in the direction given by angle $\theta_j$. 
We have $\u_C(\theta_h, \rho)=(\cos\theta_C, \sin\theta_C)$.
As for the tendency to follow the crowd, we set $\u_F=(\cos\theta_k, \sin\theta_k)$. This means that  a
candidate particle follows the direction of a field particle.

To take into account factors (F3) and (F4), we define the vector
\begin{align}\label{eq:u_p}
\u_P(\theta_h, \theta_k, \rho,q)= \frac{q\u_F+(1-q)\u_C(\theta_h, \rho)}
{||q\u_F+(1-q)\u_C(\theta_h, \rho)||} = (\cos \theta_P, \sin \theta_P),
\end{align}
where the subscript $P$ stands for \textit{pedestrians}. 
Direction $\theta_P$ is the \textit{preferred interaction-based direction}, obtained as a weighted combination between the 
tendency to follow the stream and the tendency to avoid crowded areas. 

The transition probability is given by:
\[\mathcal{B}_{hk}(i)[\rho,q]=\beta_{hk}\rho\delta_{r, i} + (1-\beta_{hk}\rho)\delta_{h,i}, \quad i=h-1,h, h+1,\]
where $r$ and $\beta_{hk}$ are defined by:
\[r=\argmin_{j \in \{h-1, h+1\}} \{d(\theta_P, \theta_j)\},\]
\[
\beta_{hk}=
\begin{cases}
1& \text{if} \,\,\, d(\theta_h, \theta_P) \geq \Delta\theta \\
\dfrac{d(\theta_h, \theta_P)}{\Delta\theta}& \text{if} \,\,\, d(\theta_h, \theta_P)< \Delta \theta.             
\end{cases}
 \]
We recall that $d(\cdot, \cdot)$ is defined in \eqref{eq:distance}.

\begin{remark}\label{rem:ped_dyn}
The kinetic approach to pedestrian dynamics in Ref.~\refcite{Agnelli2015} and \refcite{kim_quaini}, which will be used for 
comparison in Section~\ref{sec:test1}, models fear in a simplified way. The first simplification is that
the walking speed is not affected by fear. The walking direction is adapted through
a parameter $\epsilon$ that takes the place of $q$ in \eqref{eq:u_p}. 
Unlike $q$, $\epsilon$ is constant in space and time. So, the second simplification is that
all the people in the simulation have the same fear level that remains constant for the entire
time interval under consideration. \\
The kinetic type model in Ref.~\refcite{Bellomo2019_new} uses a more realistic description of stress behavior. 
In the future, it will be interesting to compare the results produced by the model in this paper with the results
obtained from the model in Ref.~\refcite{Bellomo2019_new}.
\end{remark}

\subsection{The mathematical model}

The derivation of the mathematical model can be obtained by a suitable balance of particles in an elementary volume
of the space of microscopic states, considering the net flow into such volume due to
transport and interactions.
Taking into account factors (F1)-(F4), we obtain:
\begin{align}
\frac{\partial f^i}{\partial t} &+ \nabla \cdot \left(q (\cos \theta_i, \sin \theta_i)^T f^i(t, \x,q) \right) \cl
& = \mathcal{J}^i[f](t, \x,q) + \gamma \frac{\partial((q-q^\ast) f^i)}{\partial q} \cl
& = \eta[\rho] \left( \sum_{h,k = 1}^{N_d} \int_{0}^1 \int_{0}^1 \mathcal{B}_{hk}(i)  f^h(t, \x,q)f^k(t, \x,q) dq dq - f^i(t,\x,q) \rho(t, \x)\right) \cl
& \quad + \mu[\rho] \left( \sum_{h = 1}^{N_d} \int_{0}^1 \mathcal{A}_h(i) f^h(t, \x,q) dq - f^i(t,\x,q) \right) \cl
& \quad + \gamma \frac{\partial((q-q^\ast) f^i(t,\x,q))}{\partial q}
\label{eq:model}
\end{align}
for $i= 1,2, \dots, N_d$. Functional $\mathcal{J}^i[f]$ represents the net balance of particles 
that move with direction $\theta_i$ due to interactions with the environment and with the surrounding people. 
\rev{Notice that the last term at the right-hand-side in \eqref{eq:model}
corresponds to the term at the right-hand-side in \eqref{eq:2kineticeq}, i.e.~it 
accounts for the fact that people tend to equilibrate their fear level with the average fear level
they perceive.
}
 
Equation~\eqref{eq:model} is completed with equation~\eqref{eq:rho}
for the density. 
In the next section, we will discuss a numerical method for the solution of
problem \eqref{eq:rho}, \eqref{eq:model}.

\section{Numerical method}\label{sec:num_meth}

Our approach is based on a splitting method.
The idea is to split model \eqref{eq:model} into subproblems that are easier to solve
and for which practical algorithms are readily available. The numerical method is then completed
by picking an appropriate numerical scheme for each subproblem.
Among the available operator-splitting methods, we
chose the Lie splitting scheme because it provides
a good compromise between accuracy and robustness, 
as shown in Ref.~\refcite{glowinski2003finite}.

\subsection{The Lie operator-splitting scheme}

Before applying it to problem \eqref{eq:rho},\eqref{eq:model}, we briefly
present the Lie splitting scheme for a generic method that is first-order system in time:
\begin{eqnarray}\label{LieProblem}
   \frac{\partial \phi}{\partial t} + A(\phi) &=& 0, \quad \textrm{in} \ (0,T), \cl
\phi(0) &=& \phi_0, \el
\end{eqnarray}
where A is an operator from a Hilbert space into itself. Operator A is then split, in a non-trivial decomposition, as
\begin{equation*}
 A = \sum\limits_{i=1}^I A_i.
\end{equation*}
The Lie scheme consists of the following. Let $\Delta t>0$ be a time discretization step for the time interval $[0, T]$. Denote $t^k=k\Delta t$, with $k = 0, \dots, N_t$ and let $\phi^k$
be an approximation of $\phi(t^k).$ Set $\phi^0=\phi_0.$ For $n \geq 0$, compute $\phi^{k+1}$ by solving
\begin{eqnarray}
   \frac{\partial \phi_i}{\partial t} + A_i(\phi_i) &=& 0 \quad \textrm{in} \; (t^k, t^{k+1}), \\
\phi_i(t^k) &=& \phi^{k+(i-1)/I}, 
\end{eqnarray}
and then set $\phi^{k+i/I} = \phi_i(t^{k+1}),$ for $i=1, \dots. I.$

This method is first-order accurate in time. More precisely, if~\eqref{LieProblem} 
is defined on a finite-dimensional space, and if the operators $A_i$ are smooth enough, 
then $\| \phi(t^k)-\phi^k \| = O(\Delta t)$. \cite{glowinski2003finite}

We will apply Lie splitting to problem \eqref{eq:rho},\eqref{eq:model}, which will get split into 
two subproblems:
\begin{enumerate}
 \item An advection problem that includes the term for emotional contagion at the right-hand-side in 
 eq.~\eqref{eq:model}.
 \item A problem involving the interaction with the environment and other pedestrians.
\end{enumerate}

\subsection{Lie scheme applied to problem \eqref{eq:model}}\label{sec:Lie_applied}
Let us apply the Lie operator-splitting scheme to problem \eqref{eq:rho},\eqref{eq:model}.
Given an initial condition $f^{i,0}=f^i(0, \x,q)$, for $i = 1, \dots, N_d$, the algorithm reads:
For $n=0,1,2, \dots, N_t-1,$ perform the following steps:
\begin{itemize}
\item[-] {\bf Step 1}: Find $f^i$, for $i = 1, \dots, N_d$, such that\\
\begin{equation}
\begin{cases}
 \dfrac{\partial f^i}{\partial t} + \nabla \cdot \left(q (\cos \theta_i, \sin \theta_i)^T f^i(t, \x,q) \right)= \gamma \frac{\partial((q-q^\ast) f^i(t,\x,q))}{\partial q}
   \,\,\, \text{on } (t^n, t^{n+1}), \\ \label{eq:step1}
f^i(t^n, \x,q)=f^{i,n}.
\end{cases} 
\end{equation}
 Set $f^{i,n+\frac{1}{2}}=f^i(t^{n+1}, \x)$.
 
\bigskip
\item[-] {\bf Step 2}:  Find $f_i$, for $i = 1, \dots, N_d$, such that\\
\begin{equation}
\begin{cases}
 \dfrac{\partial f^i}{\partial t} = \mathcal{J}^i[f](t, \x,q)  \,\,\, \text{on } (t^n, t^{n+1}),  \\ \label{eq:step2}
f^i(t^n, \x,q)=f^{i, n+\frac{1}{2}}.
\end{cases}
\end{equation}
 Set $f^{i,n+1}=f^i(t^{n+1}, \x)$.
\end{itemize}

Notice that once $f^{i, n+1}$ is computed for $i = 1, \dots, N_d$, we use equation~\eqref{eq:rho} to get the density $\rho^{n+1}$.

We associate with $M$ a time discretization step $\tau = (t_f - t_0)/M$
and set $t^m = t_0 + m \tau$. 
For the discretization of problem \eqref{eq:step1}, we use the scheme reported in Section~\ref{sec:discr_Bertozzi}. 
As for the approximation of the problem \eqref{eq:step2}, we use the following explicit scheme:
\[ f_{j,k,l}^{i, m+1}= f_{j,k,l}^{i,m} + \tau \Big (\mathcal{J}^m[F^{m}] \Big ), \] 
where $F^m$ is the approximation of distribution function \eqref{eq:f} at time $t^m$.

\section{Numerical results} \label{sec:num_res}

This section presents some sample simulations focused on the computation of evacuation
time from a room with one exit. We select three case studies in order to (i) compare
with the results obtained from a model that features no emotional contagion, (ii) analyze the role of 
a geometric parameter (exit size), model parameters (interaction radius and contagion interaction strength), and
initial fear distribution, and (iii) show that our model can 
roughly reproduce experimental data.

For all the simulations, we consider eight different velocity directions $N_d=8$ in the discrete set:
\[ I_{\theta}=\left \{ \theta_{i}= \frac{i-1}{8} 2\pi : i = 1, \dots, 8 \right \}. \]

\subsection{Test 1: evacuation from a room with one door of variable width}\label{sec:test1}

\begin{figure}[h!t]
\centering
\begin{overpic}[width=0.32\textwidth, grid=false]{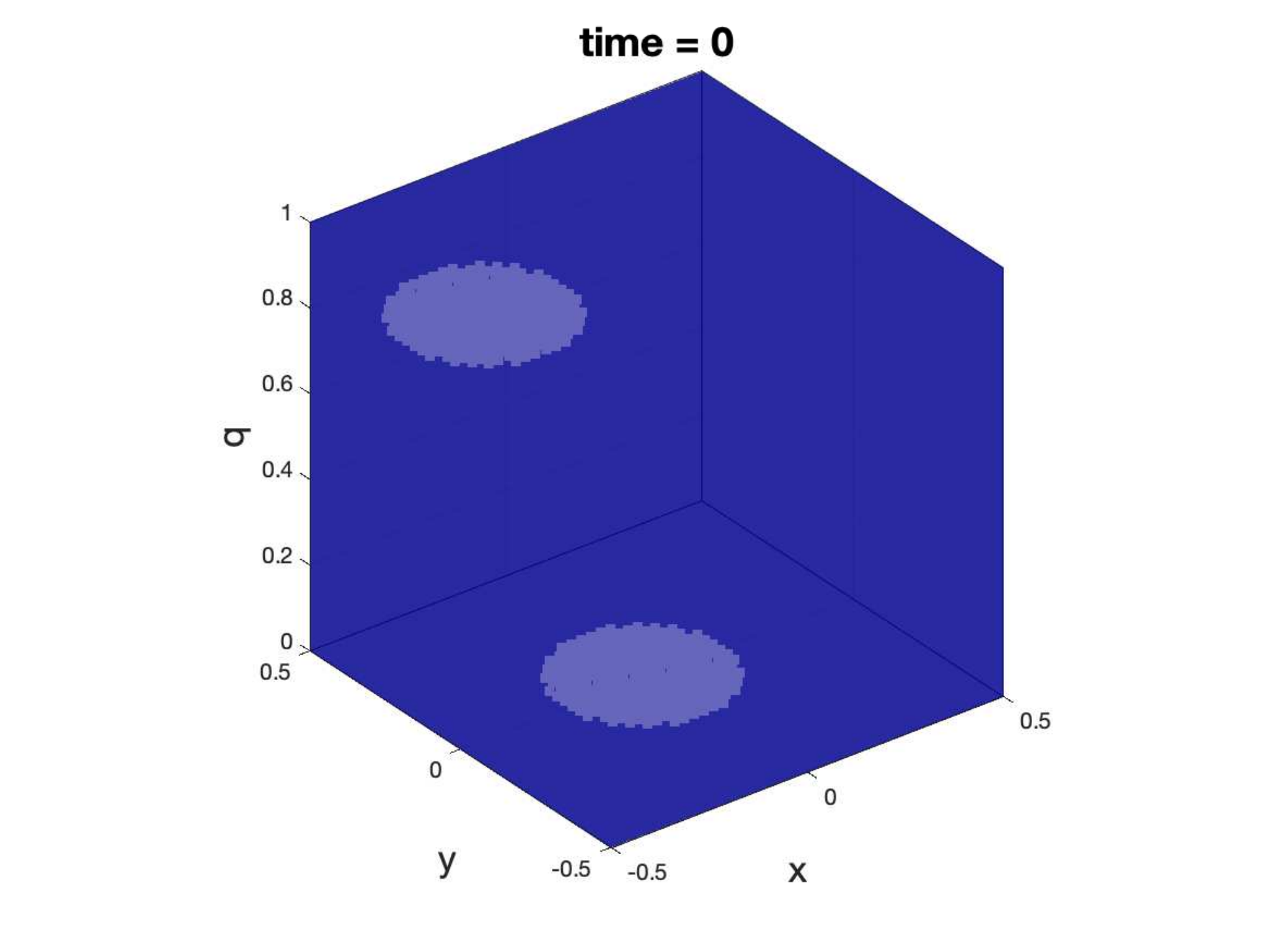}
\end{overpic}
\begin{overpic}[width=0.32\textwidth, grid=false]{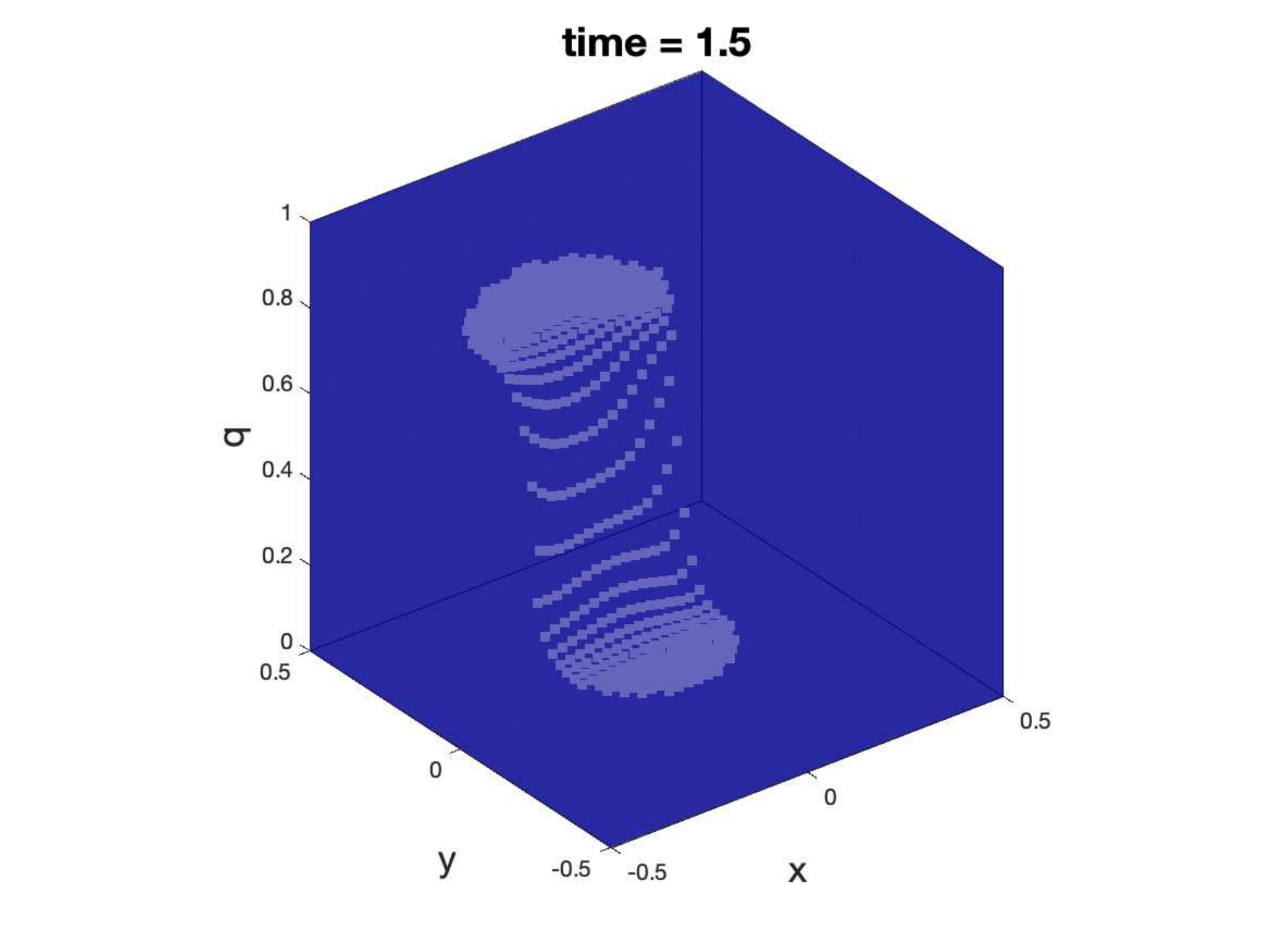}
\end{overpic}
\begin{overpic}[width=0.32\textwidth, grid=false]{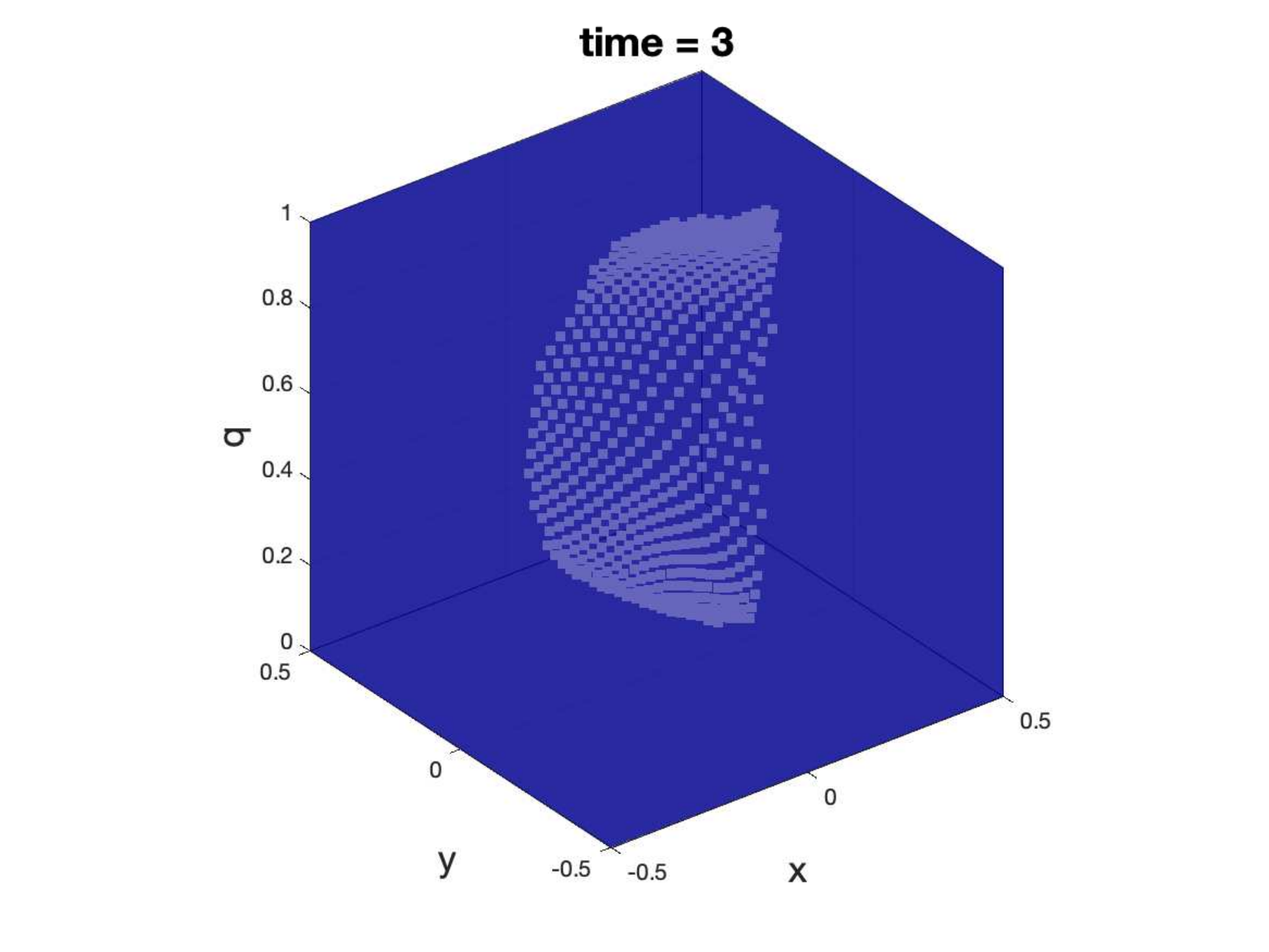}
\end{overpic}
\begin{overpic}[width=0.32\textwidth, grid=false]{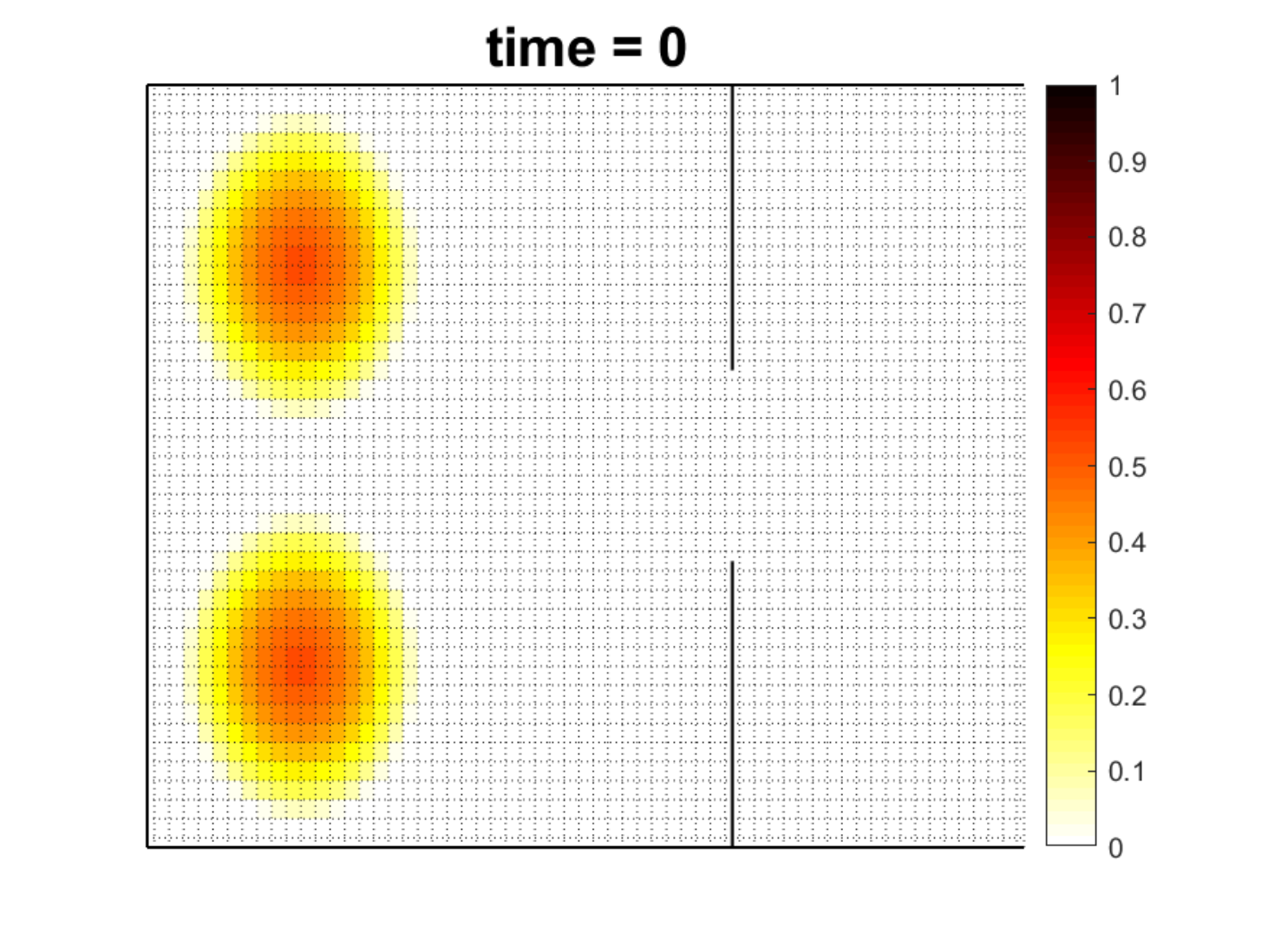}
\put(24, 54.2){\vector(0,-1){10}}
\put(24, 22){\vector(0,1){10}}
\end{overpic}
\begin{overpic}[width=0.32\textwidth, grid=false]{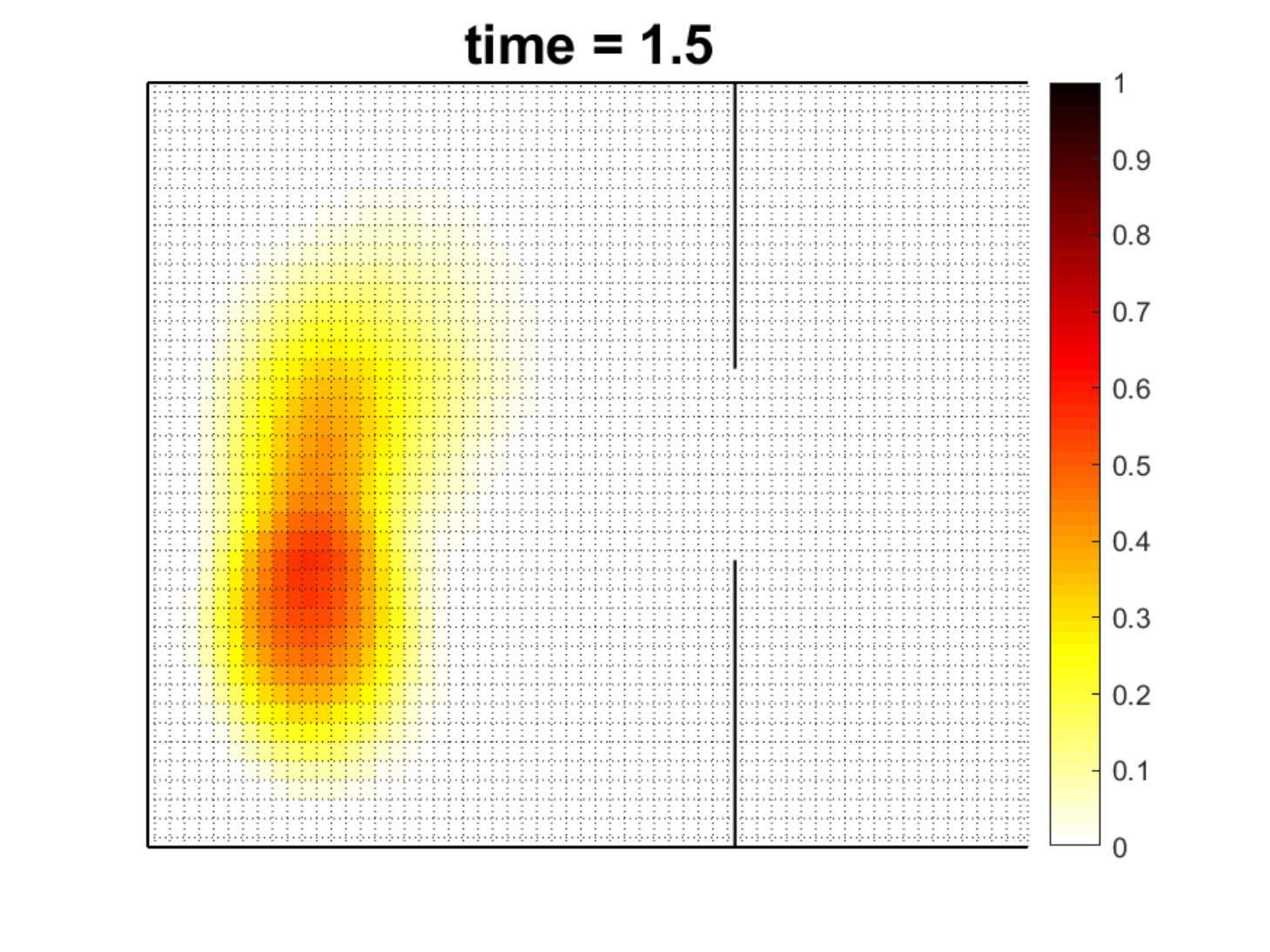}
\end{overpic}
\begin{overpic}[width=0.32\textwidth, grid=false]{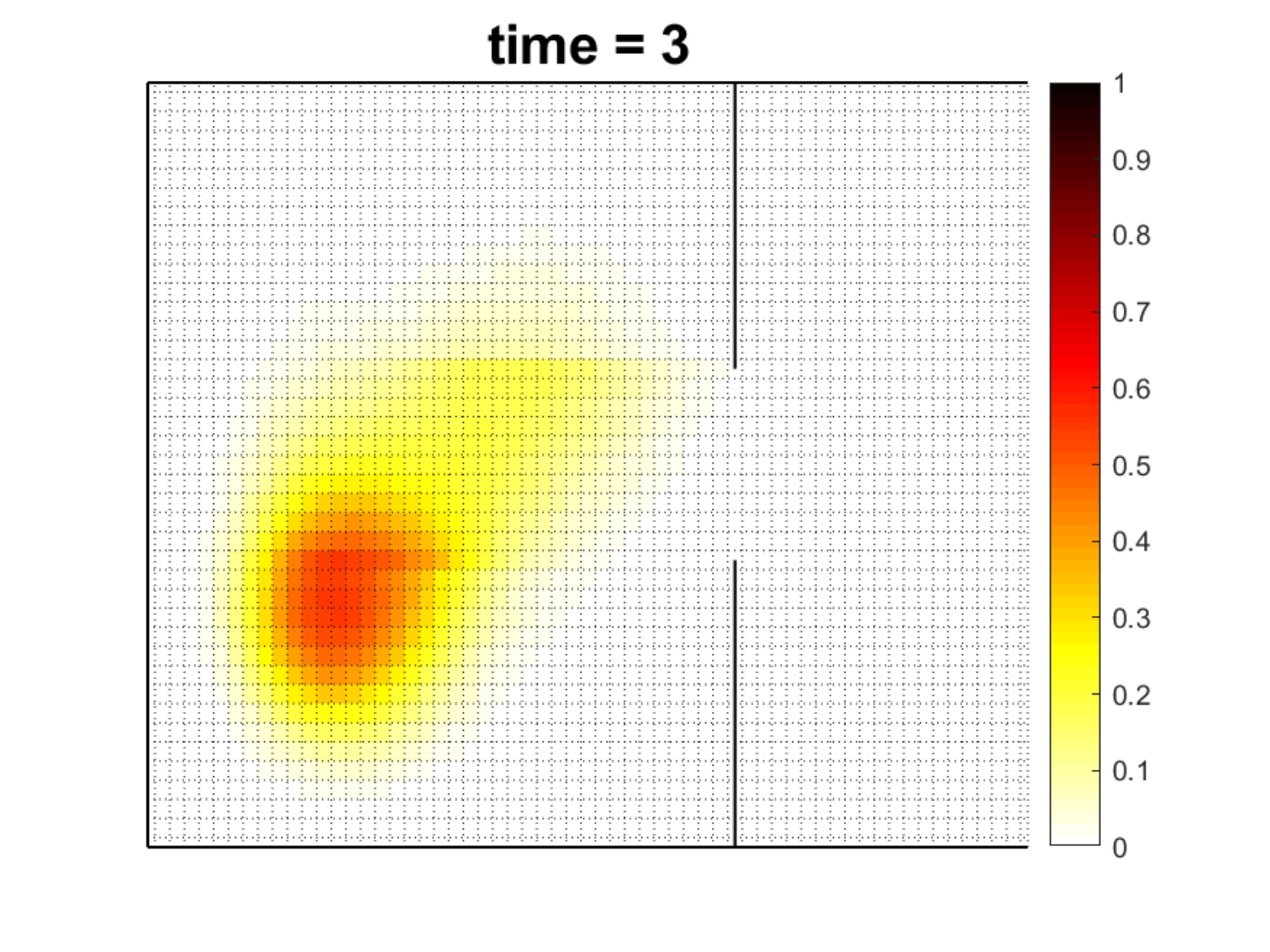}
\end{overpic}
\begin{overpic}[width=0.32\textwidth, grid=false]{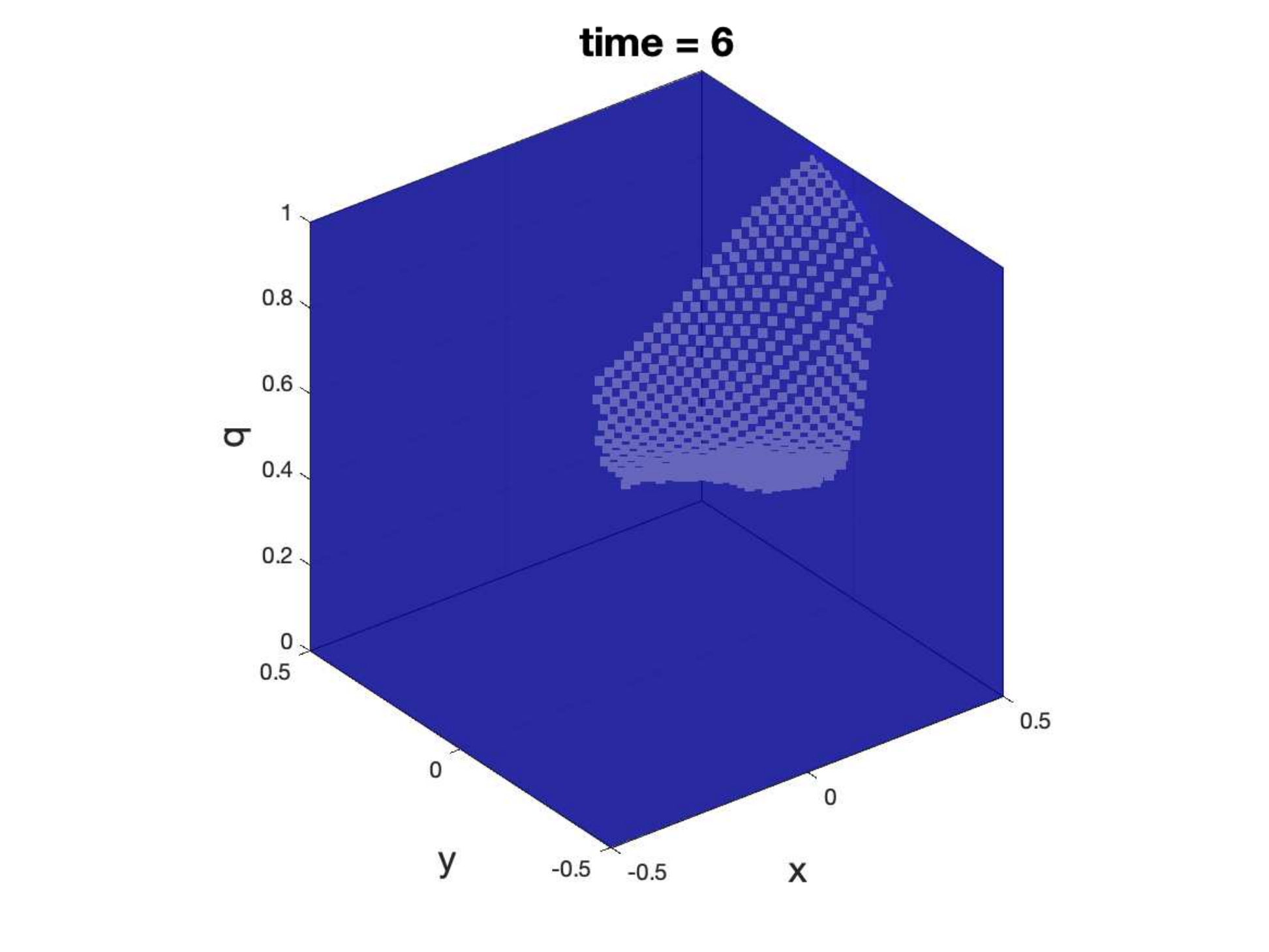}
\end{overpic} 
\begin{overpic}[width=0.32\textwidth, grid=false]{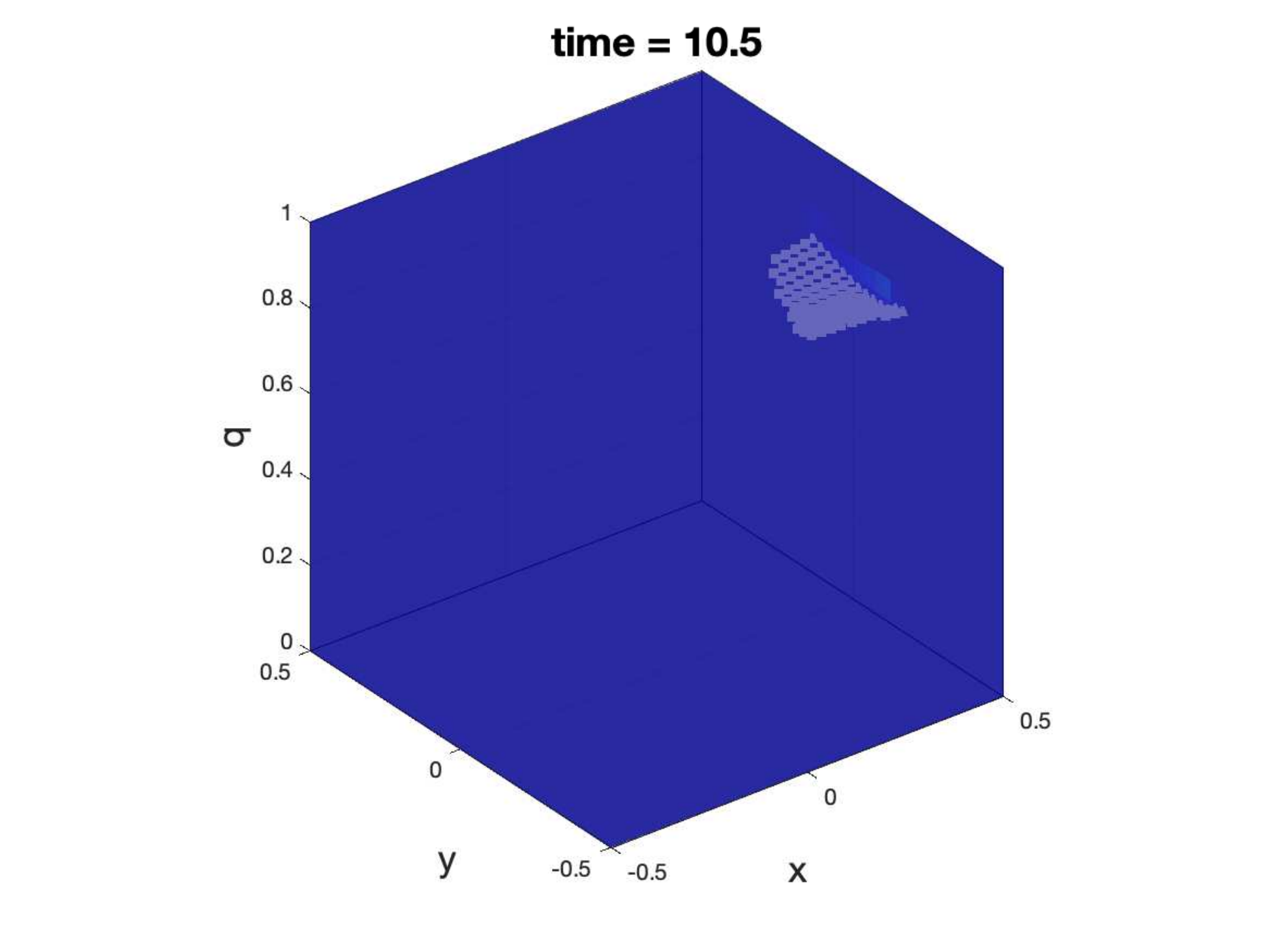}
\end{overpic}
\begin{overpic}[width=0.32\textwidth, grid=false]{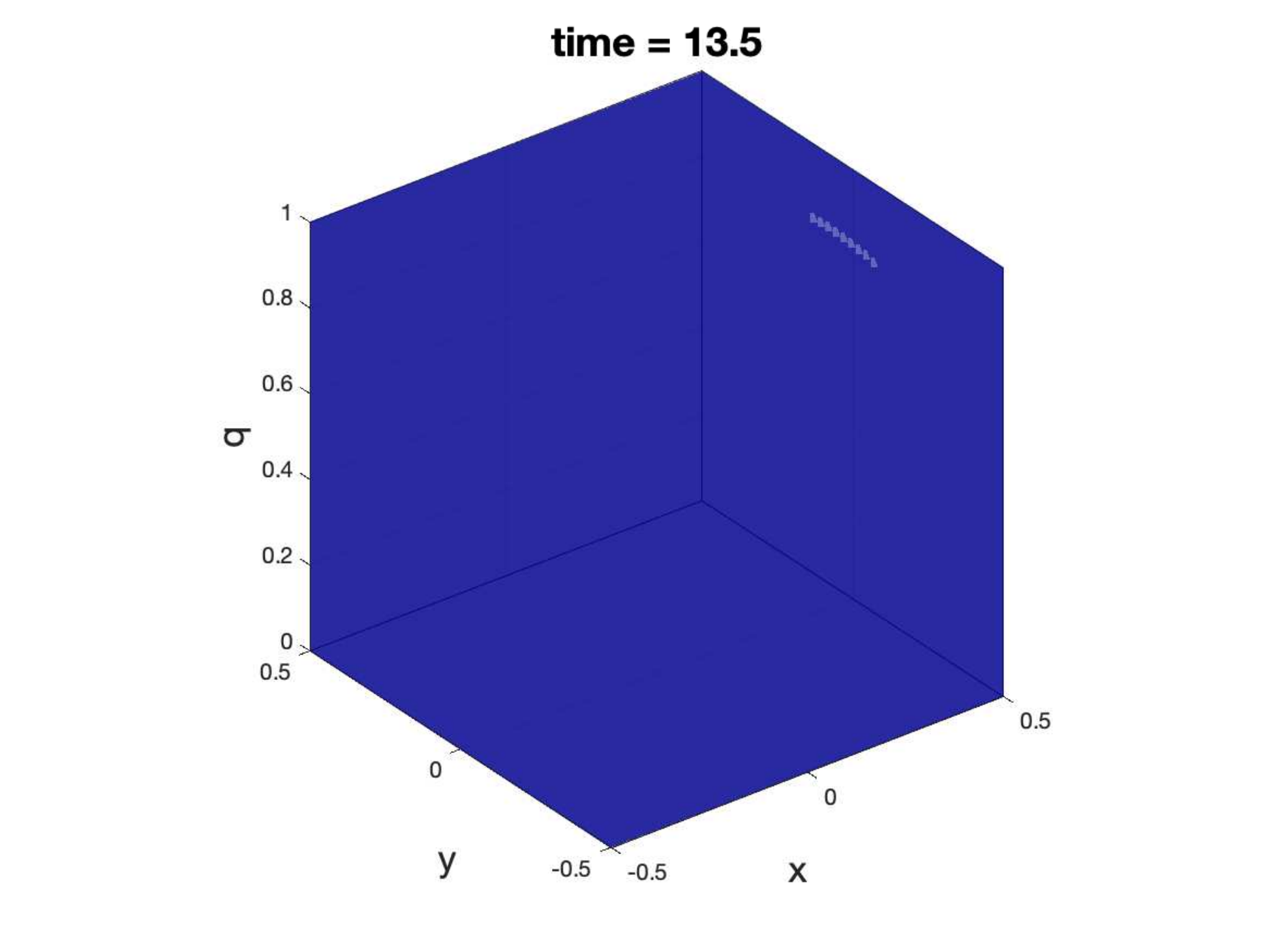}
\end{overpic}
\\
\begin{overpic}[width=0.32\textwidth, grid=false]{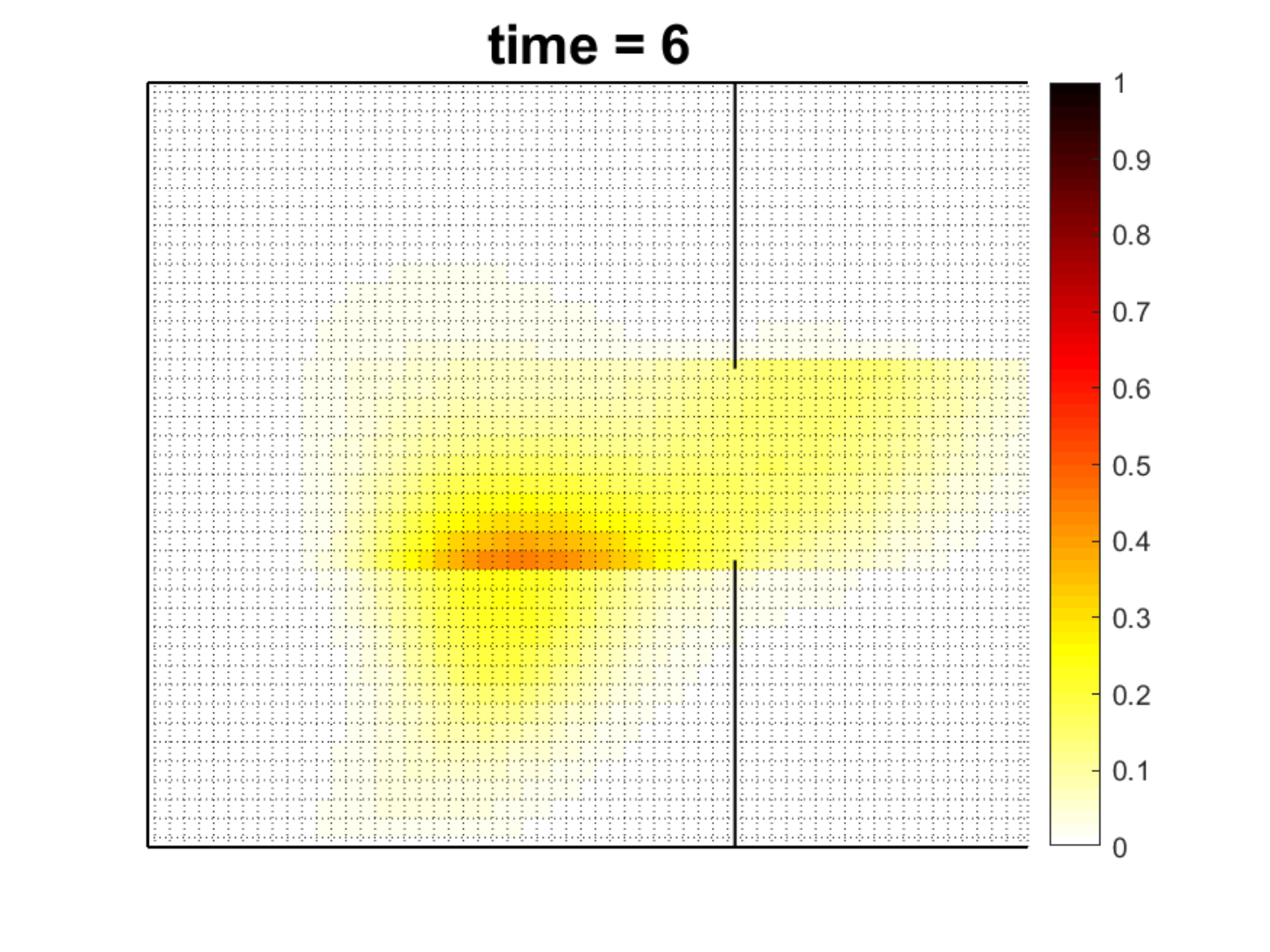}
\end{overpic}
\begin{overpic}[width=0.32\textwidth, grid=false]{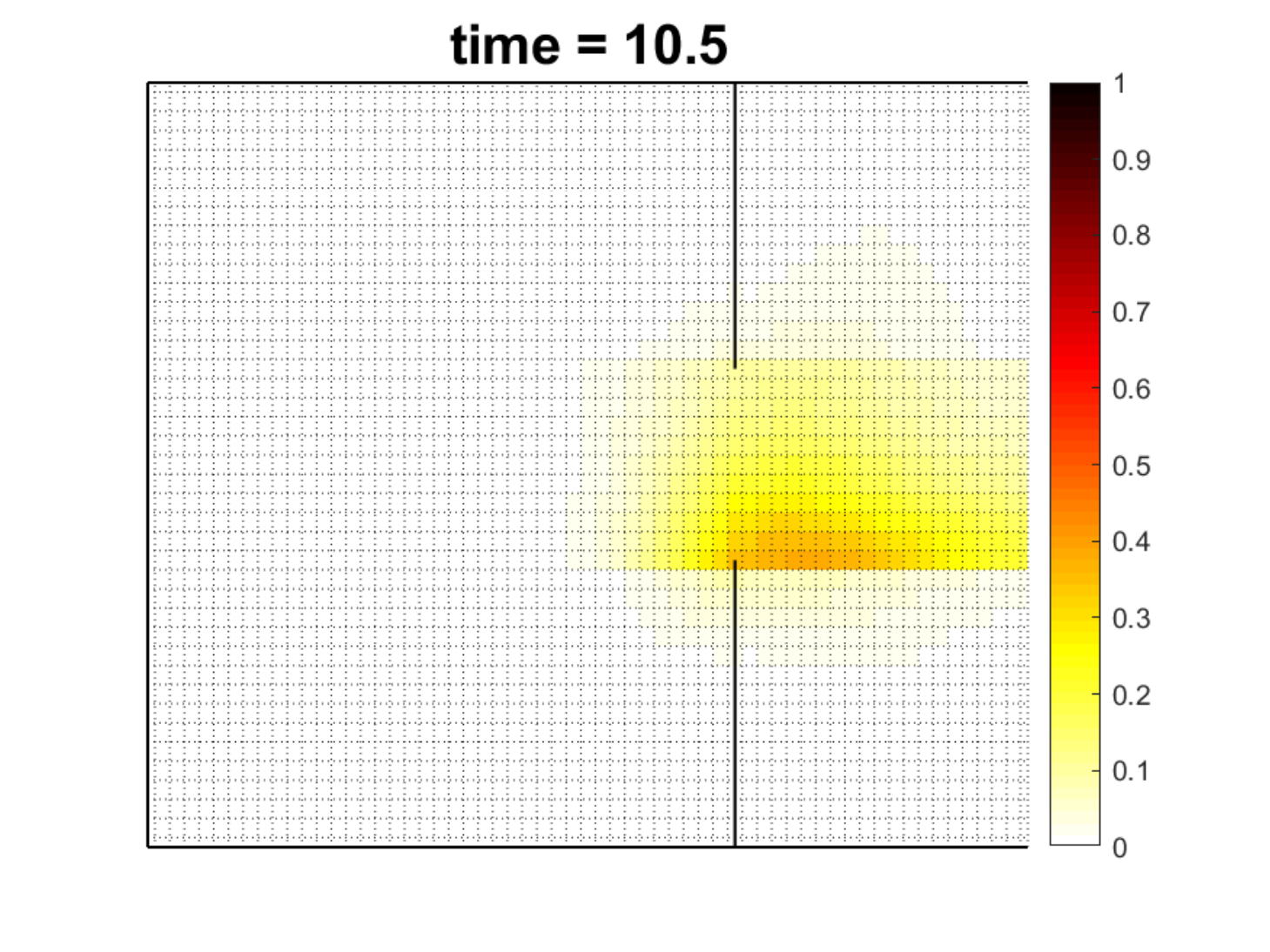}
\end{overpic}
\begin{overpic}[width=0.32\textwidth, grid=false]{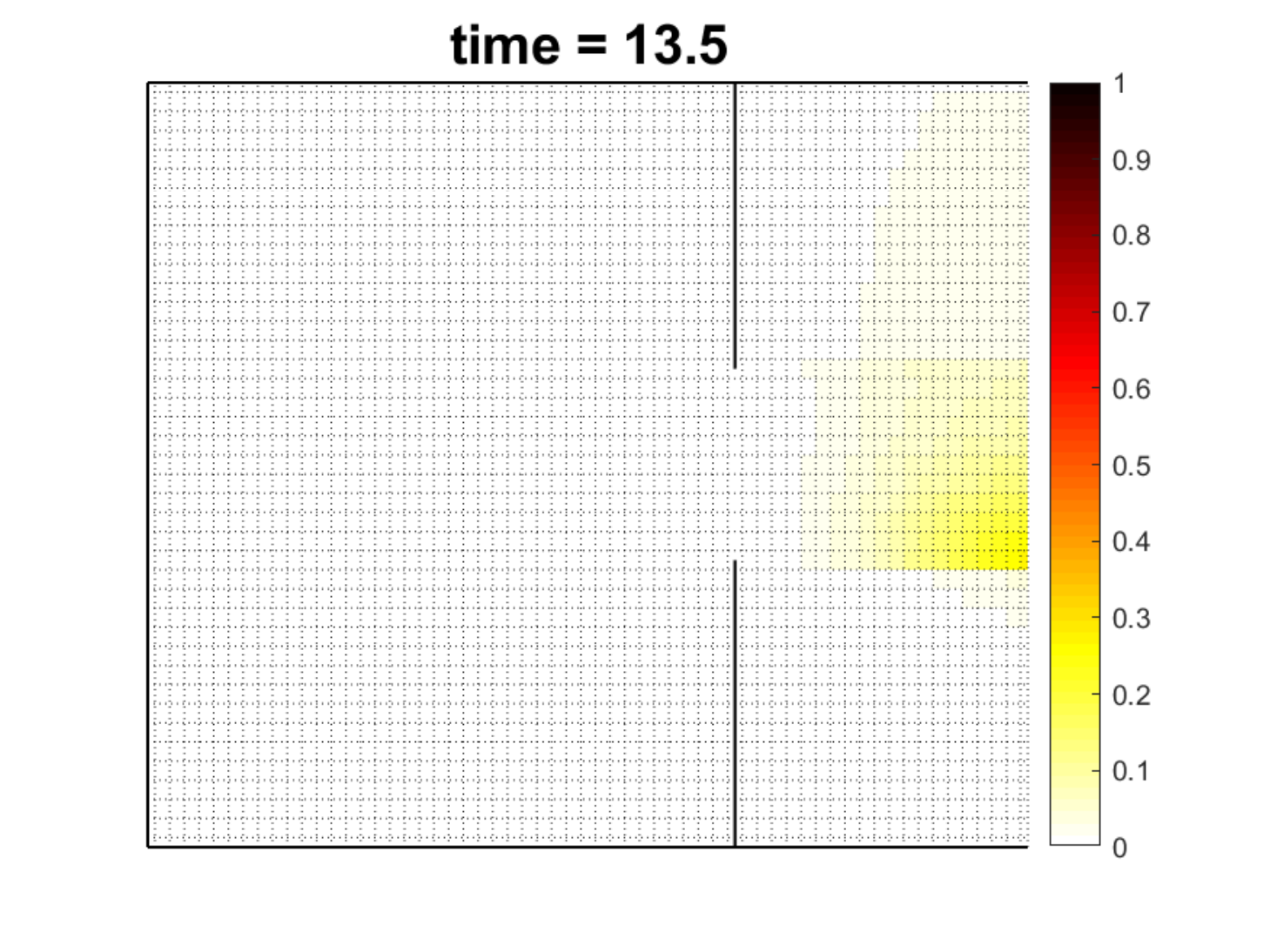}
\end{overpic}
\caption{Test 1: Evacuation process of 46 pedestrians grouped into two clusters with opposite initial directions 
$\theta_3$ and $\theta_7$ using the fine mesh: average fear level $q^\ast$ (top panel for each time) and people
density (bottom panel for each time) for $t = 0, 1.5, 3, 6, 10.5,13.5$ s.
The density plots show also the computational mesh.
} 
\label{MM}
\end{figure}

This first test case is inspired from Ref.~\refcite{Agnelli2015}.
The computational domain contains a square room with side $10$ m with
one exit door located in the middle of the right side. The exit size is $2.6$ m. 
The computational domain is larger than the room itself to follow the motion of the pedestrian
also once they have left the room. We aim at simulating the evacuation of 46 people located inside
the room and initially distributed into two equal-area circular clusters. See Figure~\ref{MM}, first panel on 
the second row. 
The two groups are initially moving against the each other with opposite initial directions  
$\theta_3$ and $\theta_7$. 
For the purpose of comparing with the crowd dynamics given by the model in Ref.~\refcite{kim_quaini}, which features no
emotional contagion (see Remark \ref{rem:ped_dyn}), in this test we assign the initial fear level as follows:
the group with initial direction $\theta_7$ has an initial high fear level ($q = 0.8$), while 
the other group has an initial low fear level ($q = 0.2$).  
We set contagion interaction strength $\gamma = 1$ and interaction distance $R = 0.5$ m.
This value of $R$ means that people adjust their fear level to the average fear
level of the people in their immediate vicinity.

In order to work with dimensionless quantities as described in Section~\ref{sec:model}, 
we define the following reference quantities: $D=10\sqrt2$ m, $V_M= 2$ m/s, $T = D/V_M= 5 \sqrt2$ s, and
$\rho_M = 7$ people/m$^2$. However, once the results are computed we convert them back to
dimensional quantities. 

We consider two different meshes: 
\begin{itemize}
\item[-] \emph{coarse mesh} with $\Delta x = \Delta y=0.5$ m and $\Delta q=0.05$;
\item[-] \emph{fine mesh} with $\Delta x = \Delta y=0.25$ m and $\Delta q=0.05$.
\end{itemize}
Notice that the level of refinement changes only in the spatial directions. 
We choose the time step to $\Delta t = 0.0375$ s for both meshes. 
This time step is such that $\tau = \Delta t/M$ satisfies the CLF condition \eqref{eq:CFL}
for $M = 3$. Figure~\ref{MM} shows the density computed with the fine mesh
at times $t = 0$, 1.5, 3, 6, 10.5, 13.5 s. At time $t = 6$ s, we observe the formation of a region of high
density next to a region of low density, with a sharp divide between the two
that is aligned with one door end. 
There are several possible reasons for this. First, the walking direction
results from the interplay of different factors and thus it may not be intuitive. 
We consider only 8 discrete directions as stated at the beginning of Sec.~\ref{sec:num_res}. 
Maybe a higher number of directions would lead to a less sharp divide, 
although it would not be realistic for the small group of people
under consideration (see Sec.~\ref{sec:model}).
Finally, the value of $R$ is small for this test, which means that it takes
longer for fear to propagate. As the group with high fear reaches the group with low fear, it takes 
a while for the emotion to spread through the entire low fear group. 

\begin{figure}[ht]
\centering
\begin{overpic}[width=0.60\textwidth, grid=false]{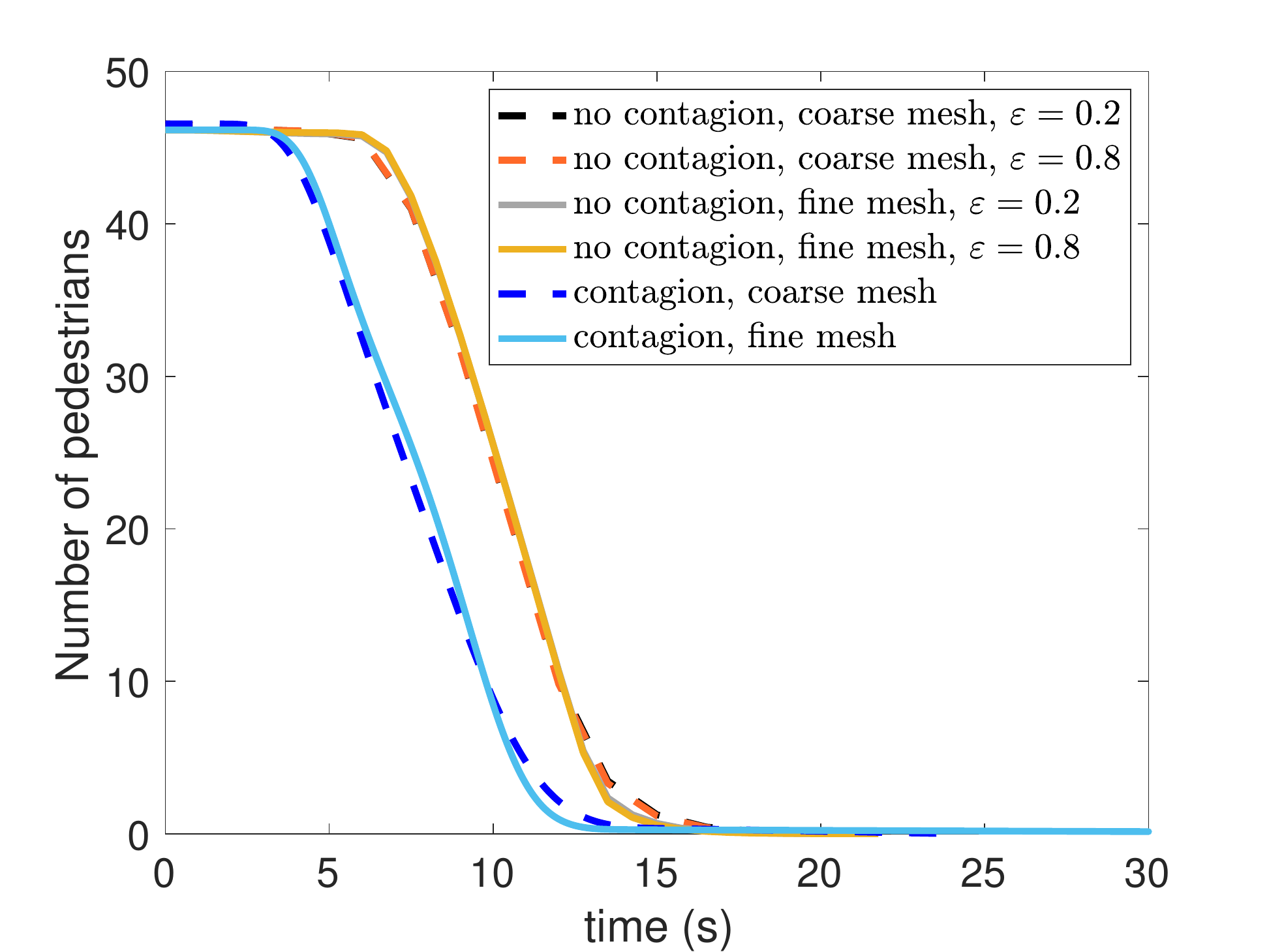}
\end{overpic}
\caption{Test 1: number of pedestrians inside the room over time computed with the model with and without emotional contagion, 
for both coarse and fine mesh. In the model with no emotional contagion, we set either low fear ($\epsilon = 0.2$)
or high fear ($\epsilon = 0.8$) in the entire domain.}
\label{Population_change}
\end{figure}

Figure~\ref{Population_change} reports the number of pedestrians left in the room 
computed with the coarse and fine mesh: we can see very good agreement for the two curves. 
In addition, in Figure~\ref{Population_change} we report the results obtained with no emotional contagion, i.e.~with 
the model in Ref.~\refcite{kim_quaini} with either low fear ($\epsilon = 0.2$)
or high fear ($\epsilon = 0.8$) in the entire domain (see Remark \ref{rem:ped_dyn}) and for both meshes under consideration. 
Since the model with no emotional contagion does not have a stringent condition like \eqref{eq:CFL} on the time step, 
we set $\Delta t = 1.5$ s for the coarse mesh and $\Delta t = 0.75$ s for the fine mesh. 
See Ref.~\refcite{kim_quaini} for more details. Fig.~\ref{Population_change} shows that the model with no emotional contagion
produces very similar results for $\epsilon = 0.2$ (low fear) and $\epsilon = 0.8$ (high fear), indicating
that its description of fear-induced behavior is too simplistic.
The model with emotional contagion predicts a faster evacuation dyanmics.
Recall that in the model with emotional contagion fear affects both the walking speed 
(which increases with the fear level) and the walking direction, while in the model with no emotional contagion it
influences only the latter.  
Our results in Figures~\ref{MM} and \ref{Population_change} 
confirm what observed in  Ref.~\refcite{Bellomo2019_new}:
the spreading of a stress condition significantly affects the overall evacuation dynamics and leads to overcrowding
in certain areas of the domain. 

\begin{figure}[ht]
\centering
\begin{overpic}[width=0.60\textwidth, grid=false]{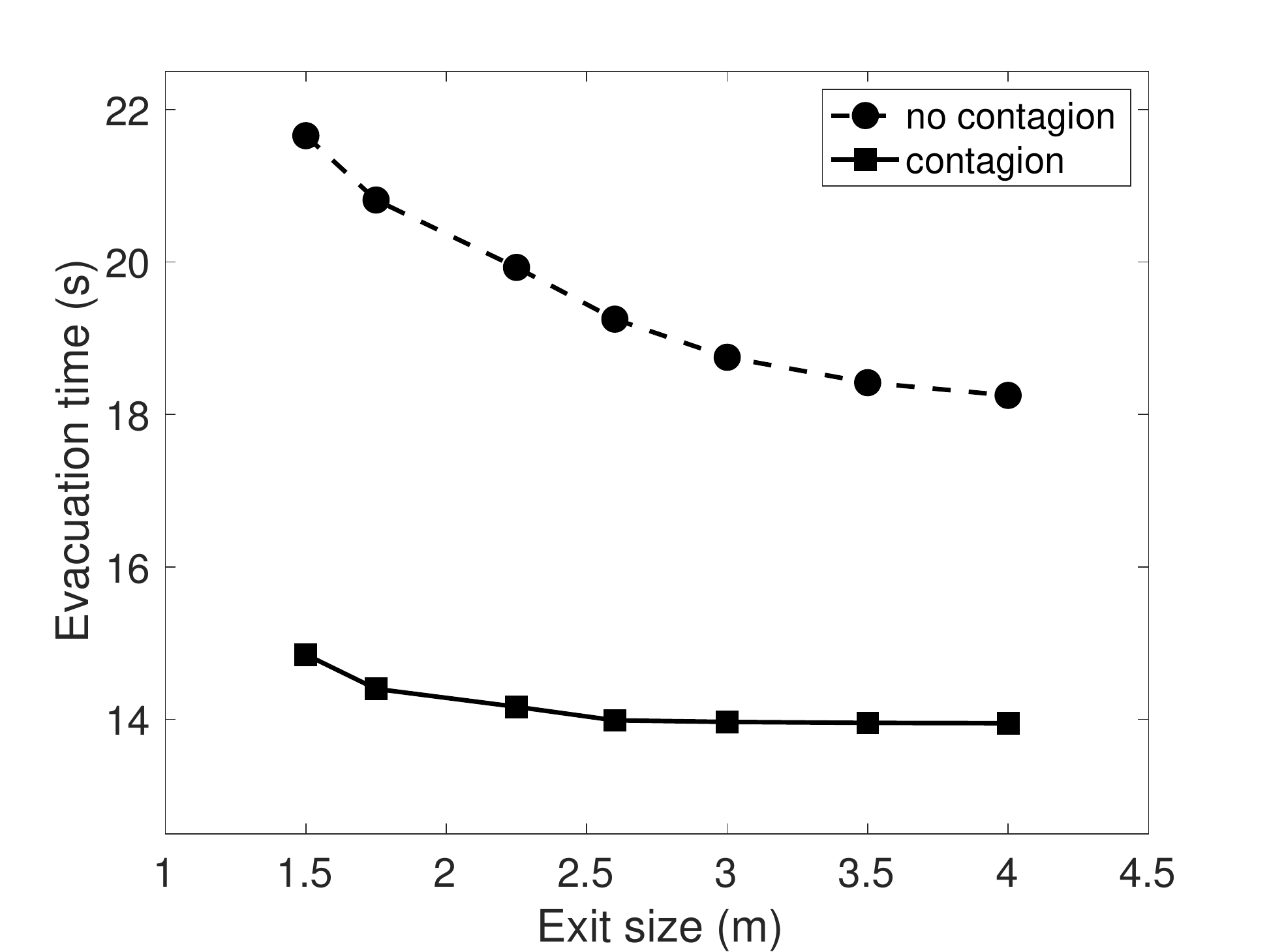}
\end{overpic}
\caption{Test 1: total evacuation time as a function of the exit size for the model with and without emotional contagion.} 
\label{Exit_sizes}
\end{figure}

Next, we consider the same room but we vary the exit size. 
We locate the exit symmetrically with respect to the room centerline and
let the exit size vary from 1.5 m to 4 m. 
We consider the coarse mesh mentioned before
since Figure~\ref{Population_change} show that is an appropriate choice 
for the problem under consideration. 
All the other model and discretization parameters are unchanged.
Figure~\ref{Exit_sizes} shows the total evacuation time as a function
of the exit size for the model with and without emotional contagion. 
We see that for all exit sizes the model with emotional contagion predicts a 
shorter evacuation time. 
The total evacuation time obtained with both models decreases with the
exit size, but that once the exit is large enough for the crowd contained in the room
the evacuation time does not change significantly if the exit is further enlarged.
This plateau is reached at a smaller exit size for the model with emotional contagion.
Finally, we notice that our results for the model with no emotional contagion are in very good agreement with the results 
reported in Ref.~\refcite{Agnelli2015}. 

The existing literature is not conclusive on how ``herding'' (i.e., the tendency to follow the stream)
influences the efficiency of collective crowd egress. \cite{Llorca_2019} 
Some works suggest that herding behavior is detriment to efficient evacuation (e.g., Refs.~\refcite{ZHENG20114627,HAGHANI201949}),
some others conclude that there may be scenarios where herding is beneficial to an escaping crowd (see, e.g., Ref.~\refcite{KIRCHNER2002260}),
while according to other works the effect of herding is still unestablished (see, e.g., Ref.~\refcite{Pan_2007,Moussaid_2016,LOVREGLIO2016421}).
The phenomenon of mixed strategy (i.e., a mixture of factors F3 and F4 in Sec.~\ref{modelinginteractions})
has been investigated by several numerical studies. Also on the efficacy of the mixed strategy, the literature is
inconclusive. Some works (e.g., Ref.~\refcite{Helbing2000}) suggest that a crowd can benefit from 
mixed strategies, while other (e.g., Ref.~\refcite{HAGHANI2019101064}) suggest that any percentage of 
herding strategy within the crowd has a negative impact on the evacuation efficiency. 
Our results agree with the former.

\subsection{Test 2:  evacuation from a room with one door and variable interaction radius}\label{sec:test2}

\begin{figure}[h!t]
\centering
\begin{overpic}[width=0.4\textwidth, grid=false]{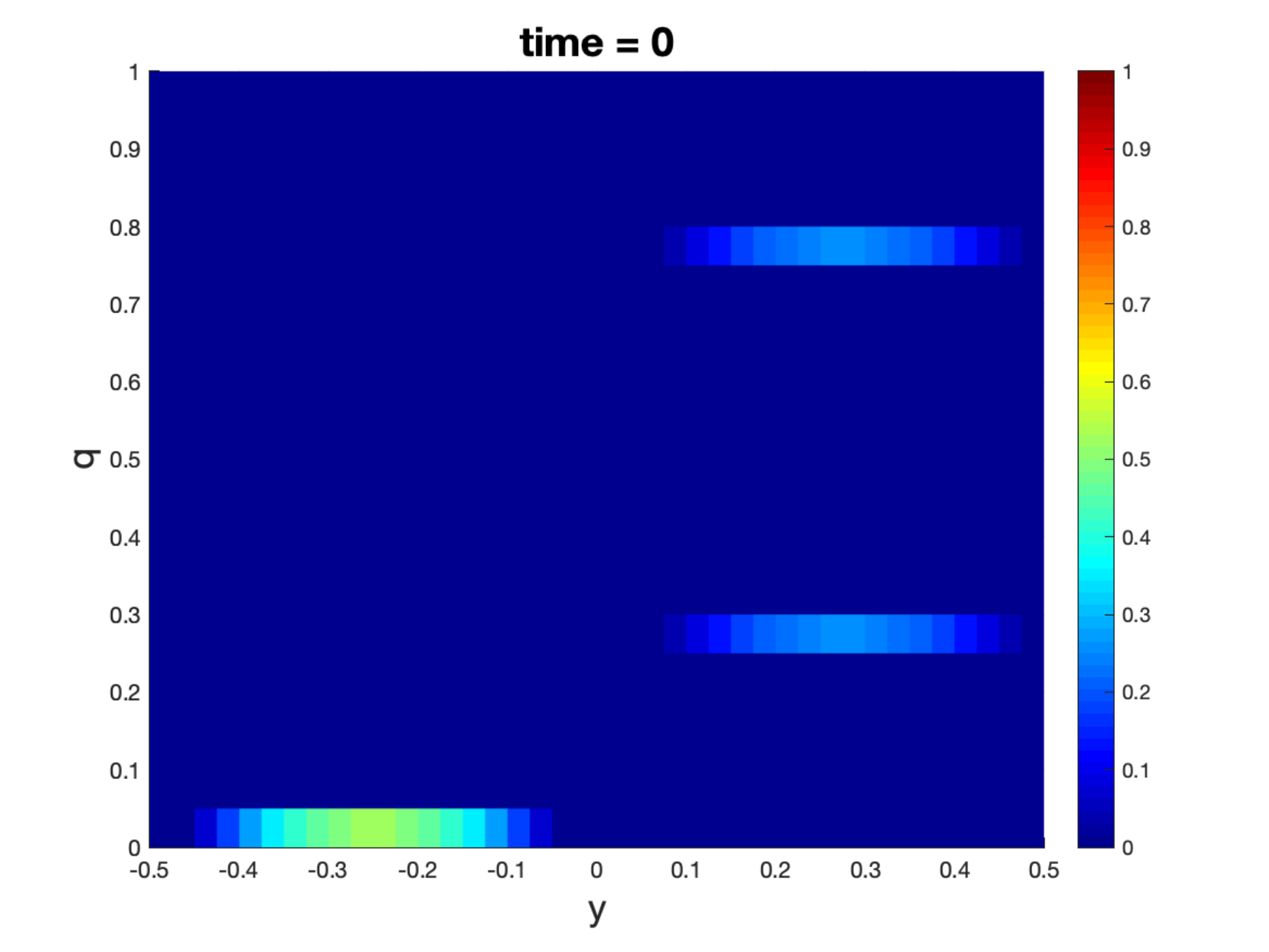}
\end{overpic}
\begin{overpic}[width=0.4\textwidth, grid=false]{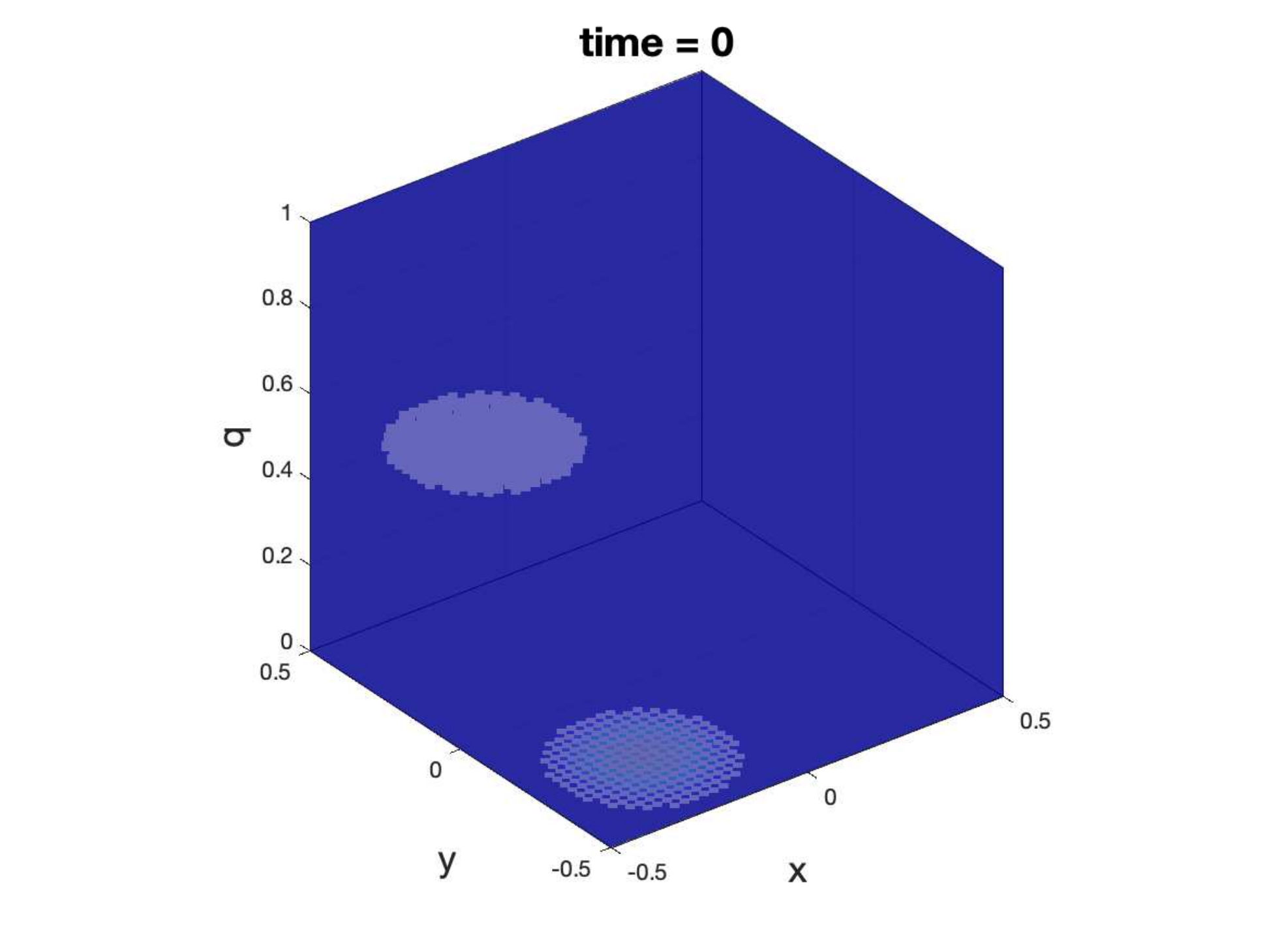}
\end{overpic}
\caption{Test 2: (left) $x = -0.25$ section of the initial distribution function $f(0, \x, q)$
and (right) initial average fear level.} 
\label{test2_initial}
\end{figure}

The computational domain is the same as for the test in Sec.~\ref{sec:test1}, with exit size $2.6$ m. 
The number of pedestrian, their initial distribution and walking direction are as in Sec.~\ref{sec:test1}.
The difference with respect to test  in Sec.~\ref{sec:test1} is in the initial fear distribution. 
The group centered at $(-0.25, -0.25)$ (see Fig.~\ref{MM}, left-most panel on the second row), 
which occupies a circle of radius 0.1414 and has initial direction $\theta=3$, 
has the following initial probability function:
\[f^3(0, \x, q) = g_1(x,y) \, \delta_{0}(q),\]
where
\[g_1(x,y)=-25.8(x+0.25)^2-25.8(y+0.25)^2+0.52, \]
and $\delta_{n}(q)$  is the Kronecker delta defined by $\delta_{n}(q)=1$ if $q=n$ and 
$\delta_{n}(q)=0$ if $q \neq n$.
As for the group centered at $(-0.25, 0.25)$, which also occupies a circle of radius 0.1414 but 
has initial direction $\theta=7$, we have the following initial probability function:
\[f^7(0, \x, q) =  g_2(x,y) \, \Big[ (1/2)\,  \delta_{1/4}(q) +  (1/2)\, \delta_{3/4}(q) \Big ], \]
where
\[g_2(x,y)=-25.8(x+0.25)^2-25.8(y-0.25)^2+0.52. \]
See the $x = -0.25$ section of the initial distribution function $f(0, \x, q)$ 
and initial average fear level in Fig.~\ref{test2_initial}. 

\begin{figure}[ht]
\centering
\begin{overpic}[width=0.60\textwidth, grid=false]{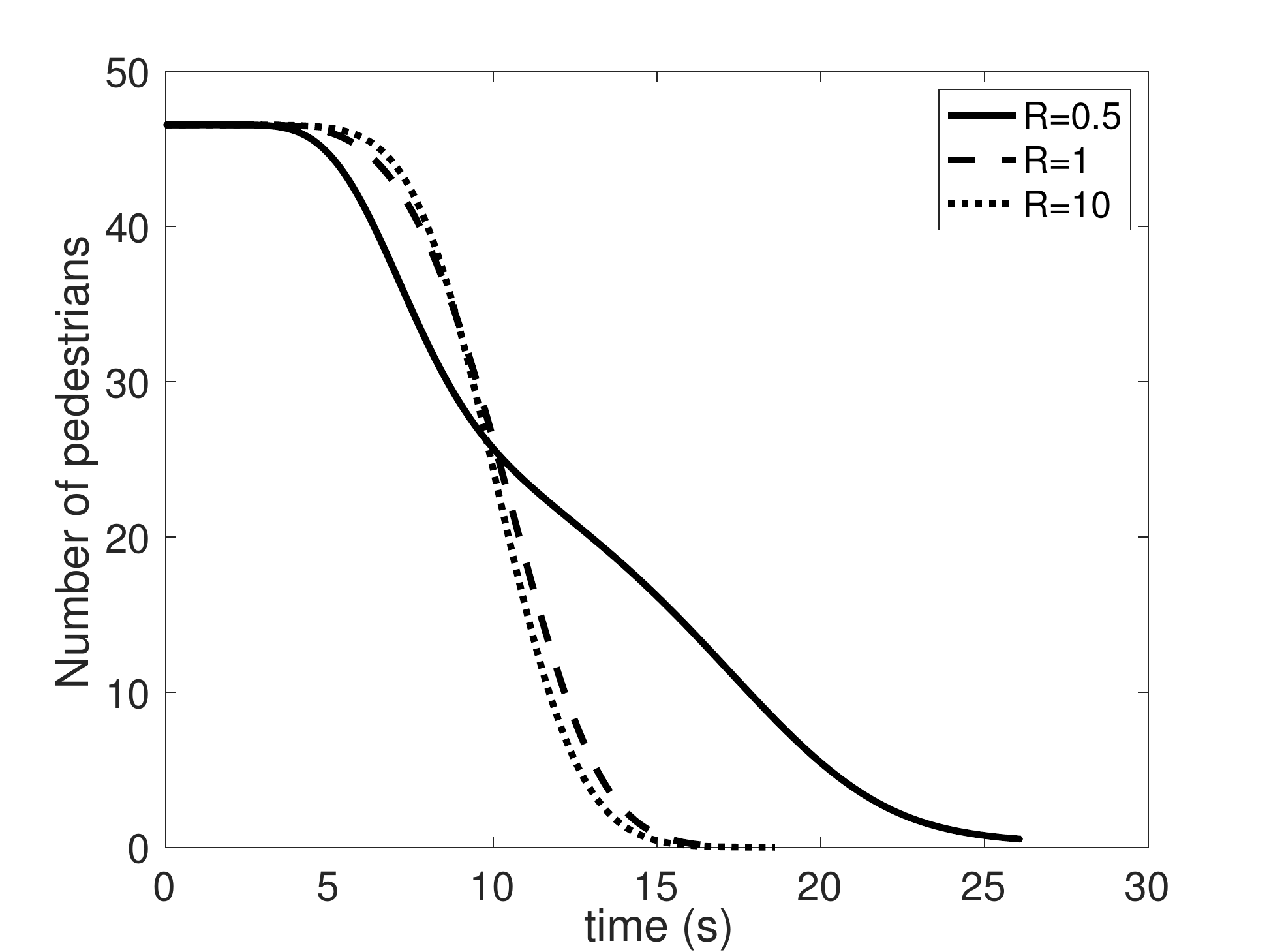}
\end{overpic}
\caption{Test 2: number of pedestrians inside the room over time computed 
for $R = 0.5, 1, 10$.}
\label{Population_change_test2}
\end{figure}

\begin{figure}[h!t]
\centering
\begin{overpic}[width=0.25\textwidth, grid=false]{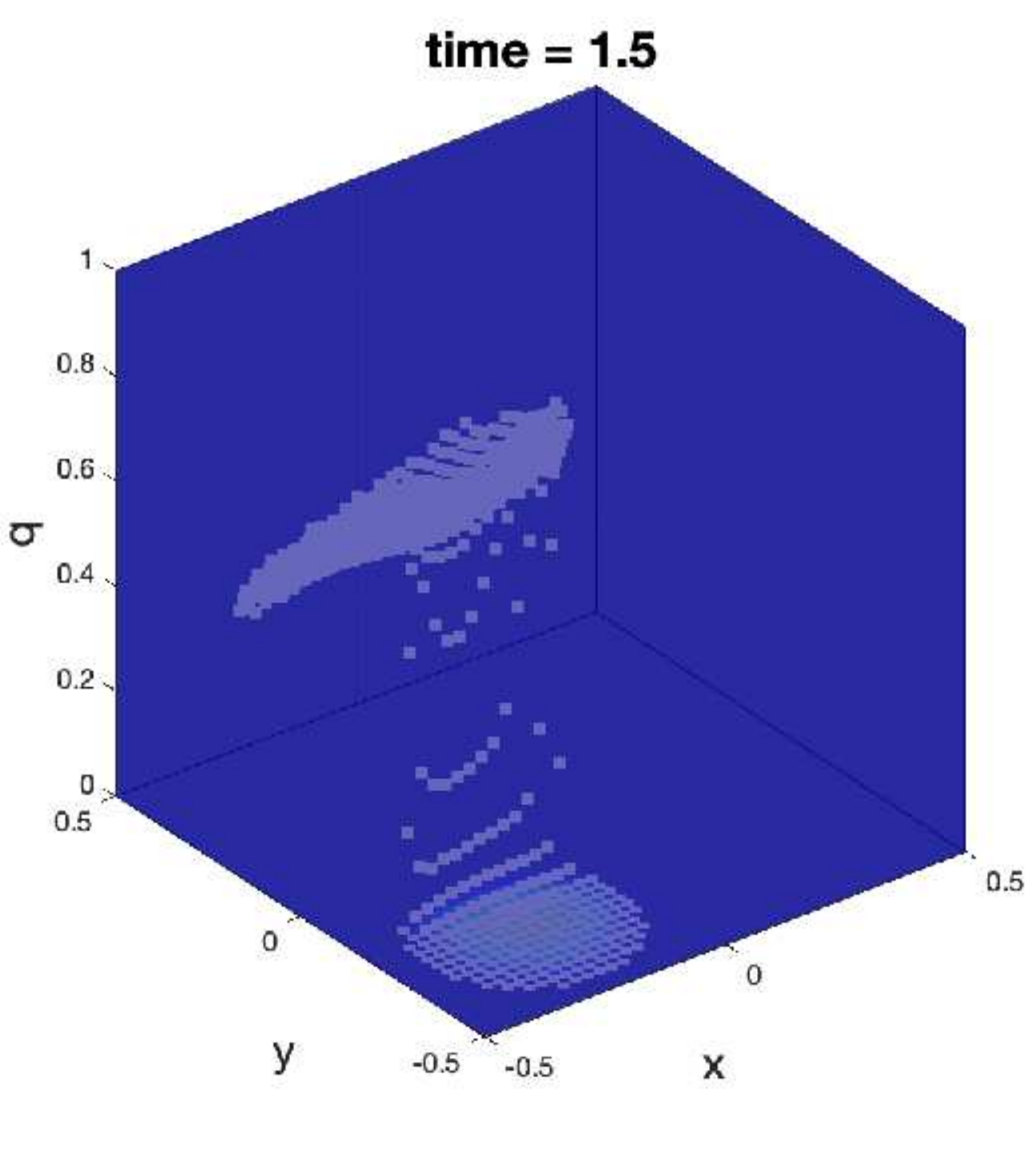}
\put(-40, 50){$R = 0.5$}
\end{overpic}
\begin{overpic}[width=0.25\textwidth, grid=false]{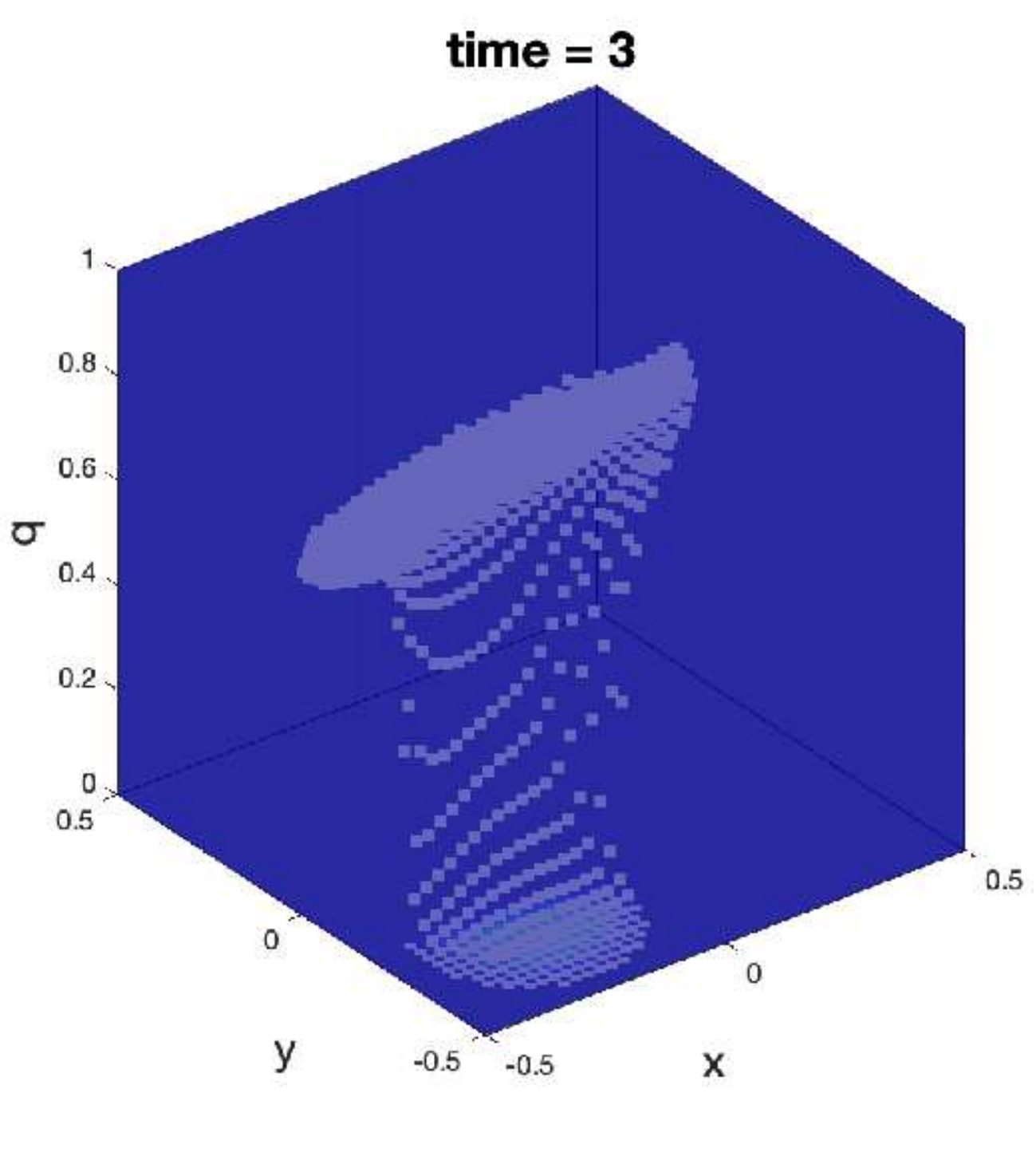}
\end{overpic}
\begin{overpic}[width=0.25\textwidth, grid=false]{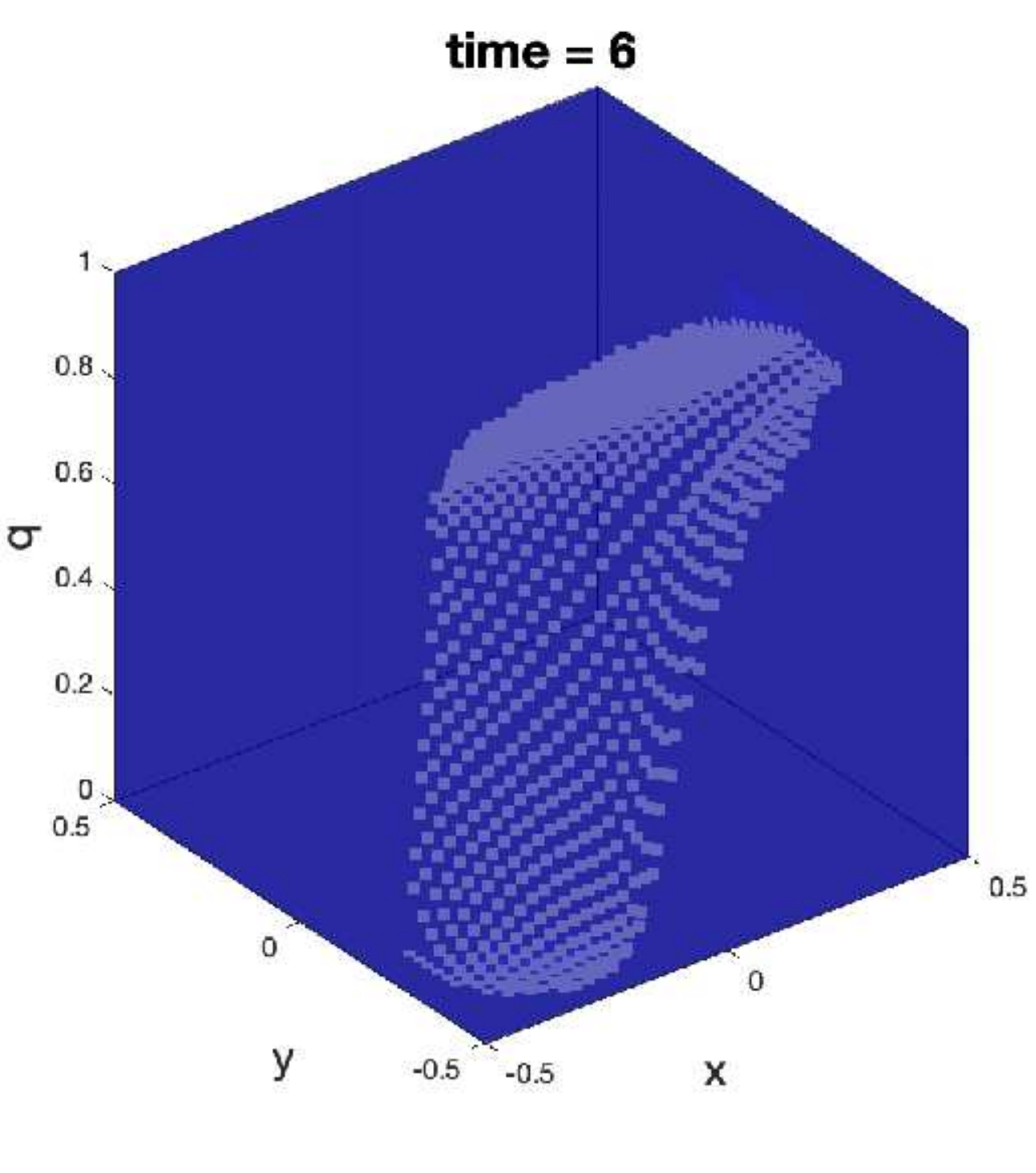}
\end{overpic}
\begin{overpic}[width=0.25\textwidth, grid=false]{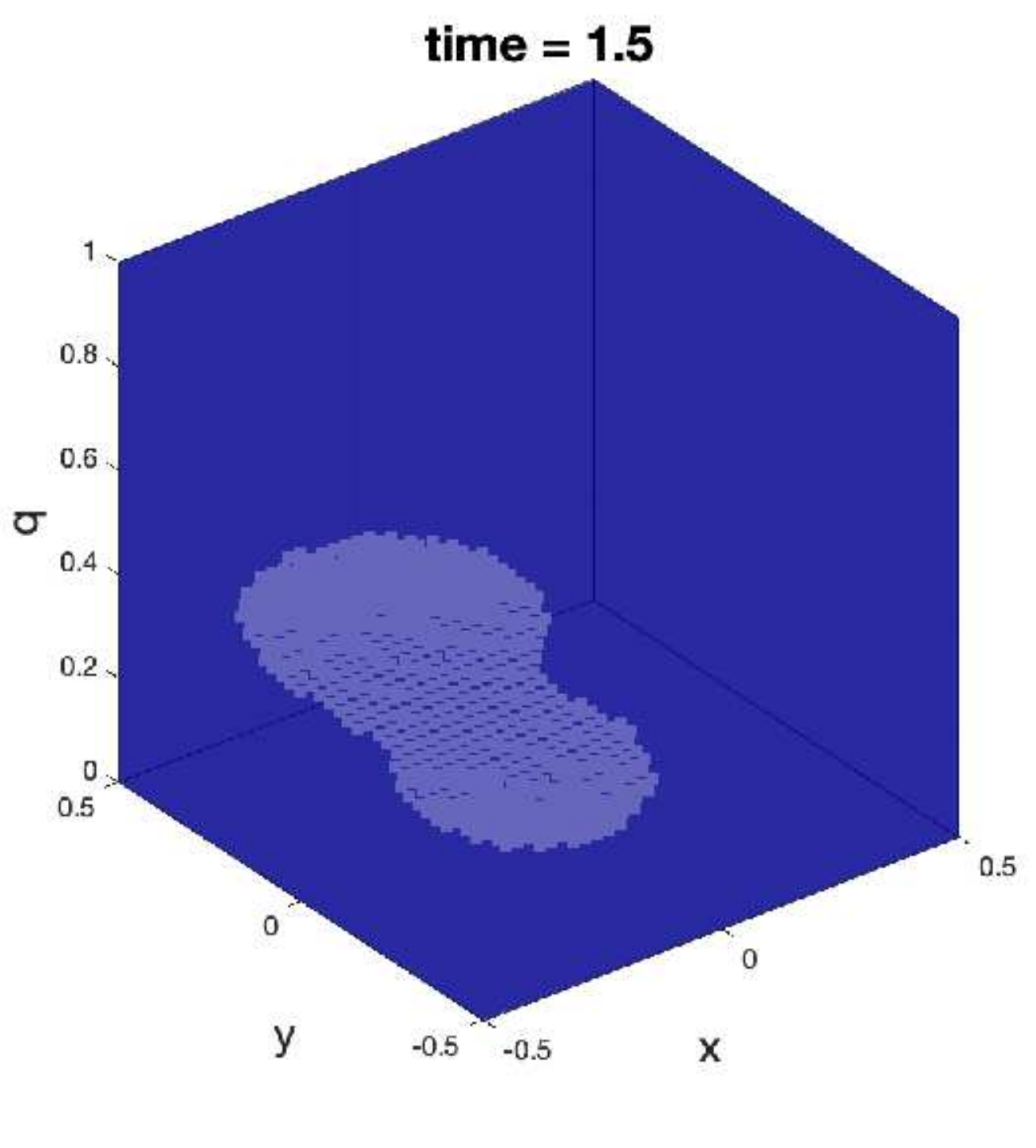}
\put(-40, 50){$R = 1$}
\end{overpic}
\begin{overpic}[width=0.25\textwidth, grid=false]{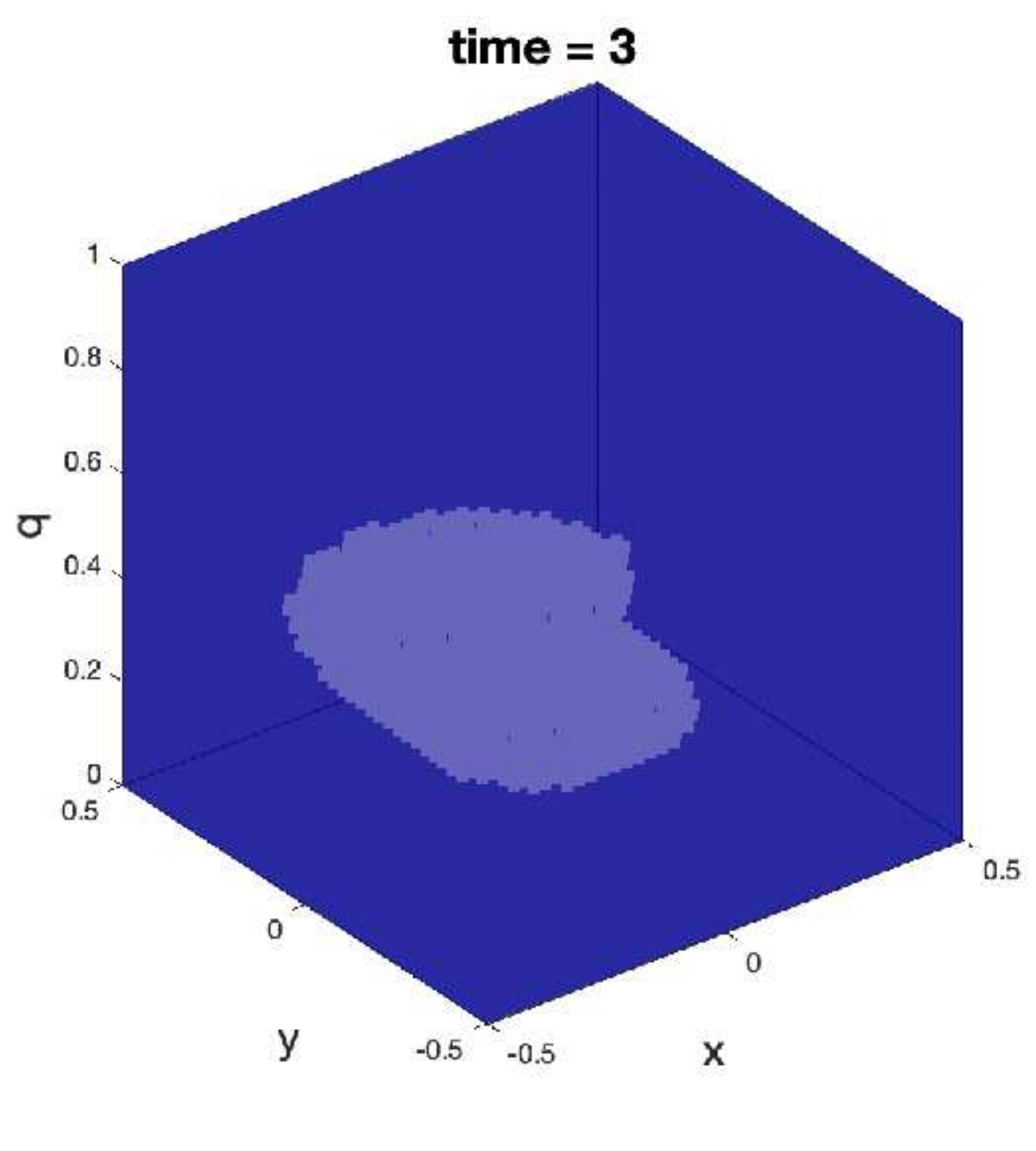}
\end{overpic}
\begin{overpic}[width=0.25\textwidth, grid=false]{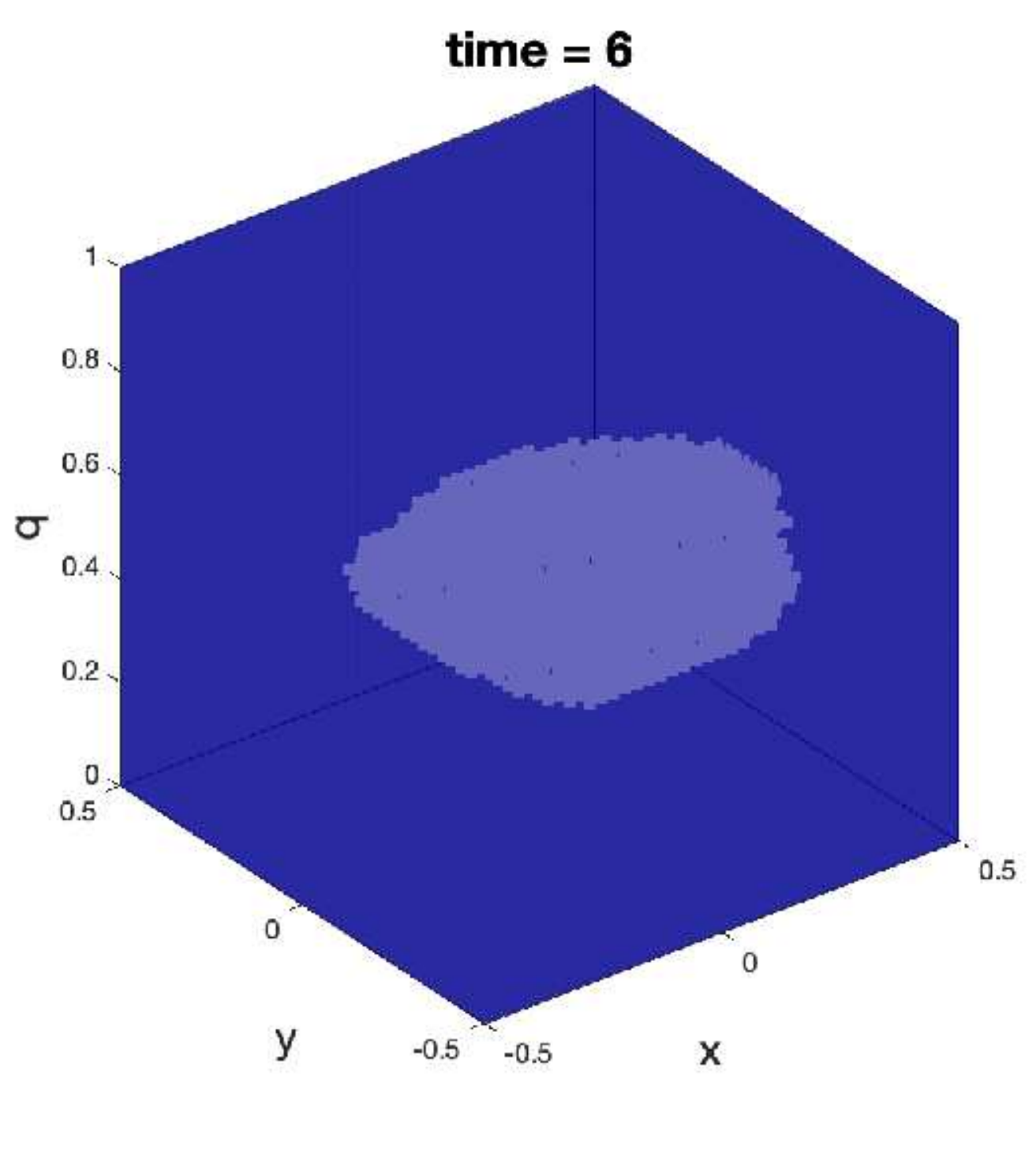}
\end{overpic}
\begin{overpic}[width=0.25\textwidth, grid=false]{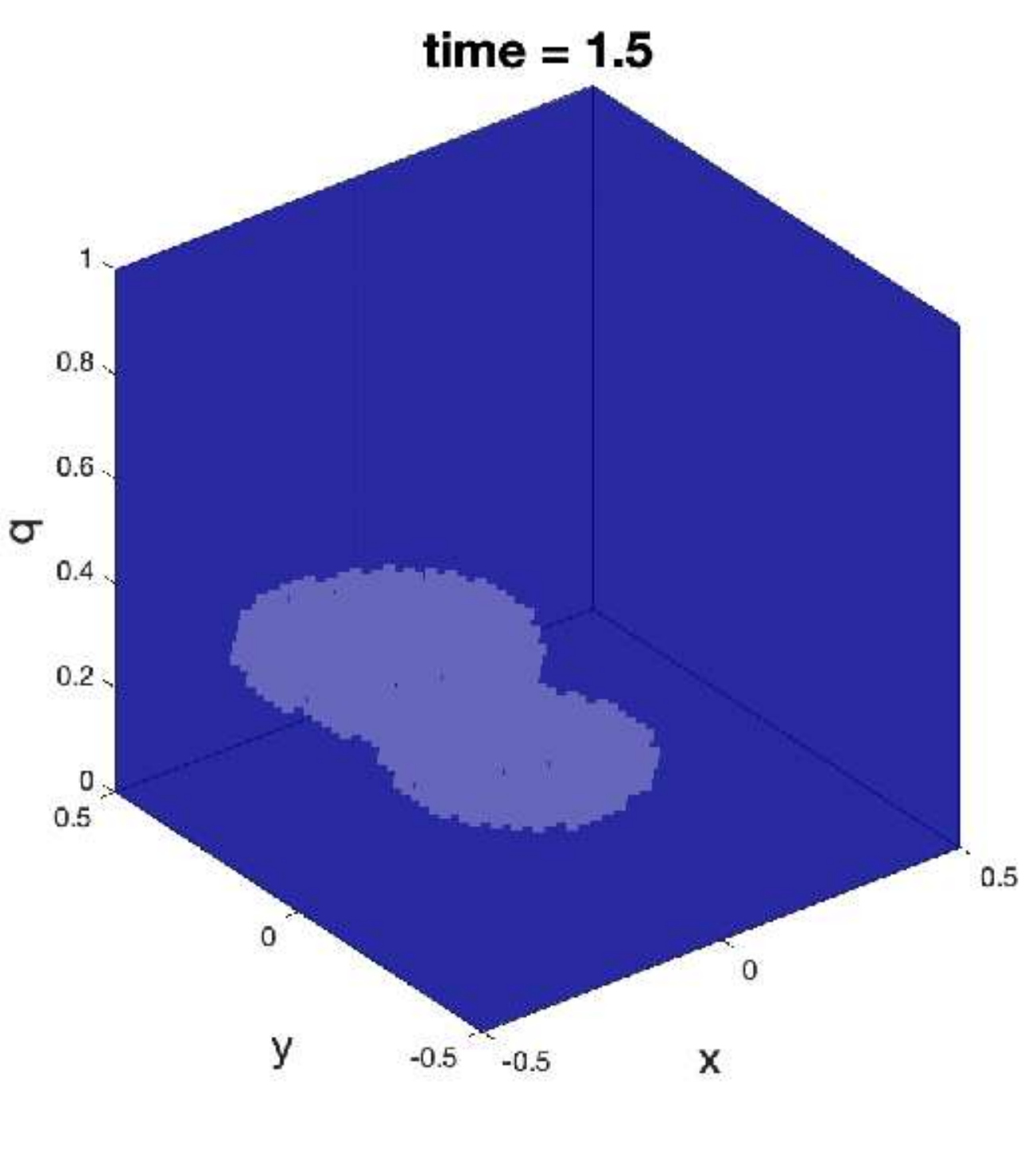}
\put(-40, 50){$R = 10$}
\end{overpic}
\begin{overpic}[width=0.25\textwidth, grid=false]{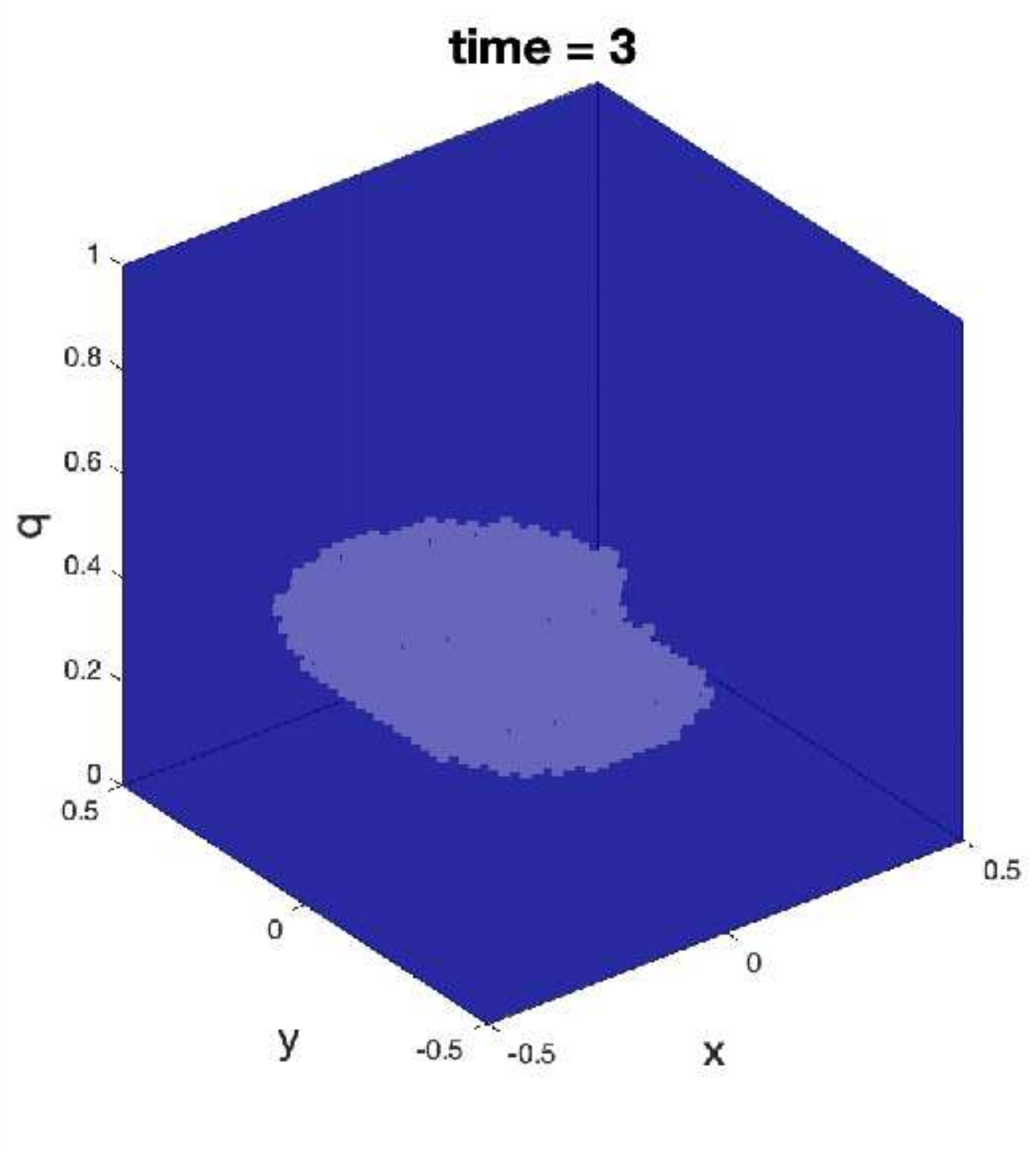}
\end{overpic}
\begin{overpic}[width=0.25\textwidth, grid=false]{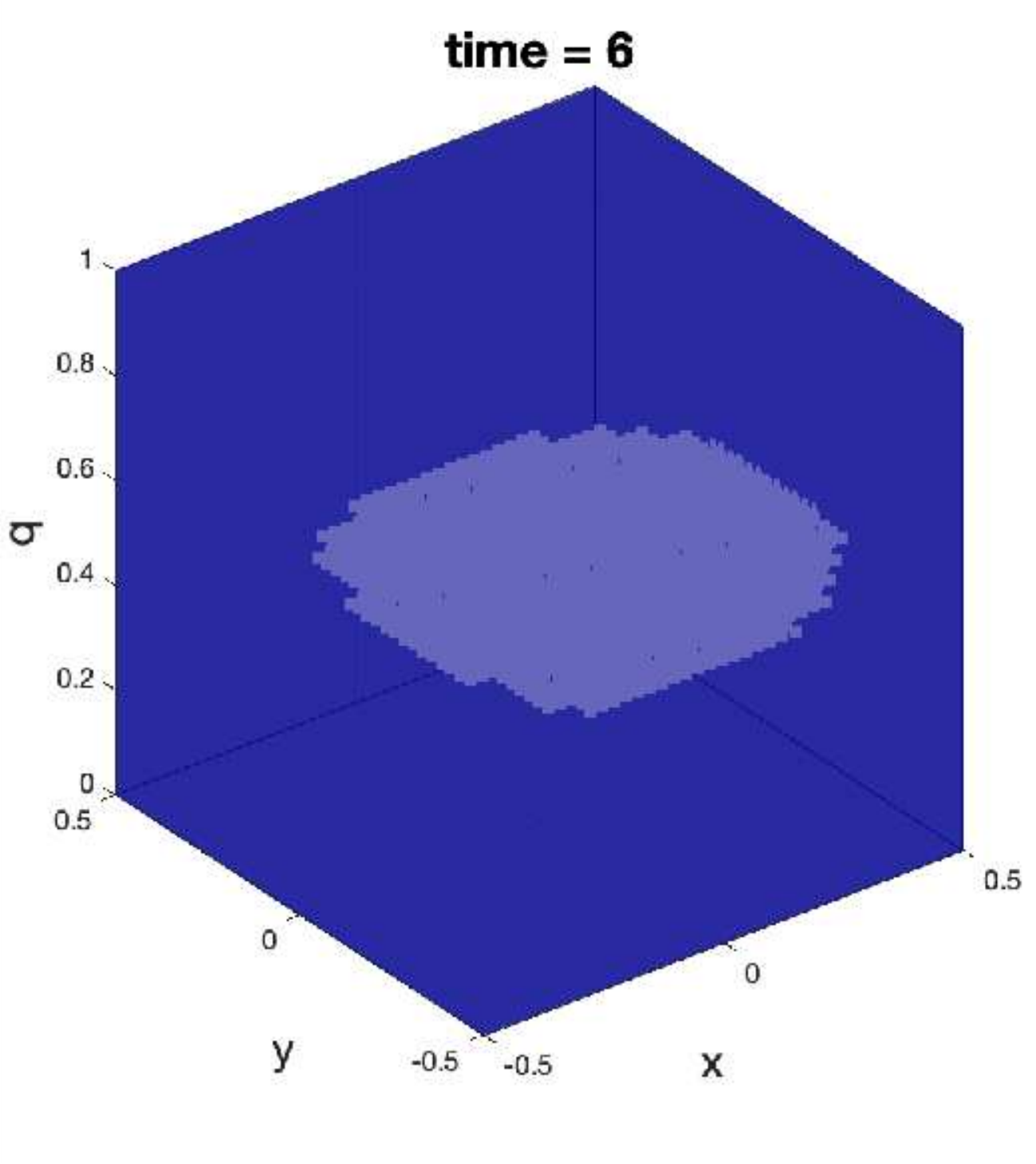}
\end{overpic}
\caption{Test 2: Evolution of the average fear level computed with the fine mesh for $R = 0.5$ (top row), 
$R = 1$ (middle row), and $R = 10$ (bottom row).
} 
\label{test2}
\end{figure}

We simulate evacuation from the room using three different choices of $R$: $R = 0.5$ (like
in Sec.~\ref{sec:test1}), $R = 1$, and $R = 10$, which corresponds
to the case where the fear level of each person 
is influenced by the fear level of everybody in the room.
Figure~\ref{Population_change_test2} reports the number of pedestrians inside the room 
over time computed with the fine mesh for the three values of $R$.
Figure~\ref{Population_change_test2} shows that there is a big difference in the evacuation dynamics
when going from  $R = 0.5$ to $R = 1$, while the change is not substantial when going
from $R = 1$ to $R = 10$. See also Fig.~\ref{test2}, which shows the evolution of the average
fear level. When $R = 0.5$, part of the group with an initially higher fear level 
that does not get in contact with the lower fear group and leaves the room quickly. This explains the initially 
fast evacuation dynamics for $R = 0.5$ seen in Figure~\ref{Population_change_test2}. A part of the
group with an initially lower fear level gets in contact with the higher fear group and thus picks up
speed. Nonetheless, the emotion spreads slowly and in fact the last 23 people in the room take longer to leave the
room. For $R = 1, 10$, the emotion spreads fast and people evacuate much faster.

\subsection{Test 3: experiment in ant colony}\label{sec:test3}

We aim at reproducing one of the experiments in Ref.~\refcite{SHIWAKOTI20111433}.
In the experimental setting, 200 ants are placed in a square chamber
(31 mm $\times$ 31 mm) with exit at the corner of the walls that is 2.5 mm wide.
At a certain time, citronella (an insect repellent)
is injected through a small hole created at the center of the square chamber
to create panic. We start our simulations right after the citronella has been injected. 
In Ref.~\refcite{SHIWAKOTI20111433}, it is observed that when citronella is introduced, ants rush to the exit in 
a manner reminiscent of humans in a crowd panic.
The evacuation time to the nearest second
and the number of ants escaped reported in Ref.~\refcite{SHIWAKOTI20111433}
were measured by manual counting from playback of digital video
recordings. The measured mean (out of 10 trials) escape time is 
11.2 s ($\pm$ 2.6 s standard deviation) for the first 50 ants. 
We aim at showing that with a suitable parameter set-up and initial condition, 
our model can match this evacuation time. 
\rev{For this study, the parameters were set using our experience with the previous tests
and by trial-and-error. More sophisticated parameter estimation techniques will be adopted
for a follow-up paper.
}

In order to work with dimensionless quantities, 
we define the following reference quantities: $D=31\sqrt2$ mm, $V_M= 1$ mm/s, $T = D/V_M= 31\sqrt{2}$ s, and
$\rho_M = 4$ ants/mm$^2$.

\begin{figure}[h!t]
\centering
\begin{overpic}[width=0.32\textwidth, grid=false]{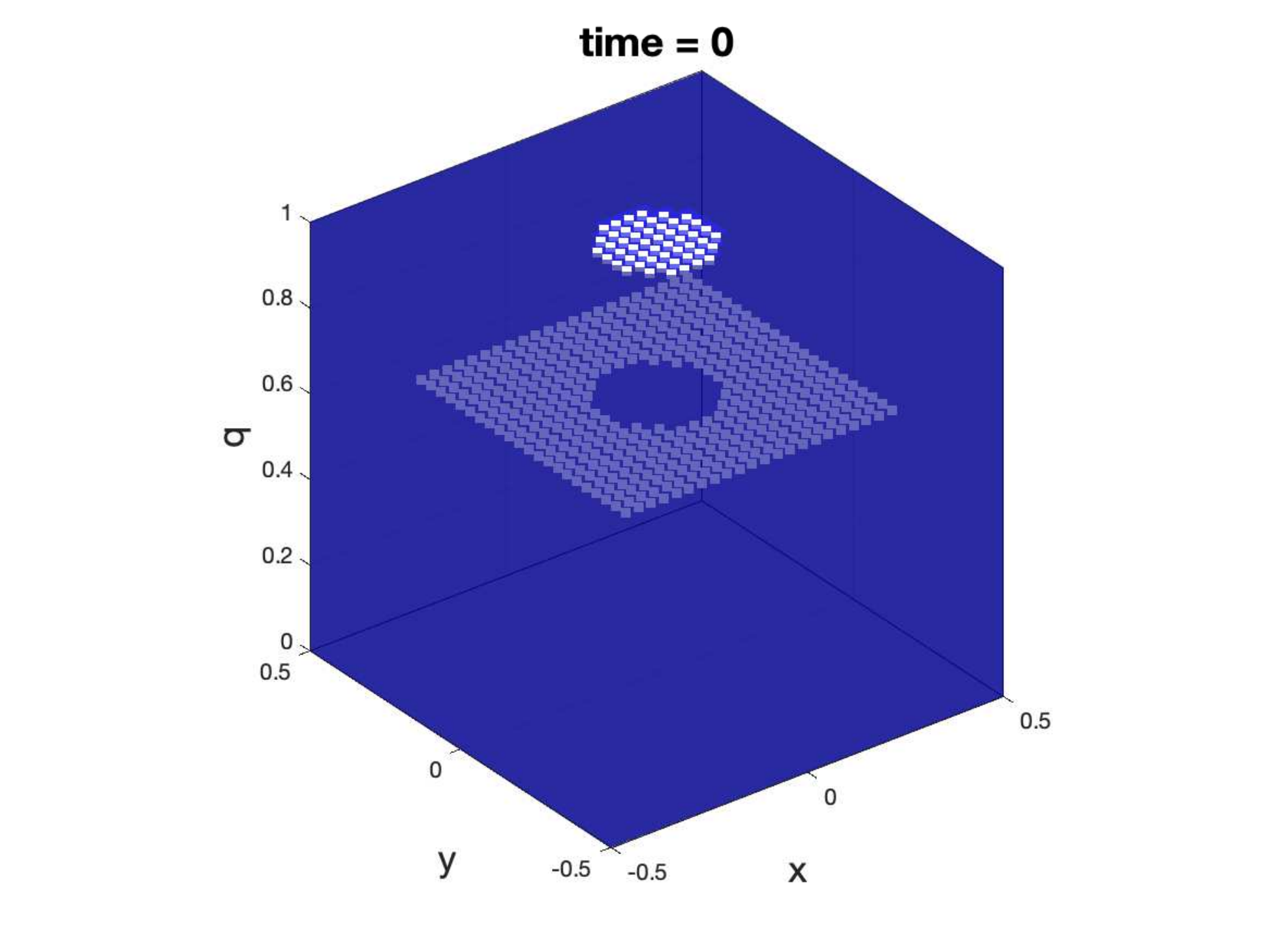}
\end{overpic}
\begin{overpic}[width=0.32\textwidth, grid=false]{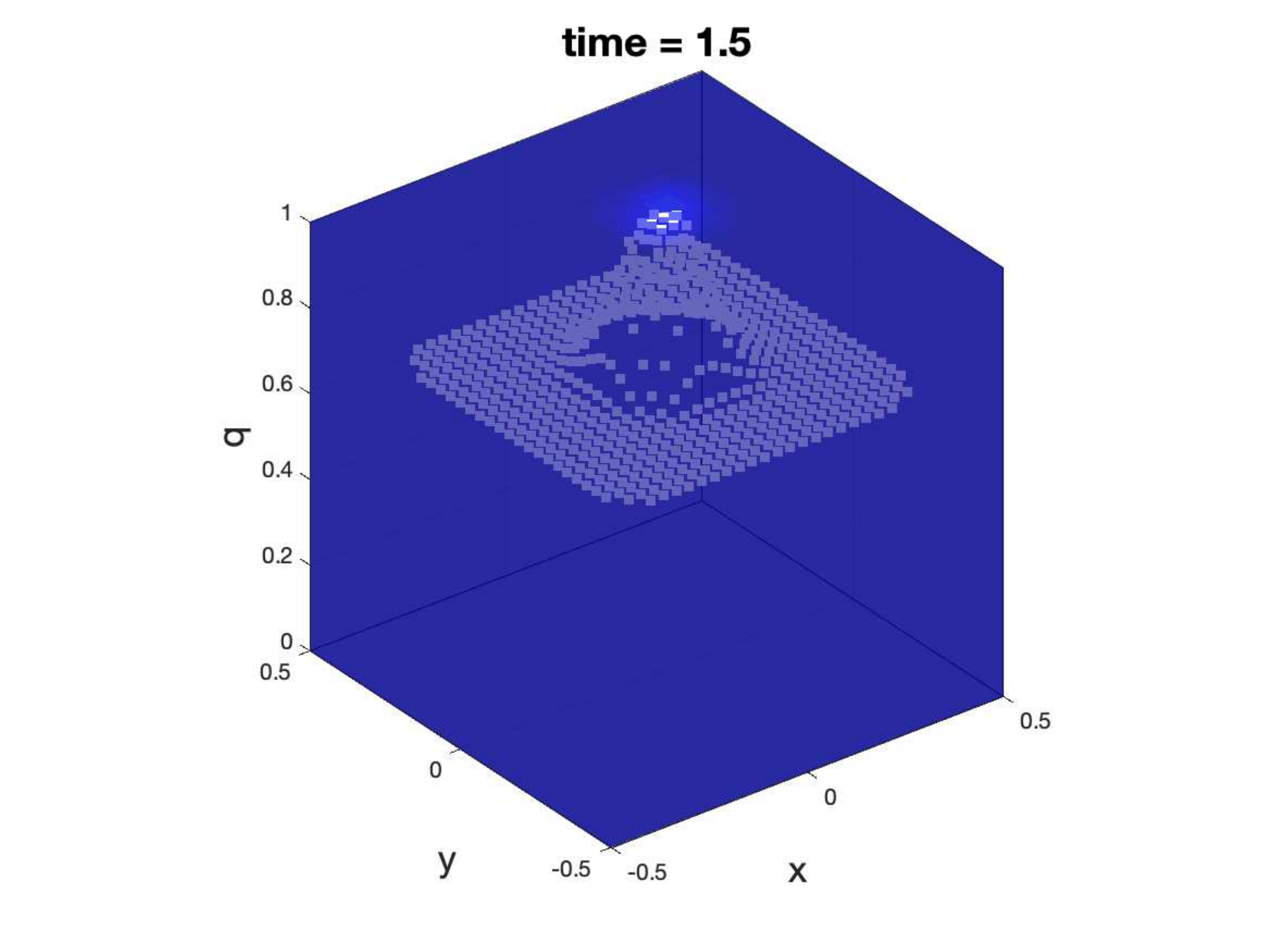}
\end{overpic}
\begin{overpic}[width=0.32\textwidth, grid=false]{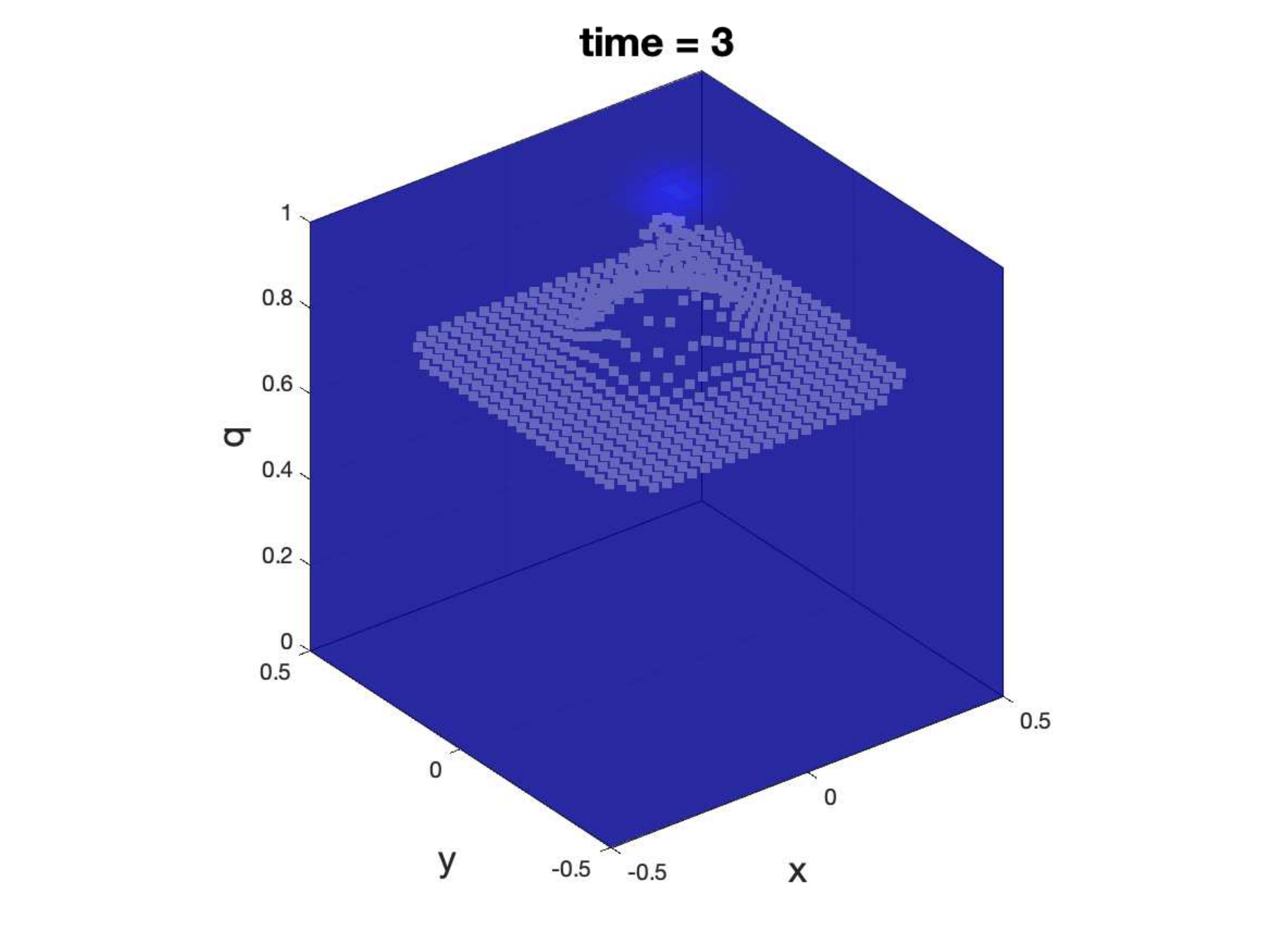}
\end{overpic}
\begin{overpic}[width=0.32\textwidth, grid=false]{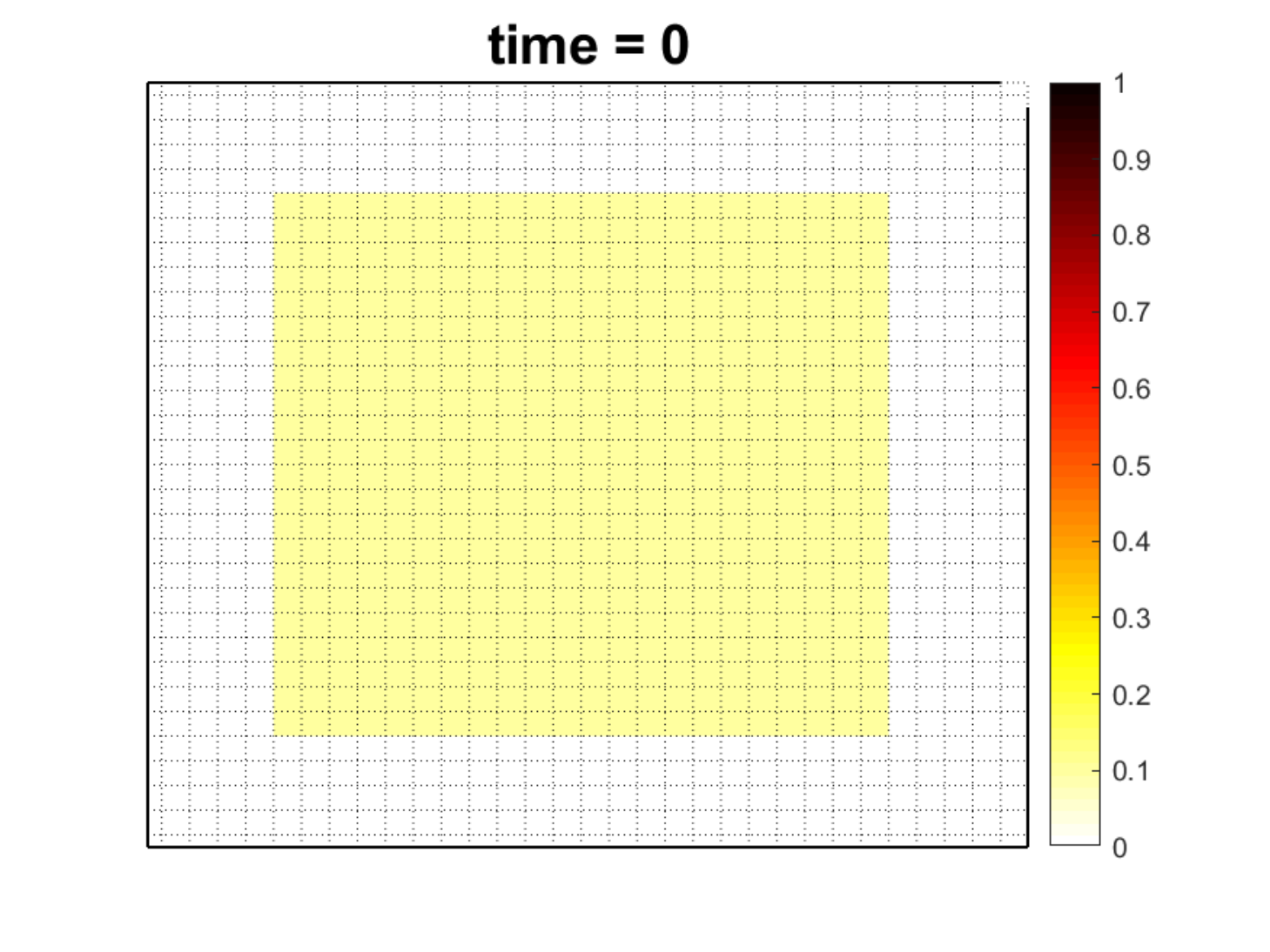}
\end{overpic}
\begin{overpic}[width=0.32\textwidth, grid=false]{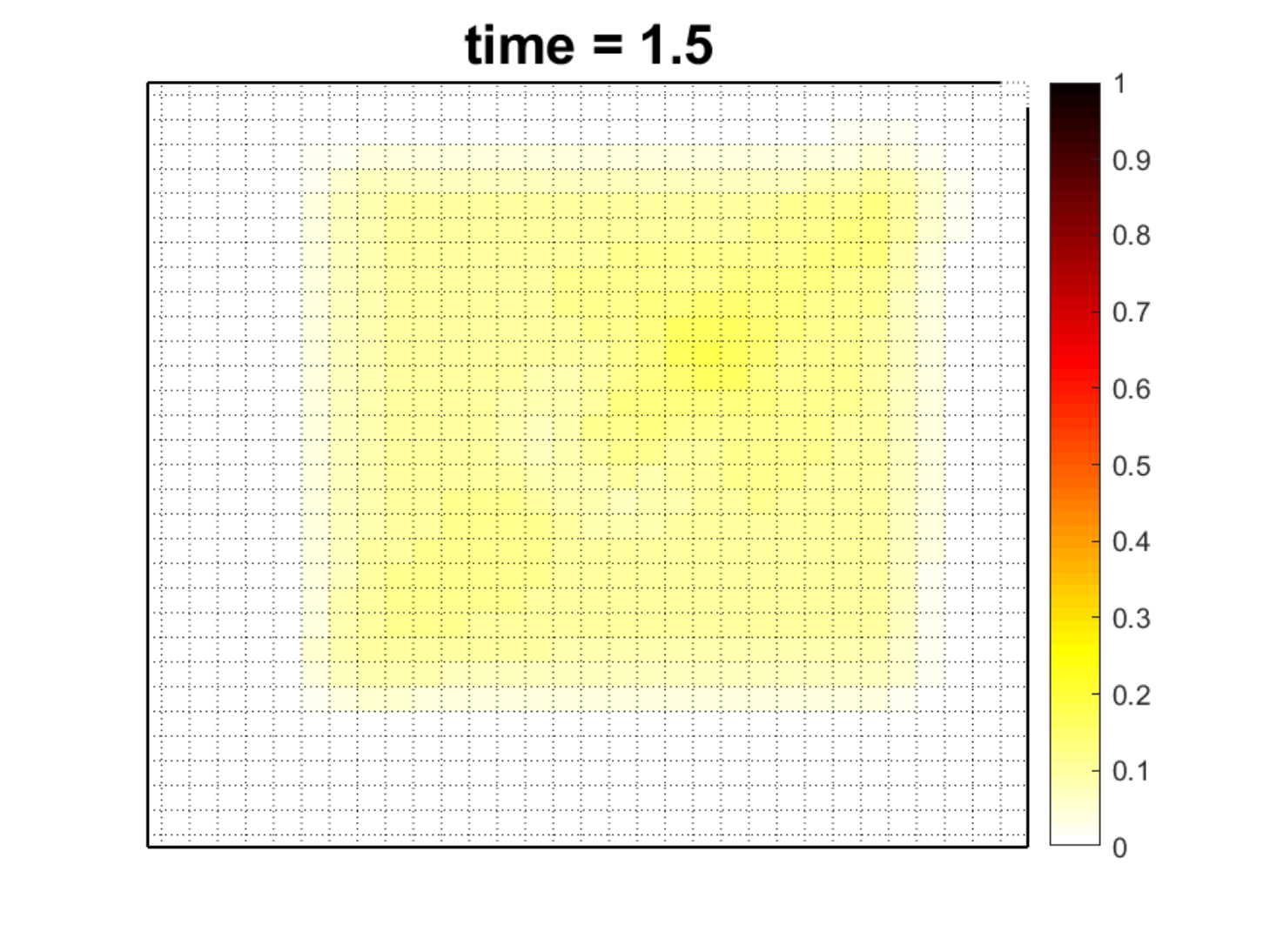}
\end{overpic}
\begin{overpic}[width=0.32\textwidth, grid=false]{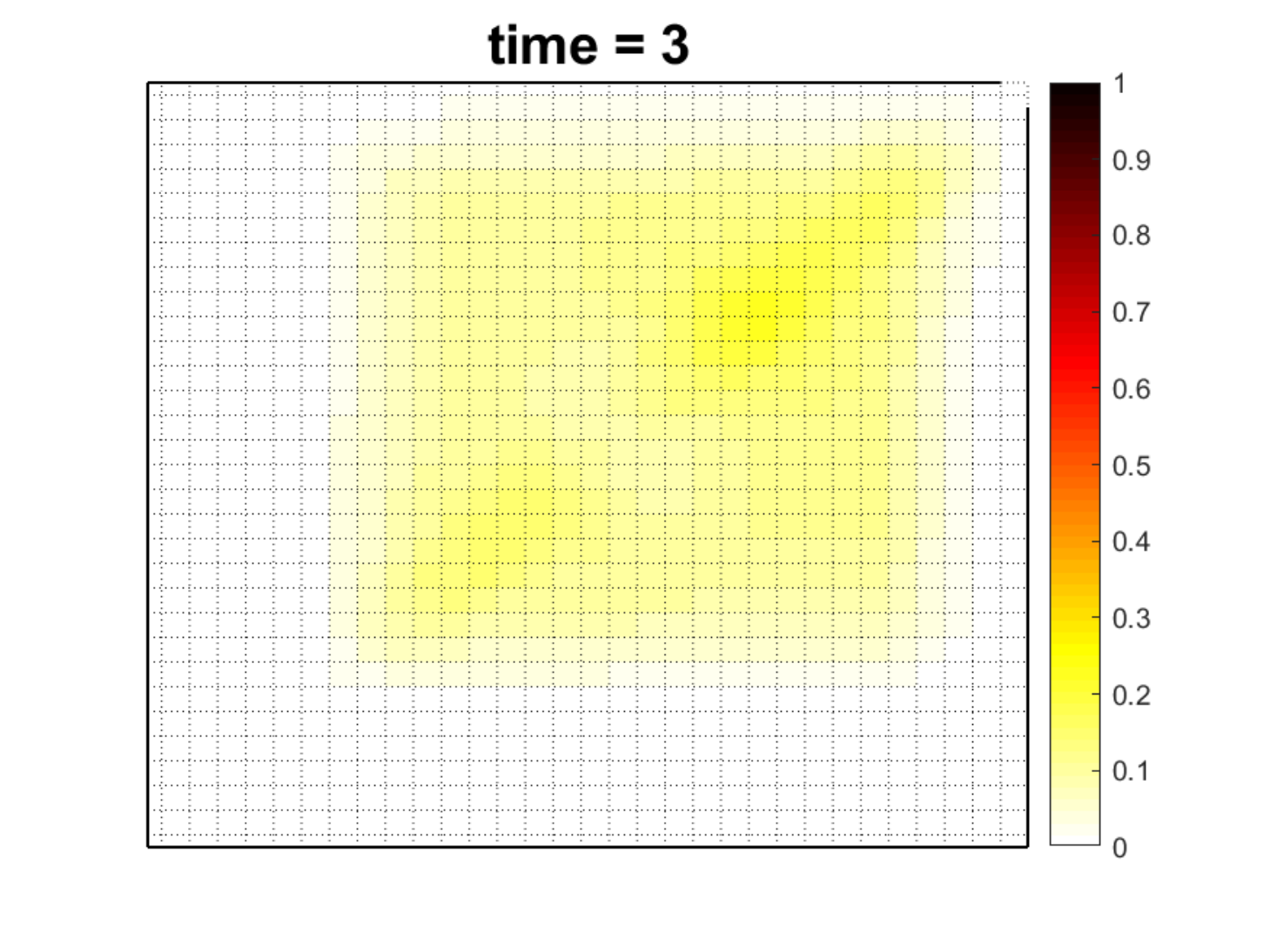}
\end{overpic}
\begin{overpic}[width=0.32\textwidth, grid=false]{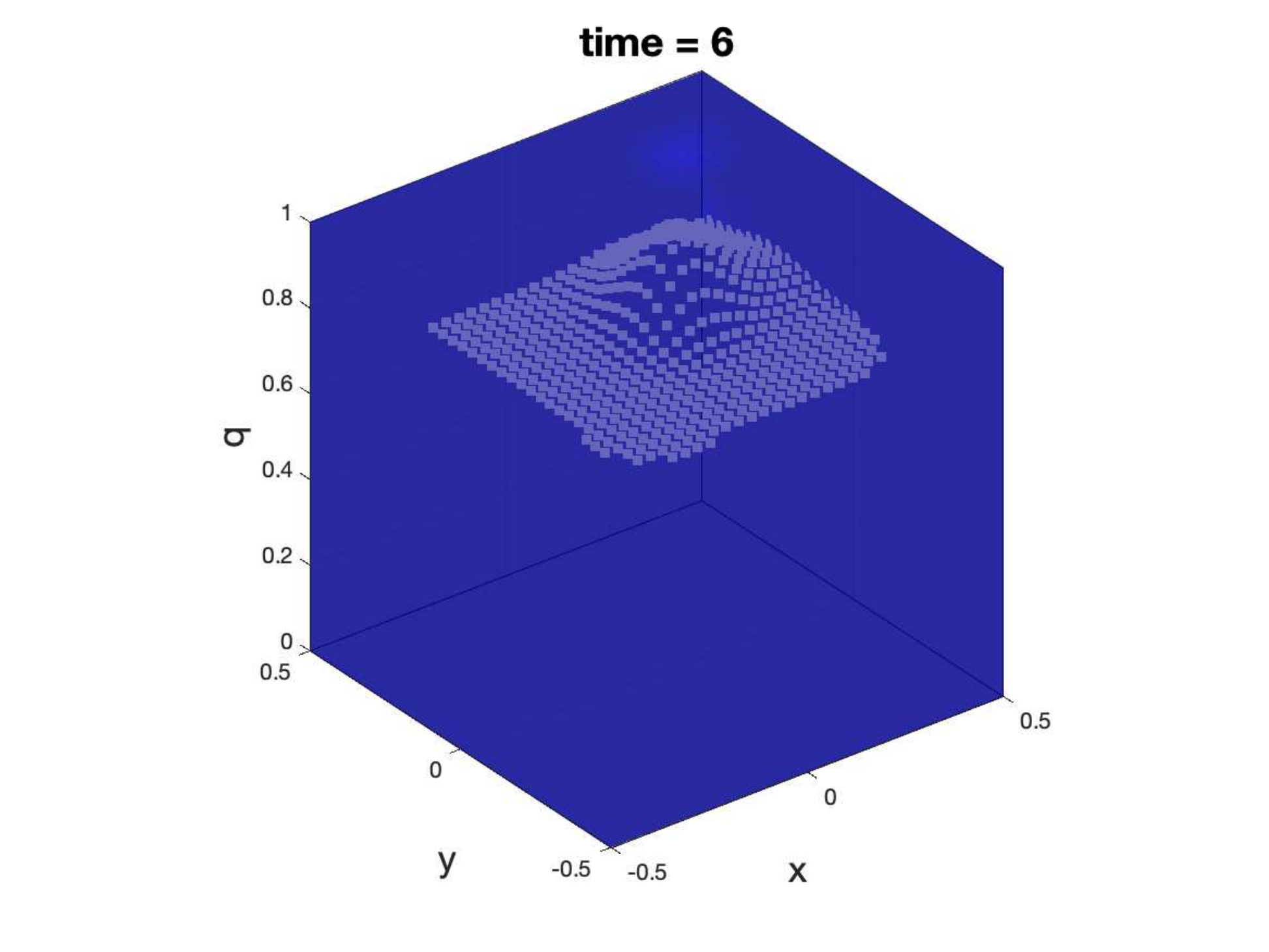}
\end{overpic}
\begin{overpic}[width=0.32\textwidth, grid=false]{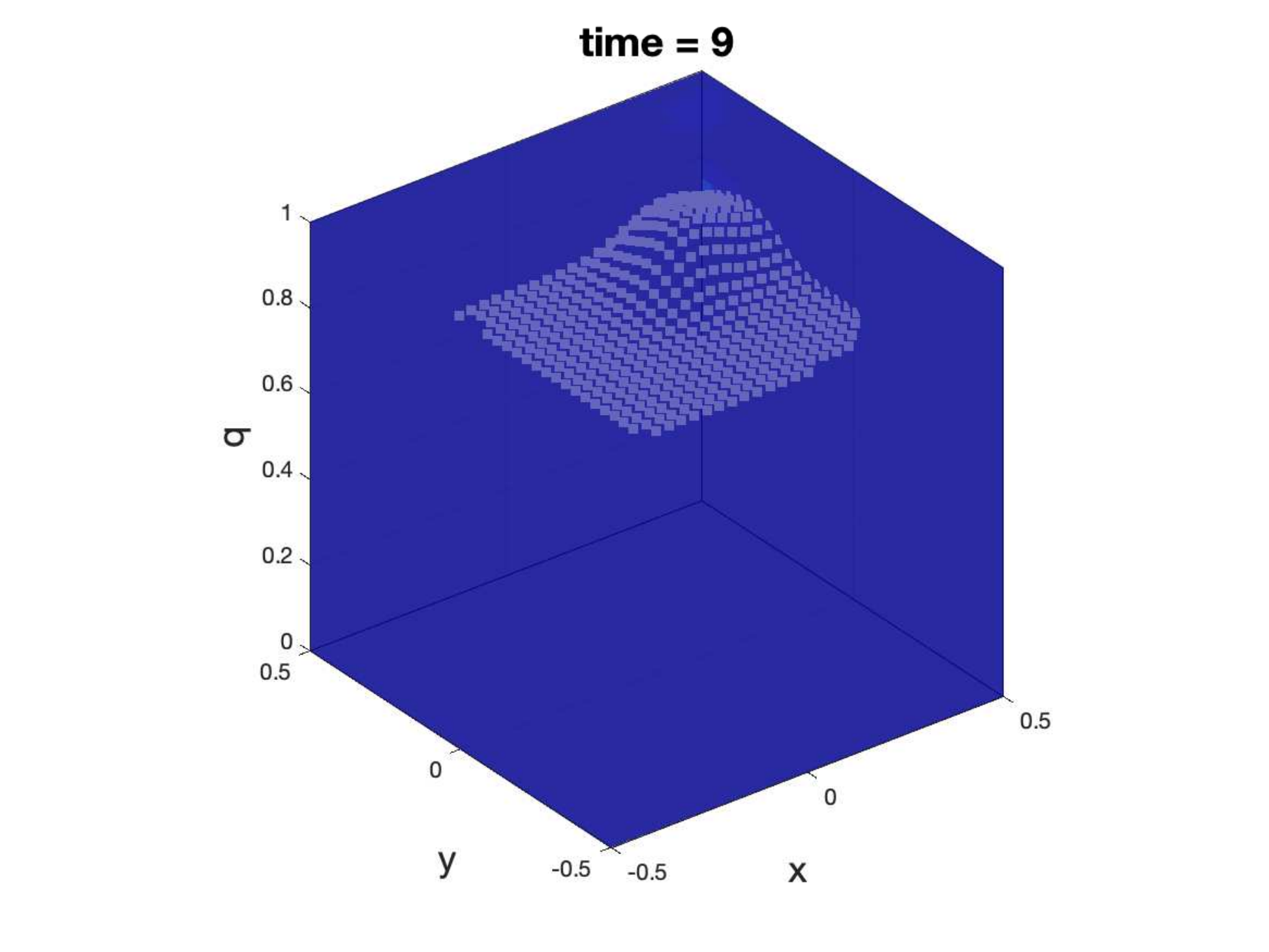}
\end{overpic}
\begin{overpic}[width=0.32\textwidth, grid=false]{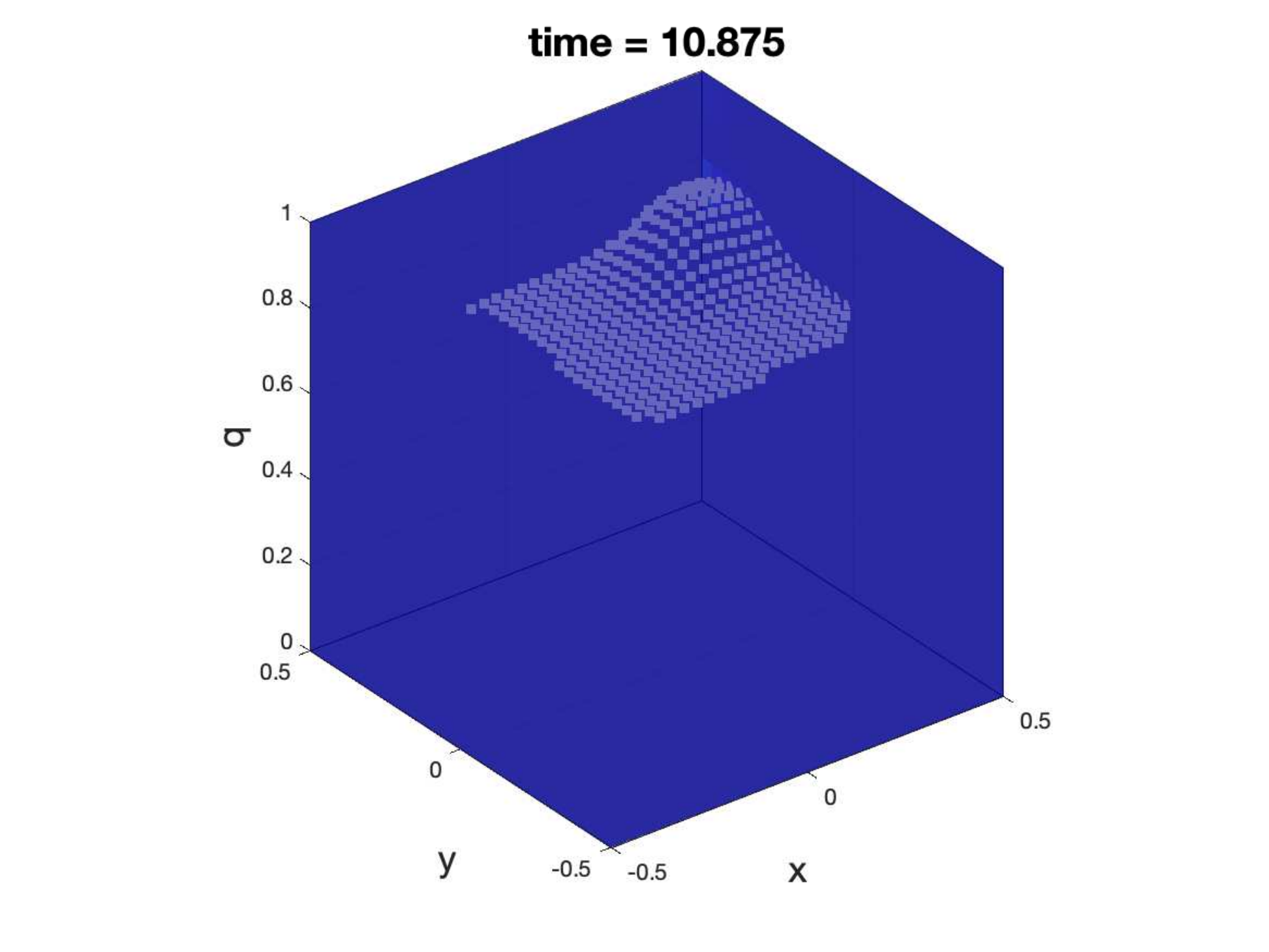}
\end{overpic}
\begin{overpic}[width=0.32\textwidth, grid=false]{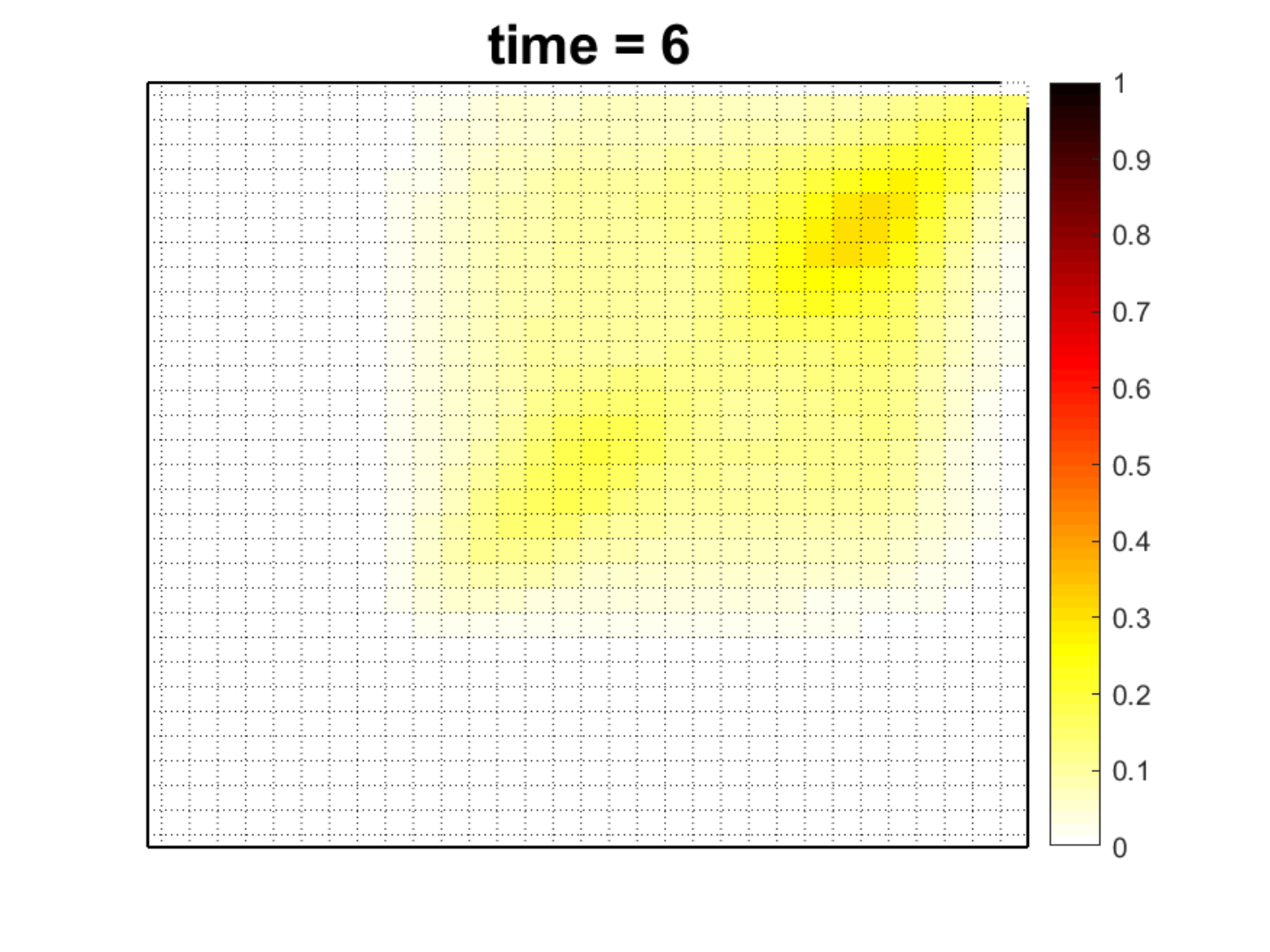}
\end{overpic}
\begin{overpic}[width=0.32\textwidth, grid=false]{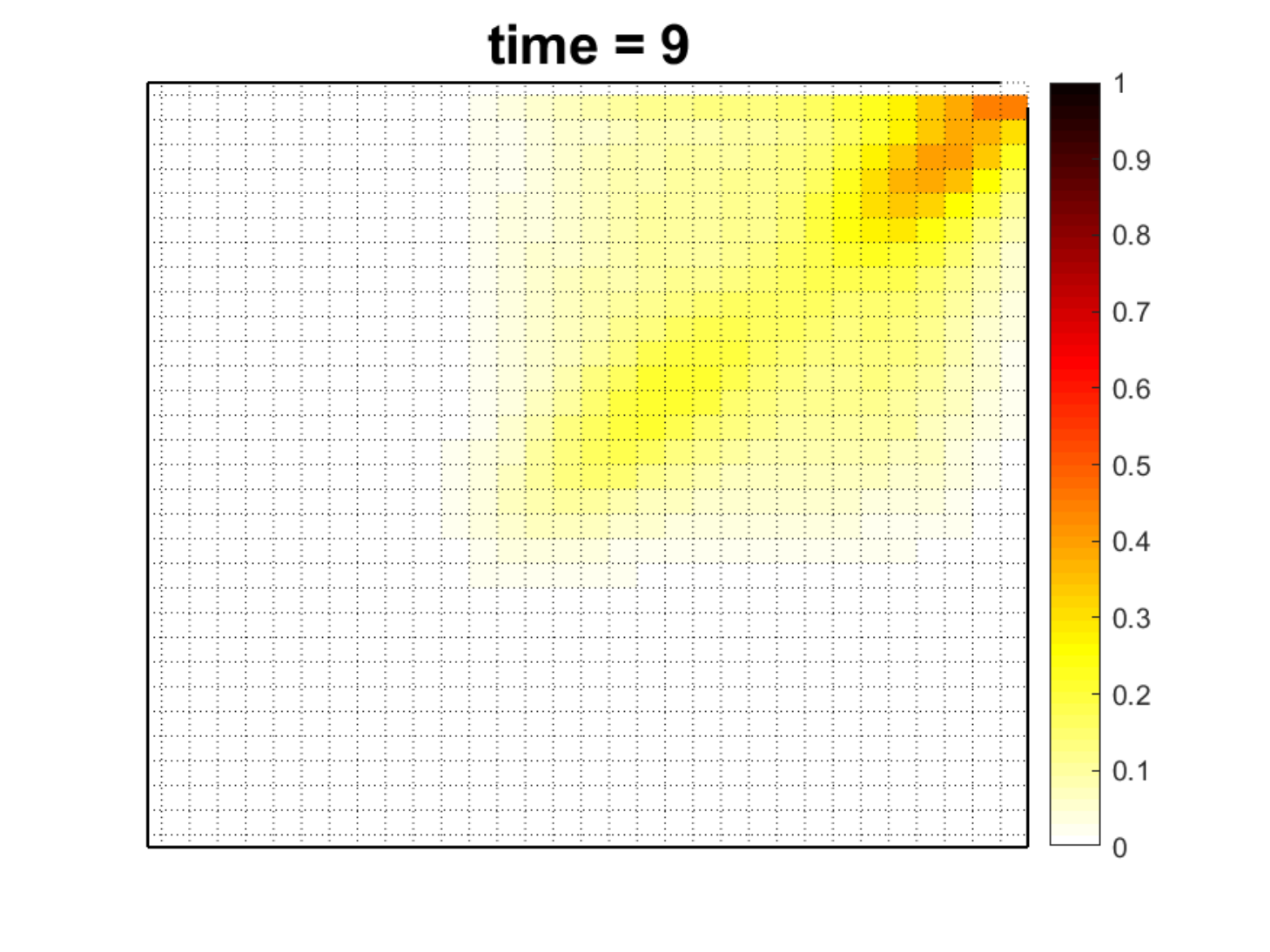}
\end{overpic}
\begin{overpic}[width=0.32\textwidth, grid=false]{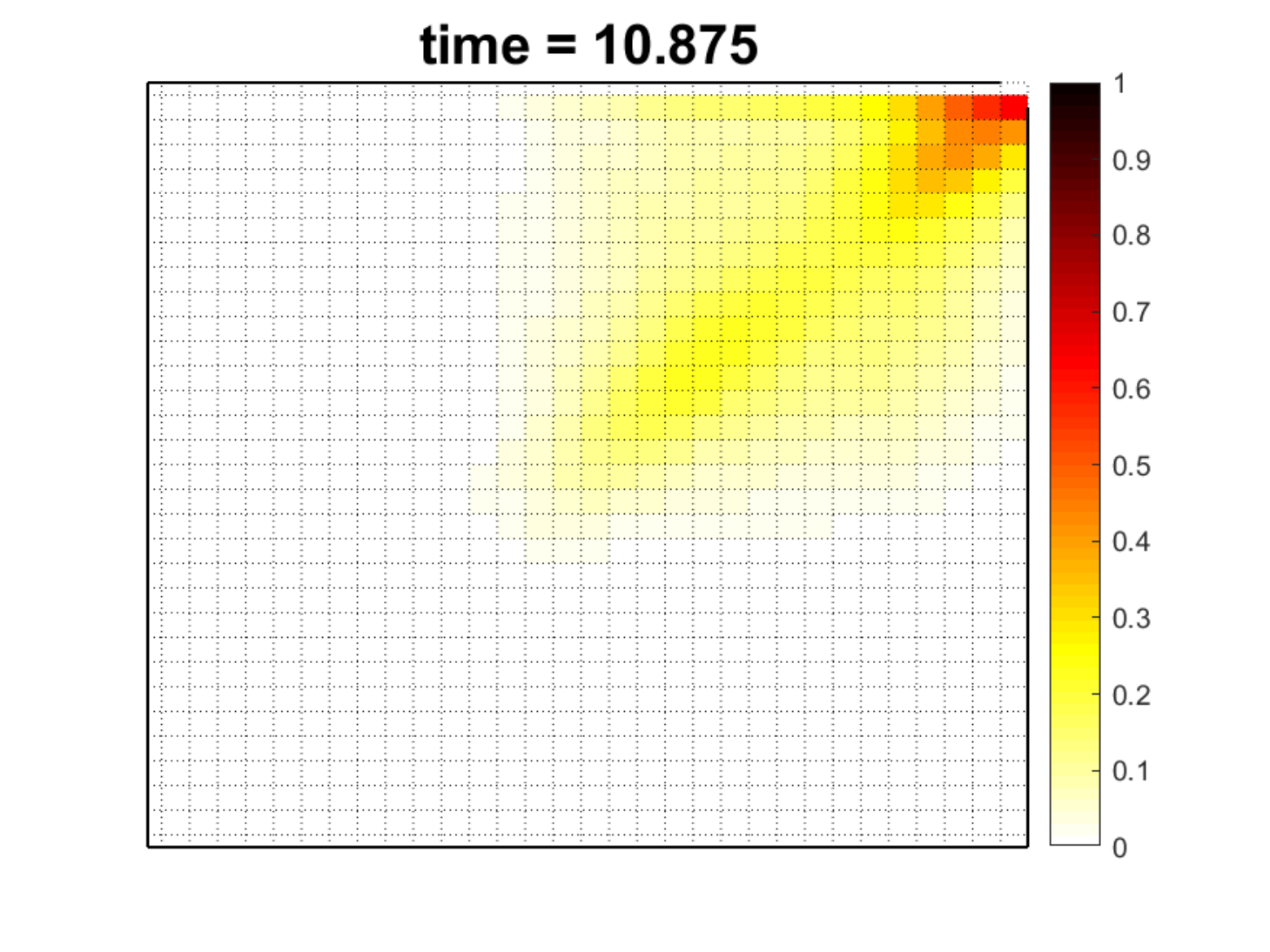}
\end{overpic}
\caption{Test 3: Evacuation process of 200 ants with initial direction 
$\theta_2$ and $\gamma = 0.1$: average fear level $q^\ast$ (top panel for each time) and people
density (bottom panel for each time) for $t = 0, 1.5, 3, 6, 9,10.875$ s.
The density plots show also the computational mesh.}\label{ant_evac} 
\end{figure}

For the numerical simulation, we choose $\Delta x = \Delta y=1$ mm
and $\Delta q=0.05$. The time step is set to $\Delta t = 0,0375$ s in order to satisfy condition
\eqref{eq:CFL}. We initially place the ants in a square with side 22 mm with uniform
density. See Fig.~\ref{ant_evac}, left-most panel on the second row.
We consider two initial conditions. Both of them have high fear level ($q = 1$) where 
the citronella is injected, but they differ in the rest of the domain. In condition 1 we
prescribe low panic ($q = 0.1$) everywhere else, while in condition 2 we assign 
$q = 0.65$ everywhere else. 
Fig.~\ref{ant_evac}, top left panel, shows the initial average fear level for condition 2.
From Ref.~\refcite{SHIWAKOTI20111433}, we could not guess
whether the repellent has some immediate effect on the entire domain and that is why
we opted for two initial conditions. Recall that the fear level corresponds to the walking speed. 
We set $\theta_2$ as initial walking direction.
Based on the results from test 2, we set $R = 1$ and try different values of the
contagion interaction strength $\gamma$.

Fig.~\ref{ant_evac} shows the evacuation process of the first 50 ants for initial condition 2
and $\gamma = 0.1$. We report the computed average fear level and the density at different
times. Around $t = 9$ s, we observe the formation of a high density region
close to the narrow exit, which gets worse as time passes.

\begin{figure}[ht]
\centering
\begin{overpic}[width=0.60\textwidth, grid=false]{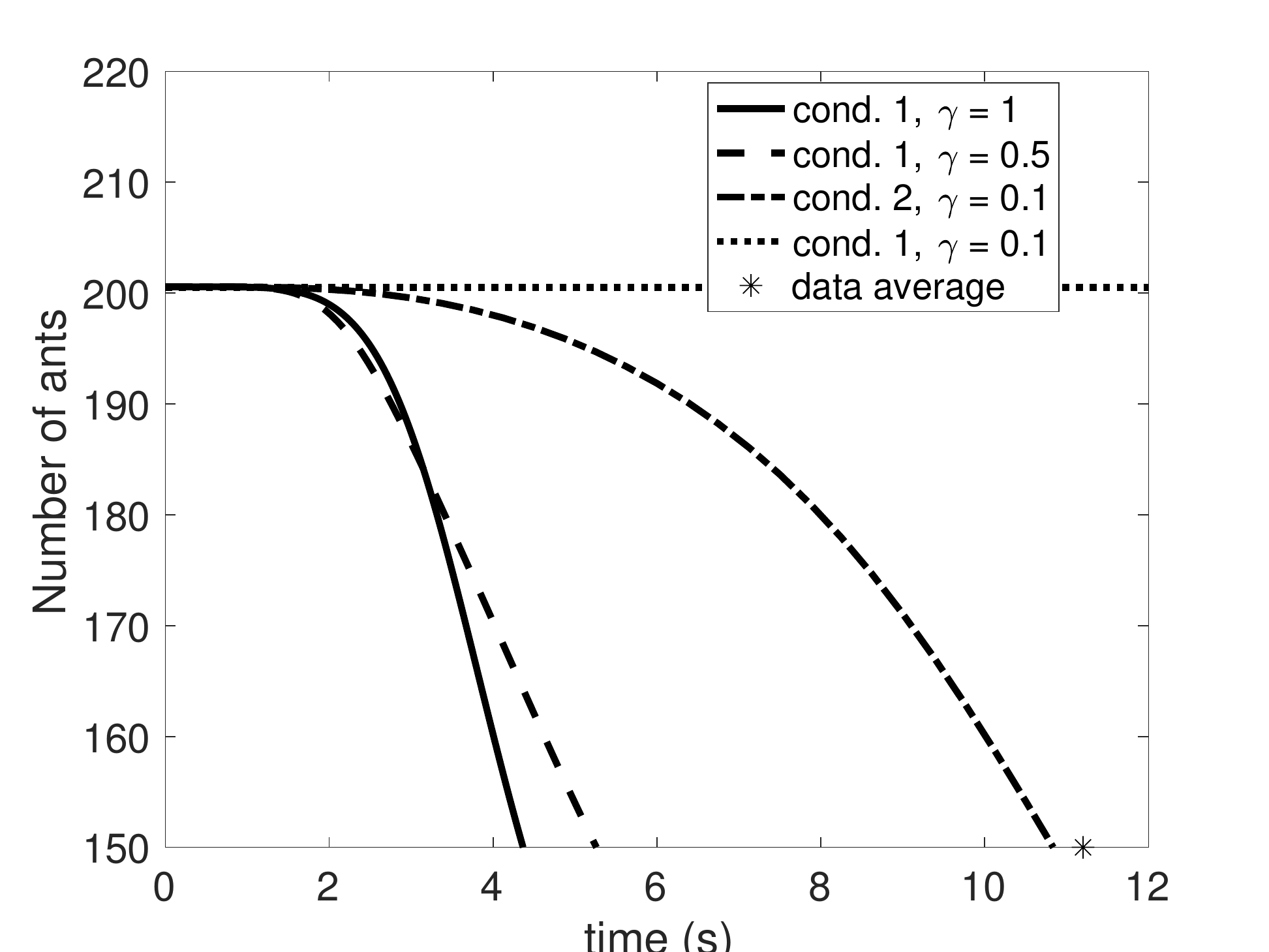}
\end{overpic}
\caption{Test 3: number of ants inside the room over time computed with two initial conditions and different
values of $\gamma$. The graph also reports the average time for the first 50 ants to leave
the room in the experiments.
}
\label{Population_change_test3}
\end{figure}

Figure~\ref{Population_change_test3} reports the number of ants inside the room 
over time computed with initial condition 2 and $\gamma = 0.1$ (i.e., same case as in Fig.~\ref{ant_evac}),
and initial condition 1 and $\gamma = 0.1, 0.5, 1$.
With condition 2 and $\gamma = 0.1$, we obtain that the first 50 ants leave the room in 10.7 s. This compares
favorably with the experimental data in Ref.~\refcite{SHIWAKOTI20111433} (11.2 s $\pm$ 2.6 s). With initial condition 1 instead, 
for no value of $\gamma$ we can match the experimental data. In addition,
notice the large difference in evacuation dynamics for the different values of $\gamma$.

\section{Conclusion}\label{sec:concl}

In this paper, we have utilized a kinetic-theoretic approach to model crowd dynamics in the presence of a propagating emotional contagion such as fear.
We have applied a Lie splitting scheme to simulate our model.
This scheme separates the overall task into an advection problem that accounts for the emotional contagion and a second problem that accounts for the physical interactions of pedestrians with the environment and one another.

We have completed three tests involving evacuation from a room.
One of our tests highlights how crowd dynamics depend on whether an emotional contagion is present or absent.
Indeed, it is known that social phenomena can modify interaction rules.
Our numerical results confirm that the propagation of an emotional contagion can significantly affect evacuation dynamics. 
We have shown that evacuation dynamics depend sensitively on two key model parameters: the interaction radius and the contagion interaction strength.
Finally, we have shown that our model can reproduce experimental data generated by distressed ants provided that we appropriately tune initial data and model parameters.

In the future, we plan to thoroughly validate our model using data generated by studies of the collective behavior of panicked animals.
Naturally there exist fewer experimental studies of collective human dynamics under emergency conditions.

\section*{Acknowledgements}
This work has been partially supported by the National Science Foundation through grant DMS-1816315 (WO) and DMS-1620384 (AQ).


\end{document}